\pdfoutput=1

\documentclass[12pt]{article}
\usepackage{float}
\usepackage{amsfonts}
\usepackage{graphicx,psfrag,epsf}
\usepackage{natbib}
\usepackage{bm}
\usepackage{url} 
\usepackage{amsthm}
\usepackage{tikz}
\usetikzlibrary{arrows.meta,calc,backgrounds,decorations.pathreplacing}
\usepackage{mathtools}
\usepackage{booktabs}
\usepackage{setspace}
\usepackage{enumitem}
\newcommand{\blind}{0}
\usepackage{algorithm}
\usepackage{algpseudocode}
\usepackage{xcolor}
\usepackage[
  colorlinks=true,
  linkcolor=blue,
  citecolor=blue,
  urlcolor=blue
]{hyperref}
\newtheorem{theorem}{Theorem}[section]
\newtheorem{assumption}{Assumption}[section]
\newtheorem{lemma}[theorem]{Lemma}
\newtheorem{proposition}[theorem]{Proposition}
\newtheorem{corollary}[theorem]{Corollary}

\newtheorem{remark}[theorem]{Remark}
\begin{document}

\def\spacingset#1{\renewcommand{\baselinestretch}%
{#1}\small\normalsize} \spacingset{1}

\if0\blind
{
  \title{\vspace{-1.0in}\bf When Do Sports Leagues Stop Being Competitive? Bayesian Detection of Competitive-Balance Eras from Match-Level Data}
  \author{Xinran Li\\
    Department of Statistics, University of Illinois Urbana-Champaign\\
    Maoran Xu\\
    Department of Statistics, Indiana University\\
    and \\
    Guanyu Hu\thanks{
    Xinran Li: \texttt{xinran28@illinois.edu}; Maoran Xu: \texttt{mx13@iu.edu};
    Guanyu Hu: \texttt{huguanyu@msu.edu}. Maoran Xu and Guanyu Hu are joint
    corresponding authors.}\\
    Department of Statistics and Probability, Michigan State University}
  \date{}
  \maketitle
  \vspace{-0.5in}
} \fi

\if1\blind
{
  \bigskip
  \bigskip
  \bigskip
  \begin{center}
    {\LARGE\bf When Do Sports Leagues Stop Being Competitive? Bayesian Detection of Competitive-Balance Eras from Match-Level Data}
\end{center}
  \medskip
} \fi

\bigskip

\begin{abstract}
Competitive balance shapes outcome uncertainty, fan interest, and league
regulation, yet standard measures summarize realized standings and cannot
separate structural ability differences from match-level randomness. We
develop a Bayesian framework that infers structural competitive balance
from longitudinal match outcomes. Team abilities are obtained by projecting
Gaussian auxiliary variables onto a graph-fused $\ell_1$ polytope whose
radius measures total pairwise ability variation. The projection places
positive probability on exact ties, so statistically indistinguishable
teams form explicit rank tiers. The radius equals realized ability
variation whenever the constraint binds, linking it to parity,
detectability, and structural inequality. An adaptive finite-regime model
lets competitive eras recur, emerge, and disappear while team hierarchies
remain season specific. Simulations show accurate recovery of regimes,
radii, abilities, and tiers. Applied to 25 English Premier League seasons,
the model identifies a persistent shift to a more stratified league
beginning in 2005/06, with 2010/11 the only later return to the compressed
regime. It also shows that a close title race can coexist with wide
league-wide separation and that head-to-head results can separate clubs
level on points. The framework gives leagues a model-based measure of
balance to inform debates on revenue sharing and financial regulation.
\end{abstract}

\noindent%
\vfill

\newpage
\spacingset{1.45} 

\section{Introduction}
\label{sec:introduction}

Rankings are used in many settings, including universities
\citep{usher2009global,goldstein1996league}, online platforms
\citep{hong2019patients,zhang2018dynamics}, forecasting
\citep{leitner2010forecasting,glickman1999parameter}, gene studies
\citep{boulesteix2009stability}, and recommendation systems
\citep{picoto2019uncovering,rendle2012bpr}. A ranking is easy to interpret,
but the comparisons behind it are often sparse, noisy, and heterogeneous.
Consequently, the inferential question is not only which item ranks higher,
but also whether the observed comparisons provide enough information to
distinguish their underlying abilities.
Probabilistic paired-comparison models provide a natural framework for this
problem. The Bradley--Terry--Luce model
\citep{Bradley1952,Luce1959} and its extensions
\citep{davidson1970extending,caron2012efficient,karle2023dynamic} form a
standard foundation for modeling relative ability. Classical formulations
treat abilities as fixed, while dynamic extensions allow strengths to vary
over time through smoothing, state-space models, and stochastic processes
\citep{glickman1999parameter,cattelan2013dynamic,tian2024spectral}.
Substantial work has also addressed scalable optimization
\citep{hunter2004mm,maystre2015fast}, Bayesian inference
\citep{mollica2021bayesian}, spectral estimation
\citep{tian2024spectral}, and high-dimensional theory \citep{Fang2026}.
A common structural feature nevertheless remains: latent abilities are
typically assigned continuous distributions on unconstrained spaces and are
therefore almost surely distinct. Such models naturally produce a strict
ordering even when the data support only a coarser hierarchy.

This issue is particularly important when similarity among competitors is
itself scientifically meaningful. Competitive balance in professional sports
is one such setting. A competitively balanced league is one in which teams
have comparable underlying strength, making outcomes less predictable and
potentially sustaining fan interest and league value
\citep{Vrooman1995,FortQuirk1995,Szymanski2001,Szymanski2003,
Zimbalist2002,FortMaxcy2003,Humphreys2002a}. From this perspective, the
central quantity is not simply the dispersion of observed standings, but the
degree of inequality in the latent abilities that generate those standings.
A model should therefore distinguish persistent structural differences from
game-level randomness and should allow genuine statistical parity to emerge
rather than rule it out by construction.

Existing measures of competitive balance largely operate on realized outcomes.
Common examples include Herfindahl--Hirschman-type indices
\citep{Owen2007}, Gini and dispersion measures
\citep{utt2002pitfalls}, and football-specific indices
\citep{Manasis2013,ManasisNtzoufras2014,Manasis2022}. These measures are
useful descriptively, but they do not provide a generative model that
separates latent ability inequality from random variation. Other work treats
competitive strength as dynamic
\citep{Cairns1987,Schmidt2001,Koop2004,PennBerridge2019,Plumley2018},
including Bayesian and state-space approaches for evolving team abilities
\citep{baio2010bayesian,Lopez2018}. These formulations capture temporal
variation but generally retain continuously distributed latent strengths, so
exact parity and rank tiers are not explicit parts of the latent structure.

The ranking literature has addressed related questions through partial
orders, clustered rankings, shared consensus structures, and sparse
difference models
\citep{marden1996analyzing,deng2014bayesian,li2022bayesian,
murphy2003mixtures,busse2007cluster,dunson2008bayesian,
pearce2025bayesian,pearce2025modeling,piancastelli2025clustered}.
The ranking lasso, for example, groups Bradley--Terry abilities by shrinking
selected pairwise differences exactly to zero
\citep{masarotto2012ranking}. These approaches show that ranking uncertainty
and latent grouping are closely related, but they do not directly provide a
single structural parameter that simultaneously measures competitive
balance, induces exact rank tiers, and evolves across historical periods.

We address this gap by restricting each period's latent ability vector to a graph-fused $\ell_1$ polytope, whose radius controls total variation across the comparison graph. Although $\ell_1$ constraints appear in Bayesian $\ell_1$-ball models \citep{xu2023bayesian} and fused regularization \citep{tibshirani2005sparsity,she2010sparse,ma2017concave}, we instead project an unconstrained Gaussian auxiliary vector onto the polytope. Because the projection can land on lower-dimensional faces, exact zero ability contrasts have positive probability and rank tiers arise directly from the generative mechanism.
The radius also measures competitive imbalance. When the constraint binds, it equals the realized total of pairwise ability gaps, making it a property of the teams rather than a season's points table. Larger radii imply greater stratification; at the parity boundary, all teams belong to one tier. The radius also determines how many matches are needed to detect a departure from parity, bounds the signal-to-noise ratio of \citet{Lopez2018}, and yields a structural Gini coefficient. It thus places these measures on a common scale with posterior uncertainty.
To allow the competitive environment to change, we let the radius follow a finite-regime process. Existing regimes can recur, new ones can appear, and unused ones can disappear, with a birth penalty inspired by mixture-of-finite-mixtures models \citep{miller2018mixture}. This estimates the number of competitive eras without fixing it in advance.
Regimes describe competitive environments, not team identities. Two seasons in the same regime share a radius but generally have different abilities: $s_t=s_{t'}$ implies $r_t=r_{t'}$, whereas typically $\theta_t\neq\theta_{t'}$. A league can therefore remain equally stratified while its leading clubs change. In the Premier League, 2015/16, won by Leicester City, and 2019/20, won by Liverpool, belong to the same high-dispersion regime. Grouping seasons by balance rather than rankings lets the model address league structure rather than particular clubs.

This paper makes three contributions to the study of competitive balance. First, we propose a model-based measure of structural balance. When the constraint binds, the polytope radius equals the realized dispersion of team abilities, so it describes the teams rather than a season’s points table. Its posterior uncertainty helps distinguish genuine changes in balance from unusual results. Near parity, the same radius links the signal-to-noise ratio of \citet{Lopez2018}, a structural Gini coefficient, and the number of matches needed to detect a departure from parity.
Second, the projection prior assigns positive probability to exact ties. Teams that the data cannot distinguish share a rank tier, with tiers estimated from the model rather than imposed afterward. This helps separate genuine ability gaps from single-season noise.
Third, we infer competitive eras without fixing their number in advance. Seasons in the same era share a level of balance but not necessarily the same leading clubs. The model therefore groups seasons by league stratification rather than by winners, and changes in league structure are estimated within a probability model rather than judged from a time series of indices.

Applied to $25$ seasons of the English Premier League, the framework
finds a persistent shift to a more stratified league beginning in
2005/06, with 2010/11 the only later return to the compressed regime.
It shows that a close title race, as in 2011/12, can coexist with wide
separation across the rest of the table, that home advantage evolves
separately from balance, and that head-to-head results can separate
clubs that finished level on points, as Arsenal and Liverpool did in
2006/07. Leading clubs change throughout the stratified era, from
Manchester United and Chelsea to Manchester City, Leicester City, and
Liverpool, while the league's overall level of separation stays the
same. A second application to the National Basketball Association
(Supplementary Material, Appendix~\ref{app:nba_application}), a closed league with very different parity mechanisms, shows recurrent
switching between regimes rather than a single lasting shift. These
results move the assessment of competitive balance from retrospective
summaries of points toward inference on parity, hierarchy, and
structural change, which is the evidence leagues need when weighing
revenue sharing, financial regulation, or format reform.

\section{Motivating Data}
\label{sec:data}

To motivate our methodological development, we consider match-level
outcomes from $25$ consecutive seasons of the English Premier League (EPL),
from 1995/96 through 2019/20 \citep{Basini2023}. The EPL provides a natural
setting for studying structural competitive balance. Each season follows
a complete double round-robin schedule, with every team playing every
other team once at home and once away, yielding a balanced and fully
connected comparison network. Although the league size remains fixed,
team membership changes through promotion and relegation. The study
period also spans substantial changes in broadcasting revenue, club
ownership, and financial resources that have motivated longstanding
debates about competitive inequality \citep{Basini2023}.
Table~\ref{tab:epl_summary} summarizes the main characteristics of the data.

\begin{table}[htbp]
\centering
\caption{Summary of the English Premier League data, 1995/96 to 2019/20.}
\label{tab:epl_summary}
\begin{tabular}{@{}ll@{}}
\toprule
Characteristic & Value \\
\midrule
Seasons & $25$ (1995/96--2019/20) \\
Teams per season & $20$ \\
Matches per season & $380$ (double round robin) \\
Total matches & $9{,}500$ \\
Distinct clubs over the study period & $47$ \\
Outcome categories & home win / draw / home loss \\
Overall proportions (home win / draw / home loss) &
  $46.5\%$ / $25.6\%$ / $27.9\%$ \\
Seasonal draw proportion (range) & $18.7\%$--$31.3\%$ \\
\bottomrule
\end{tabular}
\end{table}

\textbf{Data structure.}
For each season, we represent the match results as a complete directed
outcome network on the $20$ participating teams, with no self-loops.
The directed edge from team $i$ to team $j$ records the outcome of the
match in which $i$ hosted $j$, classified as a home win, draw, or home
loss. This representation preserves opponent-specific outcomes and home
advantage information that would otherwise be lost when results are
aggregated into final standings. Draws account for $25.6\%$ of all matches
and provide important information about competitive similarity. We
therefore retain them as a separate ordered outcome category rather than
combining them with wins or losses.

\begin{figure}[h]
\centering
\includegraphics[width=0.9\textwidth,height=0.63\textheight,keepaspectratio]{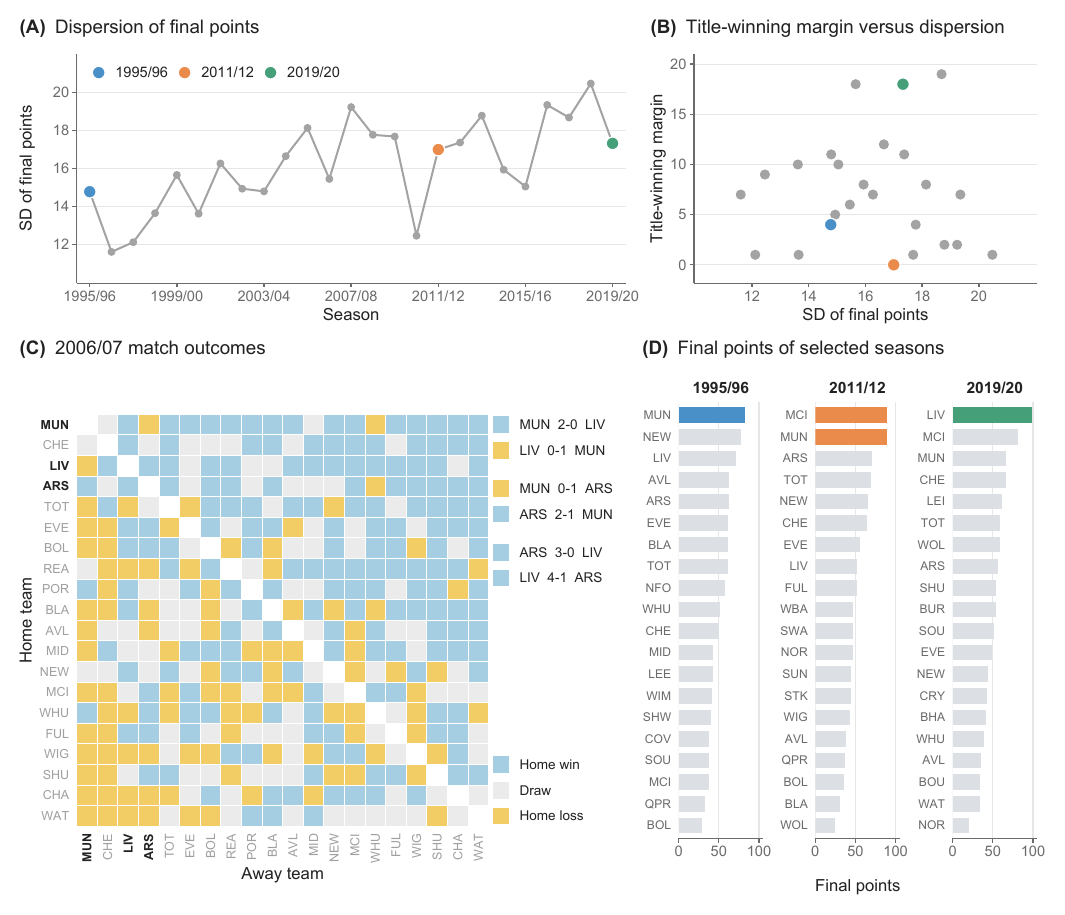}
\caption{Descriptive features of the English Premier League data, 1995/96--2019/20.}
\label{fig:epl_motivation}
\end{figure}

\textbf{Why the points table is not enough.}
Figure~\ref{fig:epl_motivation} illustrates the limits of assessing competitive balance from final standings alone. Panel~(A) shows variation in the standard deviation of final points across 25  seasons, but this may reflect either underlying team abilities or random match outcomes. Panel~(B) shows that league-wide dispersion and championship closeness need not coincide. In 2011/12, Manchester City and Manchester United finished level on 89  points, while third-placed Arsenal were 19  points behind. A close title race therefore coexisted with substantial league-wide inequality.
Panel~(C) shows that equal point totals can hide informative match results. In 2006/07, Arsenal and Liverpool both finished with 68  points, but Arsenal beat Manchester United twice, whereas Liverpool lost both meetings. Panel~(D) compares the points distributions in 1995/96, 2011/12, and 2019/20, highlighting differences in the separation between leading clubs and the rest. Final points, title margins, and individual results capture different aspects of competition; none alone identifies structural inequality in team abilities.

Together, these observations motivate three questions that
guide our framework and subsequent case study:
\vspace{-9pt}
\begin{enumerate}[label=({Q}{{\arabic*}})]
    \item \textit{Has structural competitive balance changed over time,
    and if so, when?} The variation in points dispersion shown in
    Panel~(A) motivates investigating whether the $25$ seasons contain
    persistent changes in underlying competitive inequality. We examine
    whether these seasons can be grouped into a small number of recurring
    competitive regimes without fixing their number in advance, and
    whether the timing of structural changes corresponds to developments
    in English football.\vspace{-9pt}

    \item \textit{Which teams are statistically distinguishable within
    a season?} The opponent-specific outcomes in Panel~(C) illustrate
    that similar aggregate performance can conceal different patterns
    of competition. Rather than imposing a strict ordering of all $20$
    teams, we investigate whether match outcomes support exact latent
    rank tiers and whether these tiers reveal competitive relationships
    that final standings cannot capture.\vspace{-9pt}

    \item \textit{Is structural competitive balance distinct from
    title-race closeness and home advantage?} Panels~(B) and~(D)
    illustrate that championship competitiveness and overall league
    inequality need not coincide. We investigate whether league-wide
    dispersion in latent abilities evolves separately from the
    closeness of the title race and season-specific home effects.
\end{enumerate}
\vspace{-9pt}

These questions speak directly to sports economics. The
uncertainty-of-outcome hypothesis \citep{rottenberg1956baseball}
links fan demand and league value to uncertainty about sporting
contests, motivating mechanisms such as revenue sharing, salary caps,
draft systems, and financial fair-play regulations
\citep{FortQuirk1995,Vrooman1995,Szymanski2003,Zimbalist2002}.
(Q1) asks whether changes in broadcasting revenue, club ownership, or
resource distribution coincide with persistent shifts in structural
competitive balance rather than random fluctuations in descriptive
indices \citep{utt2002pitfalls,Lopez2018}. (Q2) identifies
statistically distinguishable competitive groups, including a possible
``league within a league'' of persistently dominant teams
\citep{PennBerridge2019,Plumley2018}. (Q3) distinguishes match,
seasonal, and championship uncertainty
\citep{Cairns1987,Szymanski2003}: a close title race may coexist with
a stratified league, with different implications for attendance,
broadcasting demand, and incentives outside the leading group.

Conventional indices based on points dispersion or concentration
\citep{Humphreys2002a,Owen2007,Manasis2013} summarize realized outcomes
but cannot separate persistent ability differences from match-level
randomness or quantify uncertainty about whether clubs are genuinely
distinguishable. Figure~\ref{fig:epl_motivation} therefore motivates a
probabilistic framework that uses match-level paired comparisons to
estimate structural competitive balance, identify recurring regimes,
and assign clubs with similar latent abilities to exact rank tiers.
Section~\ref{sec:applications} revisits (Q1)--(Q3) using the EPL data;
a National Basketball Association analysis in the Supplementary
Material (Appendix~\ref{app:nba_application}) assesses broader applicability.

\section{A Geometric Framework for Structural Competitive Balance}
\label{sec:geometric_framework}

\subsection{Projection-induced paired-comparison model}
\label{subsec:projection_model}

We consider longitudinal paired-comparison data over periods
$t=1,\ldots,T$. Let $\mathcal I=\{1,\ldots,N\}$ denote the competitors and
$\mathcal D_t=\{(i_{tg},j_{tg},h_{tg},y_{tg}):g=1,\ldots,G_t\}$, where $h_{tg}$ is a contextual covariate such as home advantage and
$y_{tg}\in\{-1,0,1\}$ denotes a loss, draw, or win for competitor $i_{tg}$.
Let $\theta_t=(\theta_{t1},\ldots,\theta_{tN})^\top$ denote latent abilities,
subject to $\mathbf1^\top\theta_t=0$, and let $\eta_t$ denote a
period-specific contextual effect. Writing
$x_{tg}=e_{i_{tg}}-e_{j_{tg}}$, define
$\Delta_{tg}=x_{tg}^\top\theta_t+\eta_t h_{tg}$.

We use an ordered Bradley--Terry likelihood
\citep{Bradley1952,Luce1959,davidson1970extending}:
\begin{equation}
\Pr(y_{tg}\le c\mid\theta_t,\eta_t,\kappa)=\sigma(\kappa_{c+2}-\Delta_{tg}),
\qquad c\in\{-1,0\},
\label{eq:ordered_bt_likelihood}
\end{equation}
where $\sigma(u)=\{1+\exp(-u)\}^{-1}$. The binary Bradley--Terry model is
recovered by removing the draw category.

When $h_{tg}$ is constant across comparisons, $(\eta_t,\kappa)$ is
identified only up to a common location shift. We therefore impose
$\kappa_1=-\kappa_2$ and $\kappa_2>0$, so that the draw region is symmetric
about zero and $\eta_t$ carries the contextual location shift.

Let $E_t$ denote the multiset of scheduled comparisons and
$A_t\in\mathbb R^{|E_t|\times N}$ its oriented incidence matrix. The graph
total variation is
$\|A_t\theta_t\|_1
=\sum_{(i,j)\in E_t}|\theta_{ti}-\theta_{tj}|$.
Repeated meetings may either be counted with multiplicity or counted once,
provided the same convention is used for both $A_t$ and $r_t$.

For $r_t>0$, define
\begin{equation}
\mathcal K_t(r_t)=\{\vartheta\in\mathbb R^N:\|A_t\vartheta\|_1\le r_t,\ \mathbf1^\top\vartheta=0\}.
\label{eq:feasible_polytope}
\end{equation}
The radius $r_t$ therefore controls the amount of latent ability variation
permitted by the competitive environment.

Following the geometric idea of Bayesian $\ell_1$-ball priors
\citep{xu2023bayesian} and related fused regularization
\citep{tibshirani2005sparsity}, we introduce
$\beta_t\sim N(0,\sigma_\beta^2I_N)$ and set
\begin{equation}
\theta_t=\Pi_{\mathcal K_t(r_t)}(\beta_t)=\arg\min_{\vartheta\in\mathcal K_t(r_t)}\|\vartheta-\beta_t\|_2^2.
\label{eq:projection_map}
\end{equation}
Projection onto lower-dimensional faces of
$\mathcal K_t(r_t)$ can set graph contrasts exactly to zero, so exact ability
ties and rank tiers arise directly from the latent model.

\begin{theorem}[Posterior propriety]
\label{thm:posterior_propriety}
Suppose all primitive unknowns are assigned a proper joint prior. Then, for
every finite comparison design $\{\mathcal D_t\}_{t=1}^T$, the posterior
distribution is proper.
\end{theorem}

The bounded paired-comparison likelihood ensures posterior propriety for
a fully specified model with proper priors. Because the allocation step in
Section~\ref{sec:dynamics} is not derived from an established joint regime
prior, this theorem does not assert propriety or exactness of the
algorithm-induced regime distribution.

\begin{proposition}[Identifiability of the linear predictor and thresholds]
\label{prop:identifiability}
Fix a period and suppress $t$. Suppose the comparison graph is connected and
there are no $d\in\mathbb R^N$ and $c\neq0$ satisfying
$\mathbf1^\top d=0$ and
$x_g^\top d=-c h_g$ for every $g=1,\ldots,G$. Then
$(\theta,\eta)\mapsto(\Delta_g)_{g=1}^G$ is injective on
$\{\theta:\mathbf1^\top\theta=0\}\times\mathbb R$.

If additionally $\kappa_1=-\kappa_2$ and $\kappa_2>0$, then the map from
$(\theta,\eta,\kappa_2)$ to the collection of three-category probabilities in
\eqref{eq:ordered_bt_likelihood} is injective. Without this anchoring, when
$h_g\equiv h$, the model is identified only up to
$(\theta,\eta+c,\kappa+ch\mathbf1)$.
\end{proposition}

This anchoring is needed in both applications because every comparison has a
designated home entity, so $h_{tg}\equiv1$; the non-collinearity condition then
holds because each season contains a home-and-away pair, for which
$x_{g'}=-x_g$ forces $c=0$.

\begin{proposition}[Positive probability of exact ties]
\label{prop:exact_ties}
Fix $r_t>0$ and suppose the comparison graph is connected. Let $F$ be a
nonempty face of $\mathcal K_t(r_t)$ on which
$(A_t\vartheta)_e=0$ for some edge $e$. Then
$\mathcal R_F=\{b:\Pi_{\mathcal K_t(r_t)}(b)\in\operatorname{relint}(F)\}$
has positive Lebesgue measure and hence
$\Pr\{(A_t\theta_t)_e=0\mid r_t\}>0$. Such a face exists for $N\ge3$, so the projection prior assigns positive
probability to exact latent ties for every $r_t>0$.
\end{proposition}

\subsection{Geometry and interpretation of the structural radius}
\label{subsec:radius_geometry}

Let
$H=I_N-N^{-1}\mathbf1\mathbf1^\top$ denote the projection onto the zero-sum
subspace, write
$\theta_t(r)=\Pi_{\mathcal K_t(r)}(\beta_t)$, and define
$Z_t=\|A_tH\beta_t\|_1$.

\begin{theorem}[Radius saturation]
\label{thm:radius_saturation}
For every $r\ge0$,
\begin{equation}
\|A_t\theta_t(r)\|_1=\min\{r,Z_t\}.
\label{eq:radius_saturation}
\end{equation}
Equivalently, $\theta_t(r)=H\beta_t$ when $r\ge Z_t$, whereas
$\|A_t\theta_t(r)\|_1=r$ when $r<Z_t$.
\end{theorem}

Thus, whenever the geometric constraint binds, the radius is not merely an
upper bound: it equals the realized structural variation.

\begin{corollary}[Monotone structural ordering]
\label{cor:radius_monotonicity}
For $0\le r_1<r_2$,
$\|A_t\theta_t(r_1)\|_1\le\|A_t\theta_t(r_2)\|_1$,
with strict inequality whenever $r_1<Z_t$. Moreover, for
$M_t(r)=\mathbb E\{\|A_t\theta_t(r)\|_1\}$,
\begin{equation}
M_t(r)=\mathbb E\{\min(r,Z_t)\}=\int_0^r\Pr(Z_t>u)\,du,
\qquad
M_t'(r)=\Pr(Z_t>r)>0.
\label{eq:mean_structural_variation}
\end{equation}
Hence $M_t(r)$ is strictly increasing for every finite $r$.
\end{corollary}

\begin{proposition}[Piecewise-affine projection path]
\label{prop:piecewise_radius_path}
Fix $t$, $\beta_t$, and $R<\infty$. The path
$r\mapsto\theta_t(r)$ is continuous and piecewise affine on $[0,R]$.
Thus there exists
$0=b_0<b_1<\cdots<b_Q=R$ such that $\theta_t(r)$ is affine on each
$(b_{q-1},b_q)$. Every pairwise contrast
$\theta_{ti}(r)-\theta_{tj}(r)$ is therefore affine within each segment, so
the rank-tier partition can change only at finitely many radius values on
every compact interval.
\end{proposition}

\begin{lemma}[Two-block extreme points]
\label{lem:two_block_extreme}
Assume the comparison graph is complete and each dyad is counted once, so
$\|A\vartheta\|_1=\sum_{i<j}|\vartheta_i-\vartheta_j|$.
A vector $\theta\in\mathcal K_1$ is an extreme point if and only if it takes
exactly two distinct values. Equivalently, for some nonempty proper
$S\subset\mathcal I$,
$\theta=\pm(\mathbf1_S-|S|N^{-1}\mathbf1)/\{|S|(N-|S|)\}$.
In particular,
$\sup_{\theta\in\mathcal K_1}\theta_i=1/N$, attained at $S=\{i\}$.
\end{lemma}

\subsection{Structural radius and calibration to existing balance measures}
\label{subsec:radius_calibration}

The radius provides a structural capacity for ability inequality. A
reciprocal-radius summary $\mathbb E(r_t^{-1}\mid\mathcal D)$ is finite
only under conditions on the radius prior near zero
(Lemma~\ref{lem:index_exists} in Appendix~\ref{app:proofs_geometry}),
which the exponential prior used here does not satisfy. We therefore use
radius posterior summaries and the finite realized-variation functional
$M_t(r)=\mathbb E\{\min(r,Z_t)\}$ rather than a reciprocal index.

For comparisons across systems with different graph sizes or densities,
we report the normalized radius,
\begin{equation}
\widetilde r_t=r_t/|E_t|,
\qquad
\overline V_t(r_t)=\mathbb E\{\min(r_t,Z_t)\}/|E_t|.
\label{eq:normalized_competitiveness}
\end{equation}
The first quantity measures normalized \emph{capacity} for pairwise ability
variation, whereas the second is the expected realized mean absolute graph
contrast under the Gaussian auxiliary prior. These interpretations coincide
only when the radius constraint is active. Cross-league comparisons require
a consistent convention for repeated dyads and graph weights.

Suppress $t$ and write
$\mathcal K_r=\{\theta:\|A\theta\|_1\le r,\,
\mathbf1^\top\theta=0\}$. The following local results use the binary
Bradley--Terry model
$\Pr(Y_{ij}=1\mid\theta)=\sigma(\theta_i-\theta_j)$, with analogous local
conclusions for the ordered model.

\begin{theorem}[Single-block parity limit]
\label{thm:single_block_limit}
Suppose the comparison graph is connected and $\theta(r)\in\mathcal K_r$.
As $r\downarrow0$, $\theta(r)\to0$, and hence
$\Pr(Y_{ij}=1\mid\theta(r))\to1/2$ for every compared pair. Moreover, there
exists $L_A<\infty$ depending only on the graph such that
$\|\theta(r)\|_\infty\le L_Ar$.

For the ordered model, the limiting cell probabilities are those of
\eqref{eq:ordered_bt_likelihood} with $\Delta_{g}=\eta h_g$. They reduce to threshold-only probabilities when $h_g=0$ or $\eta=0$.
At $r=0$, $\theta_1=\cdots=\theta_N=0$, so all competitors belong to one
latent rank tier.
\end{theorem}

For entity $i$, define its expected win probability against the field and its
maximum departure from parity by
\begin{equation}
w_i(\theta)=\frac{1}{N-1}\sum_{j\ne i}\sigma(\theta_i-\theta_j),
\qquad
\delta_i(r)=\sup_{\theta\in\mathcal K_r}\bigl|w_i(\theta)-\tfrac12\bigr|.
\label{eq:structural_drift}
\end{equation}

\begin{lemma}[Local linearity of structural drift]
\label{lem:linear_drift}
Assume the graph is connected and the schedule is complete. Let
$\varrho_i=\sup_{\theta\in\mathcal K_1}\theta_i$. Then, as $r\downarrow0$,
$\delta_i(r)=c_{A,i}r+O(r^3)$ with $c_{A,i}=N\varrho_i/\{4(N-1)\}$.
For the complete graph,
$\varrho_i=1/N$, so
$c_{A,i}=\{4(N-1)\}^{-1}$ and
$\delta_i(r)=r/\{4(N-1)\}+O(r^3)$.
\end{lemma}

The preceding result determines how rapidly the amount of data required to
detect departure from parity increases as $r$ approaches zero.

\begin{theorem}[Inverse-square detection law]
\label{thm:inverse_square}
Fix entity $i$ and a complete symmetric schedule. Suppose $i$ plays $m$
conditionally independent comparisons, each won with probability
$w_i(\theta)$, and let $W_m$ denote its number of wins. Consider the
level-$\alpha$ test that rejects parity when
$|W_m-m/2|>z_{1-\alpha/2}\sqrt m/2$. For type-II error rate $\nu\in(0,1)$, assume $1-\nu>\alpha$
and let $d_{\alpha,\nu}>0$ solve
$\Phi(d-z_{1-\alpha/2})+\Phi(-d-z_{1-\alpha/2})=1-\nu$. Define
$m_i(r;\alpha,\nu)=\min\{m\ge1:\sup_{\theta\in\mathcal K_r}\Pr_\theta(\mathrm{reject})\ge1-\nu\}$.
Then, as $r\downarrow0$,
\begin{equation}
m_i(r;\alpha,\nu)=\frac{d_{\alpha,\nu}^2}{4c_{A,i}^2}r^{-2}\{1+o(1)\}.
\label{eq:inverse_square_detection}
\end{equation}
For the complete graph, the multiplicative constant is
$4(N-1)^2d_{\alpha,\nu}^2$. The familiar approximation
$d_{\alpha,\nu}\approx z_{1-\alpha/2}+z_{1-\nu}$
neglects the lower tail of the two-sided power function.
\end{theorem}

\begin{remark}[Why the horizon is fixed]
\label{rem:sequential}
A sequential rule of the form
$T_i^*=\inf\{m\ge1:|S_m|>c_\alpha\sqrt m\}$ does not provide valid
infinite-horizon type-I error control. By the law of the iterated logarithm,
$\limsup_m|S_m|/\sqrt m=\infty$ almost surely under parity. An always-valid
boundary of order $\sqrt{m\log\log m}$ or a mixture boundary restores
sequential validity, but introduces an additional logarithmic factor. We use
the fixed-horizon formulation because it matches the seasonal design and
yields the explicit rate in \eqref{eq:inverse_square_detection}.
\end{remark}

\begin{remark}[Regime of validity of the local expansions]
\label{rem:local_regime}
Lemma~\ref{lem:linear_drift}, Theorem~\ref{thm:inverse_square}, and
Theorem~\ref{thm:gini_law} are local results as $r\downarrow0$.
For a complete graph, the largest permitted contrast is of order $r/N$, so
the relevant empirical scale is
$\widetilde r_t=r_t/|E_t|$ rather than $r_t$ itself.

In the simulations with $N=20$ and
$r\in\{100,200,400\}$, the normalized radii are
$\widetilde r\in\{0.53,1.05,2.11\}$. For the two-tier configurations used
there, the linear Gini approximation overstates
$G_{\mathrm{struct}}$ by approximately $8\%$, $31\%$, and $108\%$,
respectively. The local expansions should therefore be interpreted near the
parity boundary rather than as numerical approximations at the radii observed
in the applications. In contrast, Theorem~\ref{thm:radius_saturation},
Corollary~\ref{cor:radius_monotonicity}, and
Proposition~\ref{prop:snr_bound} hold for all $r$.
\end{remark}

\subsection{Signal-to-noise and inequality functionals}
\label{subsec:snr_gini}

Following \citet{Lopez2018}, define
$\operatorname{SNR}(\theta)
=\operatorname{Var}(\theta)/\operatorname{Var}(\xi)$, where logistic noise
has $\operatorname{Var}(\xi)=\pi^2/3$ and, under
$\mathbf1^\top\theta=0$,
$\operatorname{Var}(\theta)=N^{-1}\sum_{i=1}^N\theta_i^2$.

\begin{proposition}[Radius-bounded signal-to-noise ratio]
\label{prop:snr_bound}
If the comparison graph is complete and $\theta\in\mathcal K_r$, then
\begin{equation}
\operatorname{Var}(\theta)\le\frac{r^2}{N^2(N-1)},
\qquad
\operatorname{SNR}(\theta)\le\frac{3r^2}{\pi^2N^2(N-1)}.
\label{eq:snr_bound}
\end{equation}
The bound is attained, up to permutation and sign, by a two-block
configuration with block sizes $1$ and $N-1$.

For a general connected graph,
$C_A=\sup\{\|\vartheta\|_2/(\sqrt N\|A\vartheta\|_1):\vartheta\ne0,\ \mathbf1^\top\vartheta=0\}<\infty$,
and $\operatorname{Var}(\theta)\le C_A^2r^2$ and
$\operatorname{SNR}(\theta)\le3C_A^2r^2/\pi^2$; for the complete graph,
$C_A=\{N\sqrt{N-1}\}^{-1}$.
\end{proposition}

Using $w_i(\theta)$ from \eqref{eq:structural_drift}, define
\begin{equation}
G_{\mathrm{struct}}(\theta)=\frac{1}{2N^2\bar w}\sum_{i=1}^N\sum_{j=1}^N|w_i(\theta)-w_j(\theta)|,
\qquad
\bar w=\frac1N\sum_iw_i(\theta).
\label{eq:structural_gini}
\end{equation}
For the complete comparison system, $\bar w=1/2$ exactly because
$\sigma(u)+\sigma(-u)=1$.

\begin{theorem}[Local geometric Gini law]
\label{thm:gini_law}
Assume a complete schedule with each dyad counted once. Uniformly over
$\theta\in\mathcal K_r$ as $r\downarrow0$,
\begin{equation}
G_{\mathrm{struct}}(\theta)=\frac{1}{2N(N-1)}\sum_{i<j}|\theta_i-\theta_j|+O(r^3)
\le\frac{r}{2N(N-1)}+O(r^3).
\label{eq:gini_radius_bound}
\end{equation}
Thus structural inequality is locally linear in the geometric radius.
\end{theorem}

These results put several notions of competitive balance on a common
geometric scale. As $r\downarrow0$,
$\theta\to0$,
$m_i(r;\alpha,\nu)\asymp r^{-2}$,
$\operatorname{SNR}(\theta)=O(r^2)$, and, under a complete symmetric
schedule, $G_{\mathrm{struct}}(\theta)=O(r)$. At the same time,
Theorem~\ref{thm:radius_saturation} gives $r$ an exact finite-sample
interpretation whenever the constraint is active. Thus the structural radius
links latent parity, rank tiers, statistical detectability, and structural
inequality within one probabilistic construction.

\section[Dynamic Competitiveness via Adaptive Finite Regimes]{Dynamic Competitiveness via Adaptive\\ Finite Regimes}
\label{sec:dynamics}

The geometric framework of Section~\ref{sec:geometric_framework}
characterizes structural competitive balance within a period through
the radius of the graph-fused $\ell_1$ polytope. We extend this
construction to longitudinal comparisons by allowing the radius
to vary across periods while sharing information among periods
with similar competitive environments. Rather than fixing the
number of regimes in advance, we use an adaptive occupied-state
representation in which existing regimes may recur, new regimes
may emerge, and unoccupied regimes are removed. This construction
is motivated by finite-mixture and mixture-of-finite-mixtures
methods \citep{miller2018mixture}, but the regime allocation
mechanism is specified directly below rather than through a
conventional Dirichlet transition prior.

\subsection{Dynamic geometric model}
\label{subsec:adaptive_regime_model}

Let $s_t$ denote the competitive regime assigned to period
$t=1,\ldots,T$, and define the occupied regime set
$\mathcal A=\{s_1,\ldots,s_T\}$ and its cardinality
$K_{\mathrm{act}}=|\mathcal A|$. Each occupied regime
$k\in\mathcal A$ is associated with a positive radius $r_k$,
so the structural radius for period $t$ is $r_t=r_{s_t}$.

Conditional on the regime assignment, latent abilities and
contextual effects follow
\begin{equation}
\beta_t\sim N(0,\sigma_\beta^2 I_N),\qquad\theta_t=\Pi_{\mathcal K_t(r_{s_t})}(\beta_t),\qquad\eta_t\sim N(0,\sigma_\eta^2),
\label{eq:dynamic_projection}
\end{equation}
where $\mathcal K_t(r)$ is defined in
\eqref{eq:feasible_polytope}, and match outcomes follow the
ordered Bradley--Terry likelihood
\eqref{eq:ordered_bt_likelihood}. The ordered-logit thresholds
$\kappa$ are held fixed.

The regime assignments characterize competitive environments
rather than team identities. In particular,
$s_t=s_{t'}$ implies $r_t=r_{t'}$, but generally
$\theta_t\neq\theta_{t'}$. Seasons sharing the same radius
may therefore exhibit different rankings and rank-tier
structures. This separation is important for longitudinal
sports data, where the degree of league-wide inequality may
remain stable even as the leading competitors change.

For the occupied regimes, we specify the hierarchical
radius distribution
$r_k\mid\alpha_r\sim\operatorname{Exponential}(\alpha_r)$,
$k\in\mathcal A$, with
$\alpha_r\sim\operatorname{Gamma}(a_r,b_r)$,
where the Gamma distribution is parameterized by shape and rate.
This hierarchy controls the scale of structural variation
while allowing regimes to represent distinct competitive
environments. Because the radius distribution is continuous,
the occupied radii are almost surely distinct under this prior and can be
ordered as $0<r_{(1)}<\cdots<r_{(K_{\mathrm{act}})}$.
By Corollary~\ref{cor:radius_monotonicity}, increasing
radii correspond to increasing expected latent ability
dispersion for a fixed comparison design.

\subsection{Adaptive regime allocation}
\label{subsec:adaptive_allocation}

We update the regime sequence through an occupied-state
allocation rule that balances temporal persistence with
the creation of additional competitive environments.
Immediately before updating $s_t$, let
$\mathcal A=\{1,\ldots,m\}$ denote the occupied regimes,
after temporarily removing the current assignment when
computing transition counts. Let $n_{kl}^{-t}$ denote
the number of transitions from regime $k$ to regime $l$
excluding transitions involving period $t$, and write
$n_{k\cdot}^{-t}=\sum_l n_{kl}^{-t}$.
For interior periods, let $i=s_{t-1}$ and $j=s_{t+1}$,
and define $b_m=\gamma/m$, where $\gamma>0$.

For an existing regime $k\in\mathcal A$, the incoming
and outgoing transition factors are
\begin{equation}
I_t(k)=n_{ik}^{-t}+a_{\mathrm{tr}}b_m,
\qquad
O_t(k)=\frac{n_{kj}^{-t}+\mathbb I\{t>1,\ i=k=j\}+a_{\mathrm{tr}}b_m}
{n_{k\cdot}^{-t}+\mathbb I\{t>1,\ i=k\}+a_{\mathrm{tr}}},
\label{eq:existing_transition_factors}
\end{equation}
with $I_1(k)=1$ and $O_T(k)=1$ at the boundary periods.
The indicator terms account for the dependence between
the two adjacent transitions when the proposed state
coincides with the preceding state. These factors
incorporate observed transition frequencies and favor
the reuse of previously occupied competitive environments.

For a candidate radius $r$, define the period-specific
likelihood
$f_t(r)=L_t\{\Pi_{\mathcal K_t(r)}(\beta_t),
\eta_t,\kappa\}$, where
$L_t(\theta_t,\eta_t,\kappa)
=\prod_{g=1}^{G_t}
p(y_{tg}\mid\theta_t,\eta_t,\kappa)$.
The allocation score for an existing regime is
$w_t(k)=I_t(k)O_t(k)f_t(r_k)$.

A new regime is represented by candidate $0$, with transition factors
$I_t(0)=a_{\mathrm{tr}}b_m$ and $O_t(0)=b_m$, again with
$I_1(0)=O_T(0)=1$. The propensity to introduce an additional regime is
controlled by the birth multiplier
$B_m=\lambda(m+1)\gamma/\{m(m+m\gamma)\}$,
where $\lambda>0$.

To evaluate the new-regime likelihood, we approximate
integration over the exponential radius distribution
using $J$ prior-quantile grid points,
$\widetilde f_{t,0}=J^{-1}\sum_{j=1}^Jf_t(r_j^{\mathrm{grid}})$
(Appendix~\ref{subsec:posterior_computation}). The resulting new-regime score is
$w_t(0)=B_mI_t(0)O_t(0)\widetilde f_{t,0}$.

The allocation probabilities used by the algorithm are
\begin{equation}
\Pr_{\mathrm{alg}}\{s_t=c\mid\mathrm{current}\}=\frac{w_t(c)}{w_t(0)+\sum_{k\in\mathcal A}w_t(k)},
\qquad c\in\mathcal A\cup\{0\}.
\label{eq:adaptive_allocation_probability}
\end{equation}

When the new-regime candidate is selected, its initial
radius is drawn from the same grid according to
\begin{equation}
\Pr_{\mathrm{alg}}(r_{\mathrm{new}}=r_j^{\mathrm{grid}}\mid\mathrm{new},t,\mathrm{current})
=f_t(r_j^{\mathrm{grid}})\Big/\textstyle\sum_{l=1}^J f_t(r_l^{\mathrm{grid}}).
\label{eq:new_state_radius_initialization}
\end{equation}
The radius is subsequently updated on its continuous
support using the Metropolis--Hastings step described in
Appendix~\ref{subsec:posterior_computation}. After
reassignment, regimes without occupied periods are
removed and the remaining labels are compacted.

The parameters $a_{\mathrm{tr}}$, $\gamma$, and $\lambda$
have distinct roles: they control transition-count
smoothing, baseline allocation mass, and the tendency
to introduce new regimes, respectively. Temporal
persistence is induced by the observed transition
counts rather than by an additional diagonal-specific
persistence parameter.

The allocation rule is an adaptive computational
construction. Although each update defines a proper
conditional probability distribution over its candidate
states, the rule is not asserted to be the exact
collapsed conditional of a conventional MFM-HMM prior.
Consequently, unless compatibility with a joint regime
distribution and posterior invariance are established,
uncertainty summaries involving the regime assignments
are interpreted as approximations generated by the
specified allocation procedure.

\subsection{Distinguishability of competitive regimes}
\label{subsec:regime_identifiability}

The regime representation is informative when distinct
radii correspond to distinguishable distributions of
observed comparisons. For a fixed comparison graph $A$,
let $\theta(r)=\Pi_{\mathcal K(r)}(\beta)$ and
$V(r)=\|A\theta(r)\|_1$. By
Corollary~\ref{cor:radius_monotonicity},
$\mathbb E\{V(r_1)\}<\mathbb E\{V(r_2)\}$
for $0<r_1<r_2$, so the structural radius provides
a strictly ordered measure of expected latent
ability dispersion.

To characterize the corresponding observed-data
distributions, let $\mathcal X$ denote a fixed
comparison design and define
\begin{equation}
g(y\mid r,\mathcal X)=\iint p\{y\mid\Pi_{\mathcal K(r)}(\beta),\eta,\kappa,\mathcal X\}\,
\phi_N(\beta;0,\sigma_\beta^2I_N)\,\phi(\eta;0,\sigma_\eta^2)\,d\beta\,d\eta.
\label{eq:integrated_emission}
\end{equation}
Let $\mathcal P$ be the set of home-and-away pairs of
$\mathcal X$, comparisons $g\neq g'$ with
$(i_{g'},j_{g'})=(j_g,i_g)$, and let
$\mathcal S_{\mathcal X}(r)=\sum_{(g,g')\in\mathcal P}
\Pr_r\{\mathbb I(y_g=1)\neq\mathbb I(y_{g'}=1)\}$, under
\eqref{eq:integrated_emission}, be the expected number of
such pairs in which exactly one home side wins.

\begin{assumption}[Emission separation]
\label{ass:emission_separation}
For the comparison design $\mathcal X$, the function
$r\mapsto\mathcal S_{\mathcal X}(r)$ is strictly
increasing over the radius range under consideration.
\end{assumption}

\begin{proposition}[Distinguishability of competitive environments]
\label{prop:regime_distinguishability}
Suppose Assumption~\ref{ass:emission_separation}
holds on $\mathcal R\subset(0,\infty)$.
Then, for any $r_1,r_2\in\mathcal R$, $r_1\neq r_2$ implies
$g(\cdot\mid r_1,\mathcal X)\neq g(\cdot\mid r_2,\mathcal X)$.
\end{proposition}

\begin{proof}
$\mathcal S_{\mathcal X}(r)$ is a functional of
$g(\cdot\mid r,\mathcal X)$ and is injective on $\mathcal R$.
\end{proof}

Near the parity boundary the condition holds whenever
$\mathcal P$ is nonempty, as in every season of both
applications: for any connected graph and any law of
$\eta$, each term of $\mathcal S_{\mathcal X}$ equals its
parity value plus $cr^2+o(r^2)$ with $c>0$ and is strictly
increasing on some $(0,\bar r]$
(Lemma~\ref{lem:small_radius_separation},
Appendix~\ref{app:proofs_dynamics}). Covariances of repeated
meetings are unsuitable, because integrating over $\eta$ adds
a radius-dependent term of indefinite sign
(Remark~\ref{rem:covariance_cancellation}). The lemma is local; at the
fitted radii Assumption~\ref{ass:emission_separation} remains an
assumption, with numerical support in
Appendix~\ref{app:realdata_diagnostics}.

At larger radii, the constraint binds less frequently.
By Theorem~\ref{thm:radius_saturation}, the projected
abilities approach the centered unconstrained Gaussian
draw as the radius becomes large relative to
$Z_t=\|A_tH\beta_t\|_1$. The quantity
$\Pr(Z_t>r)=M_t'(r)$ therefore characterizes the
sensitivity of expected structural variation to
changes in the radius.

The geometric model and the regime allocation
mechanism serve complementary purposes. The former
defines structural competitive balance and allows
exact rank tiers within each period, while the latter
groups periods with recurring levels of structural
inequality. The occupied regime count
$K_{\mathrm{act}}$, regime radii, and temporal
allocations provide a summary of changing competitive
environments without requiring the same team
hierarchy across periods.

\section{Bayesian Inference and Simulation Validation}
\label{sec:inference}
\subsection{Bayesian Inference}
Posterior inference uses a blockwise Markov chain Monte Carlo procedure,
given in full in Appendix~\ref{app:computation}
(Algorithm~\ref{alg:mfm_hmm_sampler}). Each sweep updates the auxiliary
abilities $\beta_t$ by elliptical slice sampling, recomputing the projection
$\theta_t=\Pi_{\mathcal K_t(r_{s_t})}(\beta_t)$, a convex quadratic
program, at every likelihood evaluation; updates the period home effects
$\eta_t$ by random-walk Metropolis--Hastings; updates each occupied radius
$r_k$ by a Metropolis--Hastings step on $\log r_k$ and draws the shared rate
$\alpha_r$ from its Gamma full conditional; and finally reallocates the
periods one at a time with the adaptive rule of
Section~\ref{subsec:adaptive_allocation}, creating new regimes through a
grid of prior radius quantiles and deleting empty ones. The threshold
$\kappa$ is held fixed. Because the allocation kernel is specified
algorithmically rather than derived as an exact collapsed update, we treat
the output as an approximate posterior: given the regime path, the updates
of $\beta_t$, $\eta_t$, $r_k$ and $\alpha_r$ leave the conditional posterior
invariant, but summaries that average over regime paths (the distribution
of $K_{\mathrm{act}}$, memberships, co-clustering and transition
frequencies, and to a lesser degree all intervals) describe this sampler,
and their calibration is checked in
Appendix~\ref{subsec:simulation_calibration}.

The retained draws are post-processed into interpretable point summaries.
State labels are first ordered by increasing radius. The number of regimes
$\widehat K$ is the most frequent value of $K_{\mathrm{act}}$, and the regime path
and radii are taken from the Dahl draw, the draw with $\widehat K$ regimes
whose period co-clustering matrix is closest in Frobenius norm to its
average over the draws. Team tiers are summarized season by season: in the draws
where the season belongs to its Dahl regime, the tiers of a draw are its
groups of exactly equal abilities, the estimated tiers are the SALSO
estimate of these partitions \citep{dahl2022search}, and abilities are posterior
means averaged within tiers. Home effects are summarized by posterior means,
and mixing is assessed with trace plots, split-$\widehat R$ and effective
sample sizes (Appendix~\ref{app:realdata_diagnostics}).

\subsection{Simulation Validation}\label{sec:simulation}

We simulate from the model to check recovery of the regime path, the radii,
the home effect and the within-season tiers, and to check that seasons with
the same radius but different tier structures are assigned to the same regime.
Eight settings cross the number of regimes (radii 200 and 400, or 100, 200 and
400), the number of seasons ($T=20$ or $40$) and the within-regime ability
structure, with 50 replicates each. Every season has 20 teams and 380
three-outcome matches in a double round robin. In the heterogeneous settings
one regime alternates between a two-tier and a three-tier configuration with
the same radius. Mean regime-path adjusted Rand index (ARI) is at least 0.985
in every setting, heterogeneous ones included, so seasons sharing a radius are
almost always placed in the same regime, whatever their tiers. This is
because regimes share only a radius: each season is allocated according to how
well its own abilities fit under each regime's radius, not according to its
tier structure. The number of
regimes is recovered in 92--98\% of two-regime replicates and in 84--94\% of
three-regime replicates. Each miss adds one regime, which in 26 of the 28
misses absorbs only one or two seasons. Radius estimates concentrate near their
true values (Figure~\ref{fig:radius_recovery}), home-effect mean squared error
is 0.013--0.016, and mean tier ARI is 0.84--0.94, lower than for the regime
path, because each regime pools information across several seasons whereas
the tiers of a season are estimated from its 380 matches alone. Equal-tailed
95\% credible intervals cover the true home effects in
93--96\% of cases and the true abilities in 97--99\%, and the smallest set of regime counts holding 95\% of the draws contains
the true number in 98--100\% of replicates (Appendix~\ref{subsec:simulation_calibration}).

Thinning the schedule separates regime identification from estimation
precision. In the two-regime setting with $T=20$, retaining each match
independently with probability $p\in\{1.0,0.9,0.8,0.7,0.6\}$ leaves regime
recovery almost unchanged: the number of regimes is recovered in 98--100\% of
replicates and mean regime-path ARI stays above 0.98 with 40\% of matches
removed. Each regime persists for five seasons, so even 228 matches per season
leave enough information to place every season. The loss falls instead on
estimation precision, with ability mean
squared error rising from 0.026 to 0.074, mean tier ARI falling from 0.94 to
0.89 and the radius estimates spreading out. Appendix~\ref{sec:simu} gives the
designs, the full results (Table~\ref{tab:simulation_results}) and the
missing-data analysis (Table~\ref{tab:missing_data_results}).
Appendix~\ref{app:sim_details} gives the computational settings.

\section{Competitive Balance in the English Premier League}
\label{sec:applications}

We fit the model to the $25$ Premier League seasons described in
Section~\ref{sec:data}, with the same model and sampler as in the
simulations. Hyperparameters were set by running short pilot chains and
then comparing neighboring configurations with match-level conditional
WAIC, which led to $(\sigma_\beta,a_r,b_r)=(2,10,200)$, where
$\sigma_\beta$ is the prior standard deviation of the auxiliary
abilities. Nearby configurations gave very similar WAIC values, so we
treat this as a stable local choice rather than a uniquely optimal one.
We ran two chains of $50{,}000$ iterations each, discarded the first
$5{,}000$ of each as burn-in, and retained $90{,}000$ draws. The
threshold was fixed at $\kappa_2=0.65$, home effects had independent
$N(0,1)$ priors, and posterior summaries follow
Appendix~\ref{subsec:post_mcmc}. Convergence diagnostics are reported in
Appendix~\ref{app:realdata_diagnostics}. Figure~\ref{fig:epl_results}
summarizes the results.

\begin{figure}[p]
    \centering
    \includegraphics[
        width=\textwidth,
        height=0.8\textheight,
        keepaspectratio
    ]{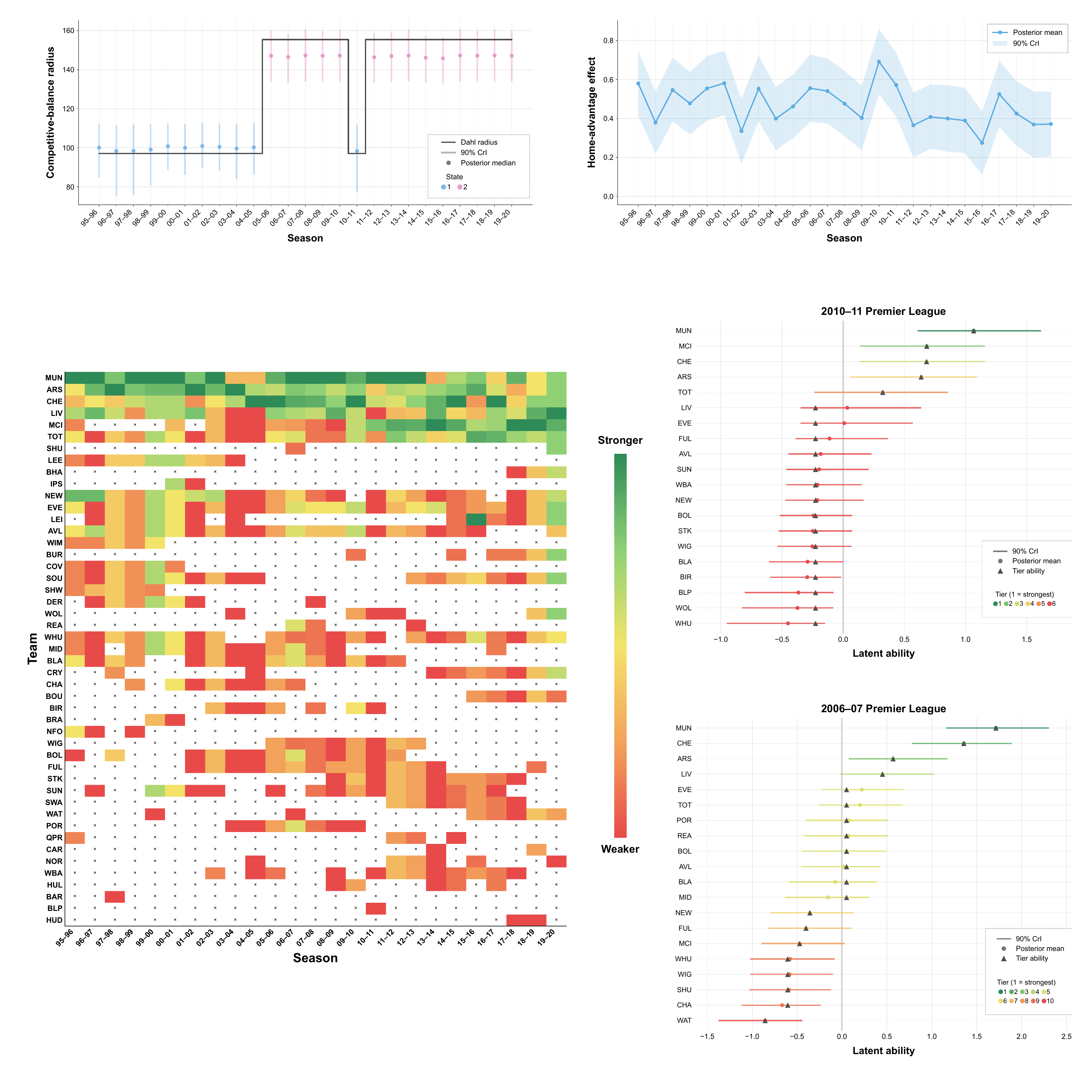}

    \caption{Posterior summaries for the English Premier League, 1995/96
    to 2019/20. Upper panels: season-specific competitive-balance radius
    and home-advantage effect with 90\% credible intervals. Lower left:
    longitudinal team tiers, with crosses marking seasons in which a club
    was not in the league. Right: team abilities and tiers for 2010/11
    (State~1) and 2006/07 (State~2); circles are posterior means,
    segments 90\% credible intervals, and triangles tier-level
    estimates.}
    \label{fig:epl_results}
\end{figure}

\paragraph{(Q1) Has competitive balance changed, and when?}
The results suggest a persistent shift toward greater structural
inequality beginning in the mid-2000s. The sampler favors two
competitive regimes, which are occupied in $67.4\%$ of retained draws.
The Dahl-representative radii are $97.1$ for the relatively balanced
State~1 and $155.4$ for the more stratified State~2. All seasons from
1995/96 through 2004/05 are assigned to State~1, followed by a
transition to State~2 in 2005/06. The league remains in State~2
for five consecutive seasons, briefly returns to State~1 in 2010/11,
and is assigned to State~2 from 2011/12 onward. Conditional on two
regimes, the estimated self-transition frequencies are $0.71$ and
$0.85$ for States~1 and~2, respectively, indicating greater persistence
in the more stratified regime. These findings suggest a sustained
change in league-wide competitive structure rather than isolated
fluctuations in seasonal outcomes. The timing is consistent with
previous evidence of increasing competitive inequality in English
football during the mid-2000s \citep{Basini2023}.

Importantly, a common competitive regime does not imply an unchanged
team hierarchy. Figure~\ref{fig:epl_results} shows that the identities
of leading clubs vary considerably within State~2, from Manchester
United, Chelsea, Arsenal, and Liverpool in earlier seasons to
Manchester City and Tottenham Hotspur in later periods. Leicester City
reaches the leading tier during its 2015/16 championship season,
while Liverpool does so in 2019/20, although both seasons belong to
the same more stratified regime.

\paragraph{(Q2) Which teams can the data tell apart?}
The inferred tiers reveal differences in the granularity of team
hierarchies across seasons. In the relatively balanced 2010/11 season,
the representative partition contains six tiers, with Manchester
United occupying the leading tier, followed by Manchester City,
Chelsea, Arsenal, and Tottenham Hotspur, and a broad lower tier
containing many of the remaining clubs. Only three points separated
second-placed Chelsea from fourth-placed Arsenal in the final standings.
This relatively compressed hierarchy is qualitatively consistent with
\citet{Basini2023}, who report a $79.9\%$ posterior probability
of a single-block structure in 2010/11, compared with probabilities
exceeding $95\%$ for two blocks in the adjacent seasons. Their
competitive blocks and our latent ability tiers, however, represent
different statistical quantities. In contrast, the more stratified
2006/07 season contains ten representative tiers, with a
$61$-point difference between champions Manchester United and
last-placed Watford.

The 2006/07 season also illustrates the information provided by
opponent-specific comparisons. Arsenal and Liverpool both finished
with $68$ points, with Liverpool placed third on goal difference.
Nevertheless, the representative hierarchy places Arsenal in a
stronger tier. Arsenal defeated champions Manchester United in both
league meetings, whereas Liverpool lost both. These results
illustrate differences in performance against a common opponent
that are not reflected in aggregate point totals. While these
head-to-head results are illustrative rather than conclusive, the model
estimates relative strengths using the full network of match outcomes.

\paragraph{(Q3) Is structural balance distinct from title-race closeness
and home advantage?}
The results suggest that these dimensions of competition evolve
differently. In 2011/12, Manchester City and Manchester United finished
level on $89$ points, with the championship decided on goal difference.
Nevertheless, the season is assigned to the more stratified regime,
while third-placed Arsenal finished $19$ points behind the leading
pair. This illustrates that a closely contested title race can coexist
with substantial league-wide inequality. The structural radius
characterizes dispersion in latent abilities across all teams rather
than the competitive gap between the leading contenders.

Home advantage exhibits a distinct temporal pattern. The estimated
home effect is positive in every season, with posterior means ranging
from approximately $0.27$ in 2015/16 to $0.69$ in 2009/10, and all
reported $90\%$ credible intervals excluding zero. Home advantage
is generally smaller after 2011/12, whereas the league remains in
the more stratified regime throughout this period. These findings
suggest that changes in home advantage and structural competitive
balance need not coincide, supporting their separate representation
within the paired-comparison framework.

\paragraph{Implications.}
The EPL analysis illustrates the value of distinguishing structural
competitive balance from descriptive measures based on final standings.
The regime-specific radius provides an interpretable measure of
league-wide latent ability dispersion, while the inferred tiers
identify groups of clubs with similar competitive strengths.
Together, these quantities provide complementary information about
when competitive inequality changes and how team hierarchies evolve.
Such evidence can inform discussions of revenue sharing, financial
regulation, and competition design, although the present findings
are descriptive rather than causal. Historical developments in
English football provide context for the inferred regimes but are
not included as explanatory variables in the model. A complementary
analysis of the National Basketball Association, a closed league
with different institutional mechanisms for promoting competitive
balance, is presented in the Supplementary Material
(Appendix~\ref{app:nba_application}) and illustrates
recurrent switching between competitive regimes rather than a
single sustained shift.

\section{Discussion}\label{sec:discuss}
We treat competitive balance as a property of teams rather than season
outcomes. Abilities are generated by projecting a Gaussian draw onto a
graph-fused $\ell_1$ polytope: when the constraint binds, its radius
equals the realized total of pairwise ability gaps, and because the
projection lands on a face with positive probability, the model
supports exact ties and posterior probabilities for rank tiers. Near
parity, the radius connects the signal-to-noise ratio of
\citet{Lopez2018}, a structural Gini coefficient, and
the number of matches needed to detect a departure from parity.
Applied to the Premier League, the fitted model favors two eras, with a
lasting shift toward greater stratification in 2005/06 and only one
later return to the compressed regime, in 2010/11; the tiers also
capture information missed by the standings, such as head-to-head
results between clubs level on points. In the NBA, regimes alternate
rather than shift once, and the home effect is smallest in the
attendance-restricted 2020/21 season.

The approach has limitations. The radius measures inequality in team
abilities, not outcomes, so outcome-based indices may remain more
useful for predictive balance; abilities are fixed within seasons;
victory margins are reduced to three ordered categories; within-era
drift is absorbed into regime changes; the regime summaries come
from an approximate allocation kernel whose invariant distribution is not
established; the number of regimes depends
on $\gamma$ and $\lambda$; and each likelihood evaluation requires a
quadratic program that will become costly for larger comparison
graphs. Natural extensions address several of these points directly.
Letting the radius depend on covariates such as revenue sharing or
spending regulation would allow associations between regulatory or
institutional changes and structural balance to be estimated, although
causal evaluation of such policies would require additional
identification assumptions. Because the likelihood is defined on an arbitrary comparison graph,
the schedule itself becomes a design variable: given an estimated
radius and ability vector, one can evaluate proposed
formats---balanced versus unbalanced schedules, playoff cut lines,
group-stage draws, seeding rules---by the distribution of terminal
ranks and title probabilities they induce, and search over formats for
a target trade-off between fairness and suspense
\citep{Szymanski2003,Groh2012,KendallLenten2017}. An ordinal or
continuous margin model would likewise recover information the
three-category outcome discards and would let the framework speak to
in-game competitive balance rather than only \emph{ex ante} balance,
which matters for broadcast economics, where suspense within a match
drives viewership \citep{ElyFrankelKamenica2015}; a richer outcome
scale would also sharpen the home-effect estimates, extracting more
signal from the attendance-restricted 2020/21 season, a rare natural
experiment on crowd effects \citep{Bryson2021}. Finally, the prior
applies unchanged beyond team sports. Chess, esports, grant review,
and language-model preference data all involve paired comparisons in
which genuine ties matter, and faster projection approximations or
variational methods would help make these applications practical.

\bibliographystyle{chicago}

\begin{spacing}{1}
\setlength{\bibsep}{7.5pt}
\bibliography{Bibliography-MM-MC}
\end{spacing}

\newpage
\appendix
\begin{center}
{\LARGE\bf Supplementary Material}
\end{center}
\noindent Appendices~\ref{app:proofs_geometry}--\ref{app:realdata_diagnostics}
form the Supplementary Material referred to in the main text.
\setcounter{figure}{0}
\renewcommand{\thefigure}{S\arabic{figure}}

\setcounter{table}{0}
\renewcommand{\thetable}{S\arabic{table}}
\setcounter{equation}{0}

\section{Proofs for Section~\ref{sec:geometric_framework}}
\label{app:proofs_geometry}

This appendix collects the proofs of the results stated in
Section~\ref{sec:geometric_framework}, together with
Lemma~\ref{lem:index_exists} on the reciprocal-radius summary. Throughout,
$H_0=\{\vartheta\in\mathbb R^N:\mathbf1^\top\vartheta=0\}$ denotes the
zero-sum subspace, $H=I_N-N^{-1}\mathbf1\mathbf1^\top$ the orthogonal
projection onto it, and the period index is suppressed where no ambiguity
arises.

\begin{proof}[Proof of Theorem~\ref{thm:posterior_propriety}]
Let $\Lambda$ denote the collection of all primitive unknowns and let
$\mathbb Q(d\Lambda)$ denote their proper joint prior. The projection map
\eqref{eq:projection_map} is the argmin of a strictly convex function over
a closed convex set depending continuously on $(\beta_t,r_t)$, hence is
single valued and continuous, and therefore Borel measurable; consequently
$\Lambda\mapsto L(\mathcal D\mid\Lambda)$ is measurable. Every factor in
\eqref{eq:ordered_bt_likelihood} lies in $(0,1)$ for finite arguments, so
\[
0< L(\mathcal D\mid\Lambda)\le1
\qquad\text{for every }\Lambda,
\]
where the upper bound holds because the likelihood is a product of
probabilities. Therefore the normalizing constant satisfies
\[
p(\mathcal D)
=
\int
L(\mathcal D\mid\Lambda)\mathbb Q(d\Lambda)
\le
\int\mathbb Q(d\Lambda)
=
1<\infty .
\]
For the lower bound, $L(\mathcal D\mid\cdot)$ is strictly positive
everywhere and $\mathbb Q$ is a probability measure, so $p(\mathcal D)>0$ by
monotone convergence applied to the sets $\{L\ge 1/n\}$, whose union has
$\mathbb Q$-measure one. Hence $0<p(\mathcal D)<\infty$ and the posterior
$\mathbb Q(d\Lambda\mid\mathcal D)=L(\mathcal D\mid\Lambda)\mathbb Q(d\Lambda)/p(\mathcal D)$ is
a well-defined probability measure.
\end{proof}

\begin{proof}[Proof of Proposition~\ref{prop:identifiability}]
Suppose $(\theta,\eta)$ and $(\theta',\eta')$ produce identical linear
predictors. Let
\[
d=\theta-\theta',
\qquad
c=\eta'-\eta.
\]
Then
\[
x_g^\top d=-ch_g
\]
for every comparison and $\mathbf1^\top d=0$.

If $c=0$, then $d_i=d_j$ for every compared pair. Connectivity implies
that $d$ is constant, and the zero-sum constraint implies $d=0$.
If $c\ne0$, this contradicts the non-collinearity condition of
Proposition~\ref{prop:identifiability}.
Thus $d=0$ and $c=0$, which proves the first claim.

For the second claim, fix a comparison $g$ and write the observed cell
probabilities as $(p_{-1},p_0,p_1)$. From
\eqref{eq:ordered_bt_likelihood},
\[
\sigma(\kappa_1-\Delta_g)=p_{-1},
\qquad
\sigma(\kappa_2-\Delta_g)=p_{-1}+p_0,
\]
and since $\sigma$ is a strictly increasing bijection from $\mathbb R$ onto
$(0,1)$ these determine $\kappa_1-\Delta_g$ and $\kappa_2-\Delta_g$
uniquely. Hence the cell probabilities identify the two differences, and
therefore identify the gap $\kappa_2-\kappa_1$ together with
$\kappa_1-\Delta_g$. Under the anchoring $\kappa_1=-\kappa_2$ we have
$\kappa_2-\kappa_1=2\kappa_2$, so $\kappa_2$ is identified, and then
$\Delta_g=\kappa_1-(\kappa_1-\Delta_g)$ is identified for every $g$.
Applying the first claim to $(\Delta_g)_{g=1}^G$ identifies
$(\theta,\eta)$.

For the final claim, suppose $h_g\equiv h$ is constant. Then
$\Delta_g(\theta,\eta+c)=\Delta_g(\theta,\eta)+ch$ for every $g$, and
replacing $\kappa$ by $\kappa+ch\mathbf 1$ leaves both differences
$\kappa_1-\Delta_g$ and $\kappa_2-\Delta_g$, and hence all cell
probabilities, unchanged. The stated family is therefore observationally
equivalent, and the anchoring $\kappa_1=-\kappa_2$, $\kappa_2>0$ selects one
member of it.
\end{proof}

\begin{proof}[Proof of Proposition~\ref{prop:exact_ties}]
Write $\mathcal K=\mathcal K_t(r_t)$ and let
$H_0=\{\vartheta:\mathbf1^\top\vartheta=0\}$, a linear subspace of
dimension $N-1$. Because the graph is connected,
$\operatorname{null}(A_t)\cap H_0=\{0\}$, so $\mathcal K$ is compact, and it
contains a relatively open neighborhood of the origin in $H_0$; hence
$\operatorname{aff}(\mathcal K)=H_0$ and $\dim\mathcal K=N-1$.

\emph{Existence of a face with a vanishing contrast.} Since $\mathcal K$ is
compact and convex it has an extreme point $v$. Suppose, for contradiction,
that $(A_tv)_e\ne0$ for every $e\in E_t$. Then the signs
$s_e=\operatorname{sign}\{(A_tv)_e\}$ are locally constant, so on a
neighborhood $\mathcal V$ of $v$ in $H_0$ the map
$\vartheta\mapsto\|A_t\vartheta\|_1$ coincides with the linear functional
$\vartheta\mapsto\sum_e s_e(A_t\vartheta)_e$. Consequently
$\mathcal K\cap\mathcal V$ coincides with the intersection of $\mathcal V$
with a closed half-space of $H_0$. Every point of such an intersection lies
in a relatively open segment contained in $\mathcal K$ whenever
$\dim H_0-1=N-2\ge1$, contradicting extremality of $v$. Hence for $N\ge3$
some contrast vanishes at $v$, and $F=\{v\}$ is a face of the required
type.

\emph{Positive measure of $\mathcal R_F$.} Let $F$ be any nonempty face of
$\mathcal K$ with $\dim F=d$, let $\mathcal L$ denote the direction space of
$\operatorname{aff}(F)$, so $\dim\mathcal L=d$, and let $\mathcal N_F$ denote the normal cone
to $\mathcal K$ at any point of $\operatorname{relint}(F)$, which does not
depend on the choice of that point. By the standard normal decomposition of
Euclidean projection onto a closed convex set
\citep[Sec.~13]{rockafellar1970convex},
\[
\mathcal R_F
=
\Pi_{\mathcal K}^{-1}\{\operatorname{relint}(F)\}
=
\operatorname{relint}(F)+\mathcal N_F ,
\]
since $\Pi_{\mathcal K}(b)=x$ if and only if $b-x\in \mathcal N_{\mathcal K}(x)$.
Because $\operatorname{aff}(\mathcal K)=H_0$, the normal cone at a point of
$\operatorname{relint}(F)$ decomposes as
$\mathcal N_F=\widetilde{\mathcal N}_F\oplus\operatorname{span}(\mathbf1)$, where
$\widetilde{\mathcal N}_F\subset H_0\cap\mathcal L^{\perp}$ is the normal cone computed within
$H_0$ and has nonempty interior relative to $H_0\cap\mathcal L^\perp$, a space of
dimension $N-1-d$. Hence $\mathcal N_F$ has nonempty interior relative to
$\mathcal L^{\perp}$, which has dimension $N-d$. Choosing a relatively open subset
$B\subset\operatorname{relint}(F)$ of dimension $d$ and a subset
$C\subset\mathcal N_F$ open in $\mathcal L^{\perp}$, the sum $B+C$ is open in
$\mathbb R^N=\mathcal L\oplus\mathcal L^{\perp}$ and contained in $\mathcal R_F$. Therefore
$\mathcal R_F$ has positive Lebesgue measure.

\emph{Conclusion.} Every $b\in\mathcal R_F$ projects into
$\operatorname{relint}(F)$ and hence satisfies $(A_t\theta_t)_e=0$. The
Gaussian auxiliary distribution has a strictly positive density on
$\mathbb R^N$, so it assigns positive probability to every set of positive
Lebesgue measure, giving $\Pr(\beta_t\in\mathcal R_F)>0$ and hence
$\Pr\{(A_t\theta_t)_e=0\mid r_t\}>0$.
\end{proof}

\begin{proof}[Proof of Theorem~\ref{thm:radius_saturation}]
Among all vectors satisfying $\mathbf1^\top\vartheta=0$, the unique
minimizer of
\[
\|\vartheta-\beta_t\|_2^2
\]
is $H\beta_t$.

If $Z_t\le r$, then $H\beta_t\in\mathcal K_t(r)$, so it remains the unique
minimizer under the additional graph-variation constraint. Hence
\[
\theta_t(r)=H\beta_t
\]
and
\[
\|A_t\theta_t(r)\|_1=Z_t.
\]

Now suppose $Z_t>r$ and write $\vartheta^\star=\theta_t(r)$. Suppose, for
contradiction, that $\|A_t\vartheta^\star\|_1<r$. The set
\[
\mathcal U
=
\{\vartheta\in H_0:\|A_t\vartheta\|_1<r\},
\qquad
H_0=\{\vartheta:\mathbf1^\top\vartheta=0\},
\]
is open in $H_0$ because $\vartheta\mapsto\|A_t\vartheta\|_1$ is
continuous, and $\vartheta^\star\in\mathcal U$. Since $\vartheta^\star$
minimizes the convex function $\vartheta\mapsto\|\vartheta-\beta_t\|_2^2$
over $\mathcal K_t(r)\supset\mathcal U$ and lies in the open subset
$\mathcal U$ of $H_0$, it is a local minimizer of that function over all of
$H_0$. A convex function has no strict local minimizer other than its
global minimizer, so $\vartheta^\star$ minimizes
$\|\vartheta-\beta_t\|_2^2$ over $H_0$, whence
$\vartheta^\star=H\beta_t$ by uniqueness. But then
\[
\|A_t\vartheta^\star\|_1=\|A_tH\beta_t\|_1=Z_t>r,
\]
contradicting $\|A_t\vartheta^\star\|_1<r$. The graph-variation constraint
is therefore active, giving
\[
\|A_t\theta_t(r)\|_1=r.
\]
Combining the two cases yields \eqref{eq:radius_saturation}. For $r=0$ and
a connected graph, $\mathcal K_t(0)=\{0\}$ and both sides of
\eqref{eq:radius_saturation} vanish, so the identity also holds at the
endpoint.
\end{proof}

\begin{proof}[Proof of Corollary~\ref{cor:radius_monotonicity}]
By Theorem~\ref{thm:radius_saturation},
$\|A_t\theta_t(r)\|_1=\min(r,Z_t)$ for the given $\beta_t$, and
$r\mapsto\min(r,Z_t)$ is nondecreasing, with $\min(r_1,Z_t)=r_1<
\min(r_2,Z_t)$ when $r_1<Z_t$. This gives the pathwise inequality
$\|A_t\theta_t(r_1)\|_1\le\|A_t\theta_t(r_2)\|_1$ and its strict version.
Moreover
\[
\min(r,Z_t)
=
\int_0^r
\mathbb I(Z_t>u)\,du ,
\]
and the integrand is nonnegative and jointly measurable, so Tonelli's
theorem gives $M_t(r)=\mathbb E\{\min(r,Z_t)\}=\int_0^r\Pr(Z_t>u)\,du$,
the identity in \eqref{eq:mean_structural_variation}.

It remains to establish the regularity claims. Since the graph has at least
one edge and $\operatorname{null}(A_t)\cap H_0=\{0\}$ under connectivity,
$A_tH\beta_t$ is a nondegenerate Gaussian vector on the contrast space, so
$Z_t=\|A_tH\beta_t\|_1$ is a norm of a nondegenerate Gaussian vector and
therefore has a continuous distribution on $[0,\infty)$ with
$\Pr(Z_t=0)=0$. Hence $u\mapsto\Pr(Z_t>u)$ is continuous, and the
fundamental theorem of calculus applied to this integral identity
yields that $M_t$ is
continuously differentiable on $(0,\infty)$ with
$M_t'(r)=\Pr(Z_t>r)$. Finally $Z_t$ has unbounded support, because
$A_tH\beta_t$ ranges over all of the contrast space with positive Gaussian
density, so $\Pr(Z_t>r)>0$ for every finite $r$, which gives
$M_t'(r)>0$ in \eqref{eq:mean_structural_variation} and strict monotonicity
of $M_t$.
\end{proof}

\begin{proof}[Proof of Proposition~\ref{prop:piecewise_radius_path}]
Introduce auxiliary variables $u\in\mathbb R^{|E_t|}$ and write the
graph-variation constraint as
\[
-u\le A_t\vartheta\le u,
\qquad
\mathbf1^\top u\le r .
\]
Together with $\mathbf1^\top\vartheta=0$, the projection problem becomes a
quadratic program in $(\vartheta,u)$ whose objective
$\tfrac12\|\vartheta-\beta_t\|_2^2$ is convex and whose constraints are
linear with a right-hand side that is affine in the scalar parameter $r$;
only the single constraint $\mathbf1^\top u\le r$ involves $r$. The objective is strictly convex in the
$\vartheta$-block and constant in $u$, so the optimal $u$ is not unique
whenever $\mathbf1^\top u<r$ at the optimum; the optimal $\vartheta$,
however, is unique, since it is the projection of $\beta_t$ onto the closed
convex set $\mathcal K_t(r)$. All statements below refer to the
$\vartheta$-block.

By the theory of parametric quadratic programming with a right-hand side
depending affinely on a scalar
\citep[see, e.g.,][]{best1996algorithm}, the optimal-value function is
piecewise quadratic and the unique optimal $\vartheta(r)$ is a piecewise
affine, Lipschitz continuous function of $r$: on the set of $r$ for which a
given optimal basis (equivalently, a given active set together with a
choice of linearly independent active constraints) remains optimal,
$\vartheta(r)$ solves a fixed linear Karush--Kuhn--Tucker system whose
right-hand side is affine in $r$, and is therefore affine in $r$. Since
there are finitely many candidate active sets, the interval $[0,R]$ is
covered by finitely many maximal subintervals on which $\vartheta(r)$ is
affine, which gives the partition $0=b_0<b_1<\cdots<b_Q=R$. Continuity
across the breakpoints follows from continuity of the projection map
$\mathcal K_t(r)\mapsto\Pi_{\mathcal K_t(r)}(\beta_t)$ in $r$, which in turn
follows from continuity of $r\mapsto\mathcal K_t(r)$ in the Hausdorff metric
together with nonexpansiveness of Euclidean projection.

Within each segment, $\theta_{ti}(r)-\theta_{tj}(r)$ is a difference of
affine functions of $r$ and hence affine. Such a function is either
identically zero on the segment or has at most one zero in its interior,
so on $[0,R]$ the rank-tier partition changes at most finitely often,
apart from ties that persist throughout a segment.
\end{proof}

\begin{proof}[Proof of Lemma~\ref{lem:two_block_extreme}]
Let $\theta\in\mathcal K_1$ be an extreme point and let $u_1>\cdots>u_p$ be
its distinct values, with multiplicities $n_1,\ldots,n_p$ summing to $N$.
On the open region
\[
\mathcal O
=
\{\vartheta:\vartheta\text{ has value pattern }
(u'_1>\cdots>u'_p)\text{ with the same block assignment}\},
\]
identified with $\{u'\in\mathbb R^p:u'_1>\cdots>u'_p\}$, both constraints
defining $\mathcal K_1$ are affine in $u'$: the zero-sum constraint reads
$\sum_l n_lu'_l=0$, and, because the complete-graph total variation is
\[
\|A\vartheta\|_1
=
\sum_{l<l'}n_ln_{l'}(u'_l-u'_{l'}) ,
\]
the constraint $\|A\vartheta\|_1\le1$ is affine on $\mathcal O$ and is
active at $\theta$, since $\|A\theta\|_1=1$ for any extreme point (a point
with $\|A\theta\|_1<1$ lies in the relative interior of $\mathcal K_1$ and
so is not extreme).

Suppose $p\ge3$. The affine set
\[
\mathcal A
=
\Bigl\{
u'\in\mathbb R^p:
\sum_l n_lu'_l=0,\;
\sum_{l<l'}n_ln_{l'}(u'_l-u'_{l'})=1
\Bigr\}
\]
is defined by two linear equations, so $\dim\mathcal A\ge p-2\ge1$. Since
$u\in\mathcal A$ and $\mathcal O$ is open in $\mathbb R^p$, there is
$\varepsilon>0$ and a direction $d\ne0$ with $u\pm\varepsilon d\in
\mathcal A\cap\mathcal O$. The corresponding vectors $\theta^\pm$ lie in
$\mathcal K_1$, are distinct, and satisfy
$\theta=(\theta^++\theta^-)/2$, contradicting extremality. Hence $p\le2$.
If $p=1$ the zero-sum constraint forces $\theta=0$, which has
$\|A\theta\|_1=0\ne1$. Therefore $p=2$.

Writing $S$ for the block on which $\theta$ takes its larger value, the
zero-sum and unit-variation constraints determine $\theta$ uniquely up to
sign and give
\begin{equation}
\theta
=
\pm\,
\frac{\mathbf 1_S-|S|N^{-1}\mathbf 1}{|S|\,(N-|S|)},
\label{eq:two_block_vertex}
\end{equation}
where $\mathbf 1_S$ is the indicator vector of $S$: with $k=|S|$ and $d$ the gap
between the two levels, $\|A\theta\|_1=k(N-k)d=1$ and the larger value is
$(N-k)d/N$. Conversely, fix a nonempty proper subset $S$ with $k=|S|$ and consider the
linear functional $\vartheta\mapsto\sum_{i\in S}\vartheta_i$ on
$\mathcal K_1$. Summing the identity
$\sum_{i\in S}\sum_{j\notin S}(\vartheta_i-\vartheta_j)
=(N-k)\sum_{i\in S}\vartheta_i-k\sum_{j\notin S}\vartheta_j
=N\sum_{i\in S}\vartheta_i$, which uses
$\mathbf1^\top\vartheta=0$, gives
\begin{equation}
\sum_{i\in S}\vartheta_i
=
\frac1N
\sum_{i\in S}\sum_{j\notin S}(\vartheta_i-\vartheta_j)
\le
\frac1N
\sum_{i\in S}\sum_{j\notin S}|\vartheta_i-\vartheta_j|
\le
\frac1N
\sum_{i<j}|\vartheta_i-\vartheta_j|
\le
\frac1N .
\label{eq:block_support}
\end{equation}
Equality throughout forces $\vartheta_i\ge\vartheta_j$ for all $i\in S$,
$j\notin S$, and $\vartheta_i=\vartheta_j$ for all pairs lying inside $S$
or inside its complement, so $\vartheta$ is constant on each block; the
zero-sum and unit-variation constraints then pin it down as the vector
\eqref{eq:two_block_vertex}. Hence that vector is the \emph{unique}
maximizer of $\vartheta\mapsto\sum_{i\in S}\vartheta_i$ over
$\mathcal K_1$, and a unique maximizer of a linear functional over a convex
set is an extreme point. The negatives are extreme by symmetry.

Finally, taking $S=\{i\}$ in \eqref{eq:block_support} gives
$\vartheta_i\le1/N$ for every $\vartheta\in\mathcal K_1$, with equality at
the vector \eqref{eq:two_block_vertex} for $S=\{i\}$. Hence
$\sup_{\theta\in\mathcal K_1}\theta_i=1/N$.
\end{proof}

\begin{lemma}[Finiteness of the reciprocal-radius summary]
\label{lem:index_exists}
Suppose the radius prior has density $p_r$ satisfying
$\int_0^\infty u^{-1}p_r(u)\,du<\infty$; a sufficient condition is
$p_r(u)=O(u^{a-1})$ as $u\downarrow0$ for some $a>1$. Then
$\mathbb E(r_t^{-1}\mid\mathcal D)<\infty$ for every finite design. The
condition fails for the exponential radius prior of
Section~\ref{subsec:adaptive_regime_model}, whose density is positive at
zero.
\end{lemma}

\begin{proof}[Proof of Lemma~\ref{lem:index_exists}]
Write $\Lambda$ for the full parameter as in
Theorem~\ref{thm:posterior_propriety}. Since
$0<L(\mathcal D\mid\Lambda)\le1$ and $p(\mathcal D)>0$,
\[
\mathbb E(r_t^{-1}\mid\mathcal D)
=
\frac{\mathbb E\{r_t^{-1}L(\mathcal D\mid\Lambda)\}}{p(\mathcal D)}
\le
\frac{\mathbb E(r_t^{-1})}{p(\mathcal D)} ,
\]
where the expectations on the right are under the prior. Conditioning on
$s_t=k$ and averaging, $\mathbb E(r_t^{-1})=\mathbb E(r_k^{-1})
=\int_0^\infty u^{-1}p_r(u)\,du<\infty$ by assumption.
\end{proof}

\begin{proof}[Proof of Theorem~\ref{thm:single_block_limit}]
Connectivity implies $\operatorname{null}(A)=\operatorname{span}(\mathbf1)$,
so the linear map $\vartheta\mapsto A\vartheta$ is injective on the
zero-sum subspace $H_0$. Since $H_0$ is finite dimensional, the two norms
$\vartheta\mapsto\|A\vartheta\|_1$ and $\vartheta\mapsto\|\vartheta\|_\infty$
on $H_0$ are equivalent, so there exists $L_A<\infty$ with
\[
\|\vartheta\|_\infty
\le
L_A\|A\vartheta\|_1
\qquad
\text{for every }\vartheta\in H_0 .
\]
Applying this to $\theta(r)\in\mathcal K_r\subset H_0$ and using
$\|A\theta(r)\|_1\le r$ gives $\|\theta(r)\|_\infty\le L_Ar$, which yields
$\theta(r)\to0$ together with the stated uniform rate; no
separate compactness or limit-point argument is needed.

For the binary model, $|\theta_i(r)-\theta_j(r)|\le2L_Ar\to0$ and $\sigma$
is continuous, so $\Pr(Y_{ij}=1\mid\theta(r))=\sigma(\theta_i-\theta_j)\to
\sigma(0)=1/2$. For the ordered model,
$\Delta_g=x_g^\top\theta(r)+\eta h_g\to\eta h_g$, and substituting into
\eqref{eq:ordered_bt_likelihood} gives the limiting cell probabilities
stated in Theorem~\ref{thm:single_block_limit} by continuity of $\sigma$.
\end{proof}

\begin{proof}[Proof of Lemma~\ref{lem:linear_drift}]
Because the constraint $\|A\vartheta\|_1\le r$ is positively homogeneous,
$\mathcal K_r=r\mathcal K_1$ on the zero-sum subspace, and $\mathcal K_1$ is
symmetric under $\vartheta\mapsto-\vartheta$ because $\|A\vartheta\|_1$ is
even. Hence
\begin{equation}
\sup_{\theta\in\mathcal K_r}\theta_i
=
r\varrho_i,
\qquad
\inf_{\theta\in\mathcal K_r}\theta_i
=
-r\varrho_i ,
\label{eq:scaled_sup}
\end{equation}
and $\varrho_i<\infty$ because connectivity makes $\mathcal K_1$ compact, while
$\varrho_i>0$ because $\mathcal K_1$ contains a relatively open neighborhood of
the origin in the zero-sum subspace.

By Theorem~\ref{thm:single_block_limit} there is $L_A<\infty$ with
$\|\theta\|_\infty\le L_Ar$ for every $\theta\in\mathcal K_r$, so all
contrasts satisfy $|\theta_i-\theta_j|\le2L_Ar$. Using the expansion
$\sigma(u)=\tfrac12+u/4+O(u^3)$, valid uniformly on bounded sets, and
$\sum_j\theta_j=0$,
\[
w_i(\theta)
=
\frac12
+
\frac{1}{4(N-1)}
\sum_{j\ne i}(\theta_i-\theta_j)
+
O(r^3)
=
\frac12
+
\frac{N}{4(N-1)}\theta_i
+
O(r^3),
\]
where the remainder is uniform over $\theta\in\mathcal K_r$. Taking the
supremum of the absolute value over $\mathcal K_r$ and using
\eqref{eq:scaled_sup} together with the symmetry of $\mathcal K_r$ gives
$\delta_i(r)=c_{A,i}r+O(r^3)$ with $c_{A,i}=N\varrho_i/\{4(N-1)\}$.

For the complete graph, the extreme points of $\mathcal K_1$ are the
two-block vectors of Lemma~\ref{lem:two_block_extreme}; among these,
$\theta_i$ is largest for the vector isolating entity $i$, namely
$v=\{e_i-N^{-1}\mathbf1\}/(N-1)$, which satisfies $\|Av\|_1=1$ and
$v_i=1/N$. Hence $\varrho_i=1/N$.
\end{proof}

\begin{proof}[Proof of Theorem~\ref{thm:inverse_square}]
Write $w=w_i(\theta)$ and $\varepsilon=w-1/2$. Under $\theta$, $W_m$ is
$\mathrm{Binomial}(m,w)$. By the Berry--Esseen theorem the normal
approximation to the distribution of $(W_m-mw)/\sqrt{mw(1-w)}$ has error
$O(m^{-1/2})$ uniformly in $w$ bounded away from $0$ and $1$, and by
Lemma~\ref{lem:linear_drift} the relevant $w$ satisfy $w\to1/2$ as
$r\downarrow0$. Hence the power of the test that rejects when
$|W_m-m/2|>z_{1-\alpha/2}\sqrt m/2$ equals
\[
\Phi\!\left(
\frac{2\sqrt m\,\varepsilon-z_{1-\alpha/2}}
{\sqrt{1-4\varepsilon^2}}
\right)
+
\Phi\!\left(
\frac{-2\sqrt m\,\varepsilon-z_{1-\alpha/2}}
{\sqrt{1-4\varepsilon^2}}
\right)
+
O(m^{-1/2}),
\]
using $w(1-w)=(1-4\varepsilon^2)/4$, so that the standard deviation of
$W_m$ is $\sqrt m\sqrt{1-4\varepsilon^2}/2$ and the two rejection
thresholds $\pm z_{1-\alpha/2}\sqrt m/2$ standardize to
$(\pm z_{1-\alpha/2}-2\sqrt m\,\varepsilon)/\sqrt{1-4\varepsilon^2}$. The power is strictly increasing in
$|\varepsilon|$, so the supremum over $\theta\in\mathcal K_r$ in the
definition of $m_i(r;\alpha,\nu)$ is attained at $|\varepsilon|=\delta_i(r)$,
the maximal departure from parity defined in \eqref{eq:structural_drift}.

Write $P(d)=\Phi(d-z_{1-\alpha/2})+\Phi(-d-z_{1-\alpha/2})$. Then
$P(0)=2\Phi(-z_{1-\alpha/2})=\alpha$, $P(d)\to1$ as $d\to\infty$, and
$P'(d)=\phi(d-z_{1-\alpha/2})-\phi(d+z_{1-\alpha/2})>0$ for $d>0$. Since
$1-\nu>\alpha$, the equation $P(d)=1-\nu$ therefore has the unique positive
root $d_{\alpha,\nu}$ of Theorem~\ref{thm:inverse_square}. With
$x_m=2\sqrt m\,\delta_i(r)$, the maximal power equals
$P\{x_m/\sqrt{1-4\delta_i(r)^2}\}+O(m^{-1/2})$. As $r\downarrow0$ we have
$\delta_i(r)\to0$, so a power of $1-\nu>\alpha$ requires $m\to\infty$, and
$\sqrt{1-4\delta_i(r)^2}=1+O(r^2)$. Along the horizons at which the power
crosses $1-\nu$, continuity and strict monotonicity of $P$ give
\[
2\sqrt m\,\delta_i(r)
=
d_{\alpha,\nu}\bigl\{1+o(1)\bigr\},
\qquad\text{that is}\qquad
m
=
\frac{d_{\alpha,\nu}^2}{4\delta_i(r)^2}
\bigl\{1+o(1)\bigr\} .
\]
Both terms of $P$ remain of order one in this limit: the lower-tail term
tends to $\Phi(-d_{\alpha,\nu}-z_{1-\alpha/2})>0$, not to zero, which is why
the constant is $d_{\alpha,\nu}$ rather than $z_{1-\alpha/2}+z_{1-\nu}$.
Since $m_i(r;\alpha,\nu)$ is the least integer $m$ with power at least
$1-\nu$ and the power is increasing in $m$, the displayed value differs
from $m_i(r;\alpha,\nu)$ by at most one, which is absorbed in the
$o(1)$ factor because the right-hand side diverges. Substituting
$\delta_i(r)=c_{A,i}r+O(r^3)$ from Lemma~\ref{lem:linear_drift} yields
\eqref{eq:inverse_square_detection}. For the complete graph
$c_{A,i}=\{4(N-1)\}^{-1}$, so
$d_{\alpha,\nu}^2/(4c_{A,i}^2)=4(N-1)^2d_{\alpha,\nu}^2$.
\end{proof}

\begin{proof}[Proof of Proposition~\ref{prop:snr_bound}]
For the complete graph, $\mathcal K_r$ is compact on the zero-sum subspace,
and the convex function $\theta\mapsto\sum_i\theta_i^2$ attains its maximum
over a compact convex set at an extreme point. By
Lemma~\ref{lem:two_block_extreme} applied to
$\mathcal K_r=r\mathcal K_1$, every extreme point takes exactly two
distinct values, say $k$ coordinates equal to $x$ and $N-k$ equal to $y$
for some $1\le k\le N-1$. The zero-sum condition gives
\[
kx+(N-k)y=0.
\]
Writing $d=x-y$, the complete-graph total variation is
\[
k(N-k)|d|=r
\]
at the maximizing boundary. Thus
\[
d
=
\frac{r}{k(N-k)},
\qquad
x
=
\frac{N-k}{N}d,
\qquad
y
=
-\frac{k}{N}d.
\]
Therefore
\[
\operatorname{Var}(\theta)
=
\frac{r^2}{N^2k(N-k)}.
\]
This is maximized when $k(N-k)$ is minimized, namely at
$k=1$ or $k=N-1$, giving
\[
\operatorname{Var}(\theta)
\le
\frac{r^2}{N^2(N-1)}.
\]
Division by $\pi^2/3$ yields the SNR bound.

For a general graph, the definition of $C_A$ implies
\[
\frac{\|\theta\|_2}{\sqrt N}
\le
C_A\|A\theta\|_1
\le
C_Ar.
\]
Squaring gives $\operatorname{Var}(\theta)\le C_A^2r^2$, and division by
$\pi^2/3$ gives the general SNR bound.
\end{proof}

\begin{proof}[Proof of Theorem~\ref{thm:gini_law}]
By Theorem~\ref{thm:single_block_limit} there is $L_A<\infty$ with
$|\theta_i-\theta_j|\le2L_Ar$ for every $\theta\in\mathcal K_r$, so all
remainders below are uniform over $\mathcal K_r$, with constants depending
on $N$ and on the graph but not on $\theta$. For small ability contrasts,
\[
\sigma(\theta_i-\theta_j)
=
\frac12+
\frac{\theta_i-\theta_j}{4}
+
O(|\theta_i-\theta_j|^3).
\]
Hence
\begin{align*}
w_i(\theta)
&=
\frac12
+
\frac{1}{4(N-1)}
\sum_{j\ne i}
(\theta_i-\theta_j)
+
O(r^3)\\
&=
\frac12
+
\frac{N}{4(N-1)}\theta_i
+
O(r^3),
\end{align*}
using $\sum_j\theta_j=0$. Therefore
\[
w_i(\theta)-w_j(\theta)
=
\frac{N}{4(N-1)}
(\theta_i-\theta_j)
+
O(r^3).
\]
Since $|a+\varepsilon|=|a|+O(|\varepsilon|)$ and $\bar w=1/2$ exactly, the
double sum in \eqref{eq:structural_gini} satisfies
\[
\sum_{i=1}^N\sum_{j=1}^N|w_i-w_j|
=
2\sum_{i<j}|w_i-w_j|
=
\frac{N}{2(N-1)}
\sum_{i<j}|\theta_i-\theta_j|
+
O(r^3),
\]
and dividing by $2N^2\bar w=N^2$ gives
\[
G_{\mathrm{struct}}(\theta)
=
\frac{1}{N^2}\cdot\frac{N}{2(N-1)}
\sum_{i<j}
|\theta_i-\theta_j|
+
O(r^3)
=
\frac{1}{2N(N-1)}
\sum_{i<j}
|\theta_i-\theta_j|
+
O(r^3).
\]
For the complete graph,
\[
\sum_{i<j}
|\theta_i-\theta_j|
=
\|A\theta\|_1
\le r,
\]
which gives \eqref{eq:gini_radius_bound}.
\end{proof}

\section{Proofs for Section~\ref{sec:dynamics}}
\label{app:proofs_dynamics}

The two lemmas below concern a single period, so the index $t$ is
suppressed. Throughout, $A$ is the oriented incidence matrix of a connected
graph on $\mathcal I$ (repeated edges allowed),
$\mathcal K=\mathcal K(1)=\{u\in H_0:\|Au\|_1\le1\}$,
$D=\max_{u\in\mathcal K}\|u\|_2<\infty$, $\theta(r)=\Pi_{\mathcal K(r)}(\beta)$
with $\theta(0)=0$, and $\mathcal K(r)=r\mathcal K$. For $u\in\mathcal K$,
$N_{\mathcal K}(u)$ denotes the normal cone of $\mathcal K$ at $u$ in
$\mathbb R^N$ and $\widetilde N(u)=N_{\mathcal K}(u)\cap H_0$. Since
$\mathcal K\subset H_0$, $N_{\mathcal K}(u)=\widetilde N(u)\oplus
\operatorname{span}(\mathbf1)$, and $x=\Pi_{\mathcal K}(b)$ if and only if
$x\in\mathcal K$ and $Hb-x\in\widetilde N(x)$. Because the graph is
connected, $\mathcal K$ is a polytope that contains a neighborhood of $0$ in
$H_0$, so it is full dimensional in $H_0$
(proof of Proposition~\ref{prop:exact_ties}).

\begin{lemma}[Small-radius vertex limit]
\label{lem:vertex_limit}
\begin{enumerate}[label=(\roman*)]
\item A point $w$ is an extreme point of $\mathcal K$ if and only if, for
some nonempty proper $S\subset\mathcal I$ such that $S$ and
$\mathcal I\setminus S$ both induce connected subgraphs (equivalently, the
edges joining $S$ to its complement form a minimal cut),
\begin{equation}
w=\pm\frac{\mathbf1_S-|S|N^{-1}\mathbf1}{|\partial S|},
\label{eq:general_vertex}
\end{equation}
where $|\partial S|\ge1$ is the number of edges, counted as in $A$, joining
$S$ to its complement.
\item For Lebesgue-almost every $\beta\in\mathbb R^N$ the linear functional
$u\mapsto\beta^\top u$ has a unique maximizer $v(\beta)$ over $\mathcal K$,
which is an extreme point, and there exists $\bar r(\beta)\in(0,\infty]$ such
that
\begin{equation}
\theta(r)=r\,v(\beta)\qquad\text{for every }0<r\le\bar r(\beta).
\label{eq:vertex_limit}
\end{equation}
Moreover $v(-\beta)=-v(\beta)$ and $v(c\beta)=v(\beta)$ for $c>0$.
\item For Lebesgue-almost every $\beta$, the path $r\mapsto\theta(r)$ is
continuous and piecewise affine on $[0,\infty)$ with finitely many pieces,
every slope has Euclidean norm at most $D$, and hence
$\|\theta(r)-\theta(r')\|_2\le D|r-r'|$ for all $r,r'\ge0$.
\end{enumerate}
\end{lemma}

By (i) and (ii), for every connected graph the latent ability vector near
parity has exactly two tiers, $S(\beta)$ and its complement. For the
complete graph with each dyad counted once, $|\partial S|=|S|(N-|S|)$ and
\eqref{eq:general_vertex} reduces to Lemma~\ref{lem:two_block_extreme},
because on the complete graph every nonempty proper $S$ satisfies the
connectivity condition.

\begin{proof}[Proof of Lemma~\ref{lem:vertex_limit}]
\emph{(i)} Let $w$ be an extreme point. Then $\|Aw\|_1=1$, since a point
with $\|Aw\|_1<1$ has a neighborhood in $H_0$ contained in $\mathcal K$. Let
$E_0$ be the set of edges $e$ with $(Aw)_e=0$, and let $C_1,\ldots,C_c$ be the
vertex sets of the connected components of the graph $(\mathcal I,E_0)$.
Every edge joining two different components has a nonzero contrast at $w$,
and $w$ is constant on each $C_l$. Let $W$ be the space of vectors in $H_0$
that are constant on each $C_l$, so $\dim W=c-1$, and let
$\varsigma_e=\operatorname{sign}\{(Aw)_e\}$ for $e\notin E_0$. On $W$, the
contrasts of the edges in $E_0$ vanish identically, and for $u\in W$
sufficiently close to $w$ the contrasts of the edges outside $E_0$ keep the
signs $\varsigma_e$, so $\|Au\|_1=\ell(u)$ with the linear functional
$\ell(u)=\sum_{e\notin E_0}\varsigma_e(Au)_e$. Since $\ell(w)=1$ and
$\ell(0)=0$, $\ell$ is not identically zero on $W$, and
$\{u\in W:\ell(u)=1\}$ is an affine subspace of dimension $c-2$ through $w$.
Its points near $w$ satisfy $\|Au\|_1=1$ and therefore lie in $\mathcal K$.
If $c\ge3$, this subspace contains a segment through $w$ in its relative
interior, contradicting extremality. If $c=1$, $w$ is constant and, being
zero-sum, equals $0$, contradicting $\|Aw\|_1=1$. Hence $c=2$, $w$ equals
$a$ on $S=C_1$ and $b$ on $C_2=\mathcal I\setminus S$ with $a\ne b$, and both
sets induce connected subgraphs because they are components of
$(\mathcal I,E_0)$. Writing $k=|S|$, the zero-sum condition gives
$a=(N-k)(a-b)/N$ and $b=-k(a-b)/N$, while $\|Aw\|_1=|\partial S|\,|a-b|=1$;
together these give \eqref{eq:general_vertex}. Connectivity of the graph
gives $|\partial S|\ge1$.

Conversely, let $w$ be given by \eqref{eq:general_vertex} with $S$ and
$\mathcal I\setminus S$ connected, so $w\in\mathcal K$ and $\|Aw\|_1=1$, and
suppose $w=(y+z)/2$ with $y,z\in\mathcal K$. Then
$1=\|Aw\|_1\le(\|Ay\|_1+\|Az\|_1)/2\le1$, so equality holds edge by edge:
$|(Aw)_e|=\{|(Ay)_e|+|(Az)_e|\}/2$ for every $e$. For an edge inside $S$ or
inside its complement, $(Aw)_e=0$, so $(Ay)_e=(Az)_e=0$. Because both sides
are connected, $y$ and $z$ are constant on $S$ and on its complement, and
being zero sum they are multiples of $\mathbf1_S-|S|N^{-1}\mathbf1$. For a
cut edge, $(Ay)_e$ and $(Az)_e$ have the sign of $(Aw)_e$, and
$\|Ay\|_1=\|Az\|_1=1$, so $y=z=w$. Hence $w$ is extreme.

\emph{(ii)} Every $c\in H_0$ attains its maximum over the polytope
$\mathcal K$ at some extreme point $w$, that is $c\in\widetilde N(w)$, so
the finitely many polyhedral cones $\widetilde N(w)$ cover $H_0$. Their
relative boundaries, and any of them with empty interior relative to $H_0$,
lie in finitely many hyperplanes of $H_0$; the preimages of these under
$\beta\mapsto H\beta$ are hyperplanes of $\mathbb R^N$ and hence
Lebesgue-null. If $c$ lies in the relative interior of $\widetilde N(w)$,
then $w$ is the unique maximizer of $u\mapsto c^\top u$: if $u\in\mathcal K$,
$u\ne w$ and $c^\top u=c^\top w$, then $c'=c+\varepsilon(u-w)\in\widetilde N(w)$
for small $\varepsilon>0$, yet $c'^\top(u-w)=\varepsilon\|u-w\|_2^2>0$, which
is impossible. Because $\beta^\top u=(H\beta)^\top u$
for $u\in H_0$, for almost every $\beta$ the vector $H\beta$ lies in the
relative interior of $\widetilde N(w)$ for exactly one extreme point
$w=v(\beta)$, which is then the unique maximizer of $u\mapsto\beta^\top u$.
The two invariances follow from $\mathcal K=-\mathcal K$ and from the
invariance of the maximizer under positive scaling of the functional.

Fix such a $\beta$ and write $w=v(\beta)$. Since
$\mathcal K(r)=r\mathcal K$, $\theta(r)=r\,\Pi_{\mathcal K}(\beta/r)$ for $r>0$,
and by the projection characterization above,
$\Pi_{\mathcal K}(\beta/r)=w$ if and only if $H\beta/r-w\in\widetilde N(w)$,
equivalently $H\beta-rw\in\widetilde N(w)$ because $\widetilde N(w)$ is a
cone. The set $\{r\ge0:H\beta-rw\in\widetilde N(w)\}$ is closed and convex,
and it contains a neighborhood of $0$ in $[0,\infty)$ because $H\beta$ is a
relative interior point of $\widetilde N(w)$. It is therefore an interval
$[0,\bar r(\beta)]$ (or $[0,\infty)$) with $\bar r(\beta)>0$, which proves
\eqref{eq:vertex_limit}. In addition, $H\beta-rw$ lies in the relative
interior of $\widetilde N(w)$ for $0\le r<\bar r(\beta)$, because it lies on
the segment between a relative interior point and a point of the cone.

\emph{(iii)} For a nonempty face $F$ of $\mathcal K$, let $L_F$ be the
direction space of $\operatorname{aff}(F)$, $P_F$ the orthogonal projection
onto $L_F$, and $q_F=\Pi_{\operatorname{aff}(F)}(0)$ the point of
$\operatorname{aff}(F)$ nearest the origin, so that
$\|q_F\|_2\le\min_{u\in F}\|u\|_2\le D$. If
$x=\Pi_{\mathcal K}(b)\in\operatorname{relint}(F)$, then
$(b-x)^\top(u-x)\le0$ for all $u\in F$, and applying this to $u$ and to
$2x-u\in F$ (valid for $u\in F$ close to $x$) gives $b-x\perp L_F$; hence
$x=\Pi_{\operatorname{aff}(F)}(b)=P_Fb+q_F$. The sets
$\mathcal R_F=\Pi_{\mathcal K}^{-1}\{\operatorname{relint}(F)\}$ partition
$\mathbb R^N$, and each is the intersection of the preimage of the relatively
open polytope $\operatorname{relint}(F)$ under the affine map
$b\mapsto P_Fb+q_F$ with the polyhedral set
$\{b:b-P_Fb-q_F\in N_{\mathcal K}(x_F)\}$, where $x_F$ is any point of
$\operatorname{relint}(F)$. Hence the boundaries of all the $\mathcal R_F$
lie in a finite union $\mathfrak H$ of affine hyperplanes of $\mathbb R^N$.

Exclude the Lebesgue-null set of $\beta$ that lie in a hyperplane of
$\mathfrak H$ passing through the origin. The open ray
$\{\beta/r:r>0\}$ then meets each hyperplane of $\mathfrak H$ in at most one
point, so $(0,\infty)$ splits into finitely many open intervals, on each of
which $\beta/r$ stays in the interior of a single $\mathcal R_F$. On such an
interval,
$\theta(r)=r\{P_F(\beta/r)+q_F\}=P_F\beta+rq_F$, which is affine with slope
$q_F$ of norm at most $D$. The map $r\mapsto\theta(r)$ is continuous on
$(0,\infty)$ because $\theta(r)=r\Pi_{\mathcal K}(\beta/r)$ and
$\Pi_{\mathcal K}$ is continuous, and it is continuous at $0$ because
$\|\theta(r)\|_2\le rD$. A continuous piecewise-affine function with
finitely many pieces and slopes bounded by $D$ is $D$-Lipschitz.
\end{proof}

For a pair $i\ne j$ write
\[
\delta_r=\theta_i(r)-\theta_j(r),\qquad
w_{ij}(\beta)=v_i(\beta)-v_j(\beta),\qquad
c_{ij}=\mathbb E\{w_{ij}(\beta)^2\},
\]
with $\beta\sim N(0,\sigma_\beta^2I_N)$. By the scale invariance in
Lemma~\ref{lem:vertex_limit}(ii), $c_{ij}$ does not depend on $\sigma_\beta$.
Since $\mathcal K=-\mathcal K$ implies $\Pi_{\mathcal K(r)}(-b)
=-\Pi_{\mathcal K(r)}(b)$ and the law of $\beta$ is symmetric, the process
$(\theta(r))_{r\ge0}$ has the same law as $(-\theta(r))_{r\ge0}$. In
particular $\mathbb E\,a(\delta_r)=0$ for odd $a$, and, writing
$\delta_r'=d\delta_r/dr$, $\mathbb E\{b(\delta_r)\delta_r'\}=0$ for even $b$.
By Lemma~\ref{lem:vertex_limit}(iii), for almost every $\beta$ the map
$r\mapsto\delta_r$ is absolutely continuous with
$|\delta_r'|\le\sqrt2D$ and $|\delta_r|\le\sqrt2Dr$.

For a period with $h_g\equiv1$, two comparisons $g,g'$ of the dyad
$\{i,j\}$ have orientations $s_g,s_{g'}\in\{-1,1\}$, with $s_g=1$ when $i$
is the home side. Since $x_g^\top\theta(r)=s_g\delta_r$, their linear
predictors are
\[
\Delta_g=\eta+s_g\delta_r,\qquad\Delta_{g'}=\eta+s_{g'}\delta_r ,
\]
so the home effect enters both with a positive sign and the ability
contrast changes sign with the orientation. A home-and-away pair has
$s_gs_{g'}=-1$. Write $q_k(\Delta)$, $k\in\{-1,0,1\}$, for the cell
probabilities in \eqref{eq:ordered_bt_likelihood}, and note that
$q_1(\Delta)=\sigma(\Delta-\kappa_2)$.

\begin{lemma}[Second-order emission separation at small radii]
\label{lem:small_radius_separation}
Fix a connected graph, a pair $i\ne j$, and $\kappa_2>0$, and let
$\beta\sim N(0,\sigma_\beta^2I_N)$ be independent of $\eta$.
\begin{enumerate}[label=(\roman*)]
\item $c_{ij}>0$, and $f(r)=\mathbb E(\delta_r^2)$ is absolutely continuous
with $f'(r)=2c_{ij}r+o(r)$ for almost every $r$ as $r\downarrow0$; in
particular $f(r)=c_{ij}r^2+o(r^2)$.
\item (Conditional covariance.) Let $g\ne g'$ be comparisons of $\{i,j\}$
in the same period, fix $\eta$, let $\psi$ be a function on $\{-1,0,1\}$,
and let $m_\psi(\Delta)=\sum_k\psi(k)q_k(\Delta)$. Then
\begin{equation}
\operatorname{Cov}_r\{\psi(y_g),\psi(y_{g'})\mid\eta\}
=s_gs_{g'}\,m_\psi'(\eta)^2\,c_{ij}\,r^2+o(r^2),
\qquad r\downarrow0 .
\label{eq:emission_cov_expansion}
\end{equation}
For the home-win indicator $m_\psi'(\eta)=\sigma'(\eta-\kappa_2)>0$, for the
away-win indicator $m_\psi'(\eta)=-\sigma'(\eta+\kappa_2)<0$, and for the
draw indicator $m_\psi'(\eta)=\sigma'(\kappa_2+\eta)-\sigma'(\kappa_2-\eta)$,
which vanishes if and only if $\eta=0$.
\item (Marginal home-and-away discordance.) Let $(g,g')$ be a home-and-away
pair of $\{i,j\}$, let $\eta$ have any distribution independent of $\beta$,
and let
\[
S_{gg'}(r)=\Pr\nolimits_r\{\mathbb I(y_g=1)\ne\mathbb I(y_{g'}=1)\}
\]
under the emission distribution \eqref{eq:integrated_emission}. With
$x=\eta-\kappa_2$, set
$Q_2(\eta)=2\sigma'(x)\{1-2\sigma(x)\}^2+4\sigma'(x)^2$ and
$\bar Q_2=\mathbb E\,Q_2(\eta)$. Then $\bar Q_2>0$,
\begin{equation}
S_{gg'}(r)=S_{gg'}(0)+\tfrac12\bar Q_2\,c_{ij}\,r^2+o(r^2),
\label{eq:discordance_expansion}
\end{equation}
and there is $r_{ij}>0$ such that $S_{gg'}$ is strictly increasing on
$[0,r_{ij}]$.
\end{enumerate}
\end{lemma}

\begin{proof}[Proof of Lemma~\ref{lem:small_radius_separation}]
\emph{(i)} Suppose $w_i=w_j$ for every extreme point $w$ of $\mathcal K$.
Because $\mathcal K$ is the convex hull of its extreme points, $u_i=u_j$ for
every $u\in\mathcal K$. But $u=\varepsilon(e_i-e_j)$ lies in $H_0$ and
satisfies $\|Au\|_1\le1$ for small $\varepsilon>0$, and $u_i-u_j=2\varepsilon$, a
contradiction. Hence some extreme point $w$ has $w_i\ne w_j$. By the proof
of Lemma~\ref{lem:vertex_limit}(ii), $v(\beta)=w$ whenever $H\beta$ lies in
the relative interior of $\widetilde N(w)$. This cone has nonempty
interior relative to $H_0$, because the normal cone of a face of dimension
$d$ of a polytope that is full dimensional in $H_0$ has dimension $N-1-d$
(proof of Proposition~\ref{prop:exact_ties}). Since $H\beta$ has a positive
density on $H_0$, the event has positive probability and
$c_{ij}\ge(w_i-w_j)^2\Pr\{v(\beta)=w\}>0$.

By absolute continuity and Fubini's theorem, with integrand bounded by
$4D^2u$,
\[
f(r)=\int_0^r\varphi(u)\,du,\qquad
\varphi(u)=\mathbb E(2\delta_u\delta_u').
\]
By Lemma~\ref{lem:vertex_limit}(ii), on the event $\{u<\bar r(\beta)\}$ the
path is $\delta_u=uw_{ij}$ near $u$, so $2\delta_u\delta_u'=2uw_{ij}^2$.
Using $w_{ij}^2\le2D^2$ and $|2\delta_u\delta_u'|\le4D^2u$,
\[
|\varphi(u)-2uc_{ij}|
\le2u\,\mathbb E\{w_{ij}^2;\bar r(\beta)\le u\}
+\mathbb E\{|2\delta_u\delta_u'|;\bar r(\beta)\le u\}
\le8D^2u\Pr\{\bar r(\beta)\le u\}.
\]
Since $\bar r(\beta)>0$ almost surely, $\Pr\{\bar r(\beta)\le u\}\to0$, so
$\varphi(u)=2c_{ij}u+o(u)$, and integrating gives
$f(r)=c_{ij}r^2+o(r^2)$.

\emph{(ii)} Given $(\theta,\eta)$ the outcomes are independent, so
$\operatorname{Cov}_r\{\psi(y_g),\psi(y_{g'})\mid\eta\}
=\operatorname{Cov}\{m(\eta+s_g\delta_r),m(\eta+s_{g'}\delta_r)\}$, with
$m=m_\psi$ and the covariance taken over $\beta$. The function $m$ is a
finite combination of logistic functions, so its derivatives of every order
are bounded; let $M_3=\sup|m'''|$. Taylor's theorem gives
$m(\eta+s\delta)=m(\eta)+X_s$ with
\[
X_s=s\,m'(\eta)\delta+\tfrac12m''(\eta)\delta^2+R_s,\qquad
|R_s|\le M_3|\delta|^3/6 .
\]
Hence $\operatorname{Cov}\{m(\eta+s_g\delta_r),m(\eta+s_{g'}\delta_r)\}
=\operatorname{Cov}(X_{s_g},X_{s_{g'}})$. Expanding, the main term is
$s_gs_{g'}m'(\eta)^2\operatorname{Var}(\delta_r)
=s_gs_{g'}m'(\eta)^2f(r)$, because $\mathbb E\delta_r=0$. The term
$\operatorname{Cov}(\delta_r,\delta_r^2)=\mathbb E\delta_r^3$ vanishes by
symmetry. Each remaining term is a covariance involving $\delta_r^2$ or
$R_s$ together with a factor of order at least $\delta_r$; since
$|R_s|\le M_3|\delta_r|^3/6$ and $|\delta_r|\le\sqrt2Dr$, each is $O(r^4)$. Part (i) then gives
\eqref{eq:emission_cov_expansion}.

For the derivatives, $q_1(\Delta)=1-\sigma(\kappa_2-\Delta)$ and
$q_{-1}(\Delta)=\sigma(-\kappa_2-\Delta)$, so $q_1'(\Delta)=\sigma'(\kappa_2-\Delta)$,
$q_{-1}'(\Delta)=-\sigma'(\kappa_2+\Delta)$ and
$q_0'=-q_1'-q_{-1}'$, using that $\sigma'$ is even. Since $\sigma'$ is
even and strictly decreasing on $[0,\infty)$,
$\sigma'(\kappa_2+\eta)=\sigma'(\kappa_2-\eta)$ if and only if
$|\kappa_2+\eta|=|\kappa_2-\eta|$, that is $\eta=0$, given $\kappa_2>0$.

\emph{(iii)} Let $m=q_1$ and $p_\pm=m(\eta\pm\delta)$, the home-win
probabilities of the two matches when $s_g=1$ (the case $s_g=-1$ swaps the
two and leaves the event unchanged). By conditional independence,
\[
\Pr\{\mathbb I(y_g=1)\ne\mathbb I(y_{g'}=1)\mid\theta,\eta\}
=Q(\eta,\delta_r),\qquad
Q(\eta,\delta)=p_++p_--2p_+p_- .
\]
The function $Q(\eta,\cdot)$ is even and smooth, and its derivatives are
bounded uniformly in $(\eta,\delta)$. Hence
$\partial_\delta Q(\eta,\delta)=Q_2(\eta)\delta+R(\eta,\delta)$ with
$|R|\le M_4|\delta|^3/6$, where $M_4=\sup|\partial_\delta^4Q|$ and
$Q_2(\eta)=\partial_\delta^2Q(\eta,0)$. Differentiating,
$\partial_\delta^2p_\pm|_{0}=m''(\eta)$ and
$\partial_\delta^2(p_+p_-)|_0=2m(\eta)m''(\eta)-2m'(\eta)^2$, so
\[
Q_2(\eta)=2m''(\eta)\{1-2m(\eta)\}+4m'(\eta)^2 .
\]
With $m(\Delta)=\sigma(\Delta-\kappa_2)$ we have $m'=\sigma'$ and
$m''=\sigma'(1-2\sigma)$ evaluated at $x=\eta-\kappa_2$, which gives the stated
form and $Q_2(\eta)\ge4\sigma'(x)^2>0$, hence $\bar Q_2>0$ for every
distribution of $\eta$.

For fixed $\eta$ and almost every $\beta$, $r\mapsto Q(\eta,\delta_r)$ is
absolutely continuous with derivative $\partial_\delta Q(\eta,\delta_r)\delta_r'$,
which is bounded. By Fubini's theorem and the independence of $\eta$ and
$\beta$,
\[
S_{gg'}(r)-S_{gg'}(0)=\int_0^r\Bigl\{\bar Q_2\,
\mathbb E(\delta_u\delta_u')+\rho(u)\Bigr\}du
=\int_0^r\Bigl\{\tfrac12\bar Q_2\varphi(u)+\rho(u)\Bigr\}du,
\]
with
$|\rho(u)|\le\mathbb E|R(\eta,\delta_u)\delta_u'|\le(M_4/6)(\sqrt2Du)^3\sqrt2D
=\tfrac23M_4D^4u^3$. By part (i), the integrand equals
$\bar Q_2c_{ij}u+o(u)$, which is positive on $(0,r_{ij}]$ for some
$r_{ij}>0$. This proves strict monotonicity on $[0,r_{ij}]$ and, after
integration, \eqref{eq:discordance_expansion}.
\end{proof}

\begin{remark}[Why covariances of repeated meetings are not used]
\label{rem:covariance_cancellation}
Integrating \eqref{eq:emission_cov_expansion} over $\eta$ does not give the
marginal covariance. With $\mu(\eta,r)=\mathbb E\{m(\eta+s\delta_r)\}$, which
by symmetry does not depend on $s$,
\[
\operatorname{Cov}_r\{\psi(y_g),\psi(y_{g'})\}
=\mathbb E_\eta\bigl[\operatorname{Cov}_r\{\psi(y_g),\psi(y_{g'})\mid\eta\}\bigr]
+\operatorname{Var}_\eta\{\mu(\eta,r)\}.
\]
By the expansion in part (ii), $\mu(\eta,r)=m(\eta)+\tfrac12m''(\eta)f(r)+O(r^4)$
uniformly in $\eta$. The leading $r^2$ coefficient of the marginal
covariance is therefore
$c_{ij}[s_gs_{g'}\mathbb E\{m'(\eta)^2\}+\operatorname{Cov}_\eta\{m(\eta),m''(\eta)\}]$.
For a home-and-away pair ($s_gs_{g'}=-1$), the sign of this coefficient
depends on the law of $\eta$ and on $\kappa_2$, so a sum of such covariances
need not be monotone in $r$. The discordance probability in part (iii) is a
function of $\delta_r$ that is even, with positive curvature at zero for
every value of $\eta$, and so avoids this cancellation.
\end{remark}

The dynamic model inherits deterministic control of the latent abilities
from the projection constraint.

\begin{assumption}[Comparison-graph regularity]
\label{ass:graph}
Let $\tau_t$ denote the smallest constant such that
\begin{equation}
\|\theta\|_1
\le
\tau_t\|A_t\theta\|_1
\label{eq:graph_poincare}
\end{equation}
for every $\theta\in\mathbb R^N$ satisfying $\mathbf1^\top\theta=0$. Assume
$\tau_{\max}=\sup_t\tau_t<\infty$.
\end{assumption}

\begin{proposition}[Regime-controlled ability bound]
\label{prop:dynamic_ability_bound}
Under Assumption~\ref{ass:graph},
$\|\theta_t\|_1\le\tau_tr_{s_t}\le\tau_{\max}r_{s_t}$ for every period $t$.
Consequently,
\begin{equation}
\max_{1\le t\le T}
\|\theta_t\|_1
\le
\tau_{\max}
\max_{1\le k\le K_{\mathrm{act}}}r_k,
\label{eq:dynamic_ability_bound}
\end{equation}
with $K_{\mathrm{act}}=|\mathcal A|$ the number of occupied regimes
(Section~\ref{subsec:adaptive_regime_model}). The bound is
deterministic given the regime radii.
\end{proposition}

\begin{proof}[Proof of Proposition~\ref{prop:dynamic_ability_bound}]
By construction,
\[
\theta_t
\in
\mathcal K_t(r_{s_t}),
\]
so
\[
\|A_t\theta_t\|_1
\le
r_{s_t}.
\]
Applying \eqref{eq:graph_poincare} gives
\[
\|\theta_t\|_1
\le
\tau_t\|A_t\theta_t\|_1
\le
\tau_tr_{s_t}.
\]
Taking the maximum over periods yields
\eqref{eq:dynamic_ability_bound}.
\end{proof}

\subsection{Posterior tail bound on the ability process}
\label{app:ability_tail}

Proposition~\ref{prop:dynamic_ability_bound} bounds the abilities in terms
of the realized regime radii. Averaging that bound over the prior turns it
into an explicit posterior tail bound. Fix the state sequence $s_{1:T}$,
with $K_{\mathrm{act}}\le T$ occupied regimes, so that the remaining
unknowns have the proper prior of Section~\ref{subsec:adaptive_regime_model},
and suppose Assumption~\ref{ass:graph} holds and $a_r>1$. Under the prior,
$\mathbb E(r_k)=\mathbb E(\alpha_r^{-1})=b_r/(a_r-1)$; set
$R_\star=K_{\mathrm{act}}\,b_r/(a_r-1)$. Then for every fixed dataset and
every $B>0$,
\begin{equation}
\Pr\left(
\max_{1\le t\le T}
\|\theta_t\|_1
>B
\;\middle|\;
\mathcal D,s_{1:T}
\right)
\le
\frac{\tau_{\max}R_\star}{B\,p(\mathcal D\mid s_{1:T})} ,
\label{eq:dynamic_posterior_tightness}
\end{equation}
where $p(\mathcal D\mid s_{1:T})\in(0,1]$ is the marginal likelihood given
the state sequence. That $\max_t\|\theta_t\|_1$ is finite with posterior
probability one already follows from Theorem~\ref{thm:posterior_propriety};
the content of \eqref{eq:dynamic_posterior_tightness} is the explicit rate in
$B$, which exhibits the dependence of the posterior tail on the number of
regimes and the prior mean radius, inflated by the graph constant
$\tau_{\max}$. The bound involves the unknown $p(\mathcal D\mid s_{1:T})$ and
is therefore not directly computable.

\begin{proof}[Proof of \eqref{eq:dynamic_posterior_tightness}]
By Proposition~\ref{prop:dynamic_ability_bound},
$\max_t\|\theta_t\|_1\le \tau_{\max}\max_{k\le K_{\mathrm{act}}}r_k$, so for
any $c>0$
\[
\Pr\Bigl(\max_t\|\theta_t\|_1>\tau_{\max}c\;\Big|\;\mathcal D,s_{1:T}\Bigr)
\le
\Pr\Bigl(\max_{k\le K_{\mathrm{act}}}r_k>c\;\Big|\;\mathcal D,s_{1:T}\Bigr).
\]
Since $0<L(\mathcal D\mid\Lambda)\le1$, for any event $\mathcal E$ we have
$\Pr(\mathcal E\mid\mathcal D,s_{1:T})\le\Pr(\mathcal E\mid s_{1:T})/
p(\mathcal D\mid s_{1:T})$, where the probability on the right is under the
prior. Bounding the maximum by the sum,
\[
\mathbb E\Bigl(\max_{k\le K_{\mathrm{act}}}r_k\Bigm| s_{1:T}\Bigr)
\le
\sum_{k=1}^{K_{\mathrm{act}}}\mathbb E(r_k)
=
\frac{K_{\mathrm{act}}\,b_r}{a_r-1}
=
R_\star<\infty .
\]
Markov's inequality then gives
$\Pr(\max_{k\le K_{\mathrm{act}}}r_k>c\mid s_{1:T})\le R_\star/c$, so
\[
\Pr\Bigl(\max_t\|\theta_t\|_1>\tau_{\max}c\;\Big|\;\mathcal D,s_{1:T}\Bigr)
\le
\frac{R_\star}{c\,p(\mathcal D\mid s_{1:T})} .
\]
Setting $c=B/\tau_{\max}$ yields \eqref{eq:dynamic_posterior_tightness}.
\end{proof}

\section{Posterior Computation and Summaries}
\label{app:computation}

\subsection{Posterior computation}
\label{subsec:posterior_computation}

Each sweep of the sampler updates, in turn, the auxiliary abilities, the
home effects, the occupied-state radii and their shared rate, and the state
sequence. The projected abilities are deterministic functions of the
auxiliary variables and the radii, so they are recomputed when needed rather
than sampled. For season $t$, write
$L_t(\theta_t,\eta_t,\kappa)=\prod_{g=1}^{G_t}p(y_{tg}\mid\theta_t,\eta_t,\kappa)$
for the likelihood contribution, with cell probabilities from
\eqref{eq:ordered_bt_likelihood} and the threshold $\kappa$ held fixed, and
\begin{equation}
\ell_t(\beta_t;\eta_t,\kappa,r)
=
\log L_t\{\Pi_{\mathcal K_t(r)}(\beta_t),\eta_t,\kappa\}
\label{eq:projected_loglikelihood}
\end{equation}
for the log-likelihood as a function of the auxiliary abilities.

The auxiliary abilities have prior $\beta_t\sim N(0,\sigma_\beta^2I_N)$ and are
updated by elliptical slice sampling with target proportional to
$\exp\{\ell_t(\beta_t;\eta_t,\kappa,r_{s_t})\}\,
\phi_N(\beta_t;0,\sigma_\beta^2I_N)$, recomputing the projection at every
likelihood evaluation with the solver of
Appendix~\ref{subsec:computation_projection}. Each home effect has an
independent $N(0,\sigma_\eta^2)$ prior and is updated by a Gaussian random
walk, $\eta_t^\star=\eta_t+\zeta_t$ with $\zeta_t\sim N(0,\varsigma_\eta^2)$,
accepted with probability
\[
\min\left\{1,\,
\frac{L_t(\theta_t,\eta_t^\star,\kappa)}
     {L_t(\theta_t,\eta_t,\kappa)}
\exp\left[-\frac{(\eta_t^\star)^2-\eta_t^2}
                 {2\sigma_\eta^2}\right]\right\}.
\]

The occupied radii follow $r_k\mid\alpha_r\sim\operatorname{Exponential}(\alpha_r)$
with shared rate $\alpha_r\sim\operatorname{Gamma}(a_r,b_r)$ in the
shape--rate parameterization. Writing $\mathcal T_k=\{t:s_t=k\}$ for the
seasons in state $k$, each radius is updated by a random walk on the log
scale, $\log r_k^\star=\log r_k+\zeta_k$ with
$\zeta_k\sim N(0,\varsigma_r^2)$, and accepted with probability
$\min(1,A_r)$, where
\begin{equation}
A_r=
\exp\{-\alpha_r(r_k^\star-r_k)\}
\frac{r_k^\star}{r_k}
\prod_{t\in\mathcal T_k}
\frac{L_t\{\Pi_{\mathcal K_t(r_k^\star)}(\beta_t),\eta_t,\kappa\}}
     {L_t\{\Pi_{\mathcal K_t(r_k)}(\beta_t),\eta_t,\kappa\}}
\label{eq:radius_mh_ratio}
\end{equation}
and $r_k^\star/r_k$ is the Hastings correction for the log-scale proposal
(Appendix~\ref{app:radius_hastings}). Given the $K_{\mathrm{act}}$ occupied
radii, the rate has the conjugate update
\begin{equation}
\alpha_r\mid r_{1:K_{\mathrm{act}}}
\sim
\operatorname{Gamma}\left(
a_r+K_{\mathrm{act}},\,
b_r+\sum_{k=1}^{K_{\mathrm{act}}} r_k
\right).
\label{eq:radius_rate_update}
\end{equation}

Opening a new state requires the season likelihood averaged over the
unknown radius,
$\int_0^\infty f_t(r)\,\alpha_re^{-\alpha_r r}\,dr
=\int_0^1 f_t\{F_{\alpha_r}^{-1}(q)\}\,dq$, where
$F_{\alpha_r}^{-1}(q)=-\log(1-q)/\alpha_r$ is the exponential quantile
function. After $\alpha_r$ is updated, we take $J$ equally spaced
probabilities $q_j=0.05+0.90(j-1)/(J-1)$, $j=1,\ldots,J$, and the
corresponding quantiles
\begin{equation}
r_j^{\mathrm{grid}}
=
\max\left\{
10^{-6},
-\frac{\log(1-q_j)}{\alpha_r}
\right\},
\label{eq:radius_quantile_grid}
\end{equation}
which stay fixed for the state sweep that follows. The new-state likelihood
\begin{equation}
\widetilde f_{t,0}
=
\frac{1}{J}\sum_{j=1}^J
f_t(r_j^{\mathrm{grid}})
\label{eq:new_state_integrated_score}
\end{equation}
is an equally weighted approximation of the integral over $q$, so no
additional density or Jacobian factor enters; the outer 5\% of prior
probability on each side is omitted. It enters the score
$w_t(0)=B_mI_t(0)O_t(0)\widetilde f_{t,0}$ of
Section~\ref{subsec:adaptive_allocation}, and the same grid supplies the
initial radius in \eqref{eq:new_state_radius_initialization}. We use $J=15$
in all analyses. Once a state is occupied, its radius is updated by
\eqref{eq:radius_mh_ratio} and is no longer restricted to grid values.

The seasons are then reallocated one at a time by
\eqref{eq:existing_transition_factors}--\eqref{eq:new_state_radius_initialization}.
The candidates for season $t$ are all currently occupied states, including
the current state of $t$ even when $t$ is its only member, together with one
new state. Empty states are deleted after each reassignment and the
remaining labels compacted, so no fixed state capacity is kept. This
allocation kernel is not the exact collapsed update of a finite hidden
Markov model with symmetric Dirichlet transition rows: its cell
pseudo-counts $a_{\mathrm{tr}}b_m$ sum to $a_{\mathrm{tr}}\gamma$ while the
outgoing denominator contains $a_{\mathrm{tr}}$, and the birth multiplier
$B_m$ is set algorithmically rather than derived from a prior on the number
of states. Posterior invariance has not been established for this step, and
we treat the sampler output as an approximate posterior.
Algorithm~\ref{alg:mfm_hmm_sampler} summarizes one sweep.

\begin{algorithm}[H]
\caption{Blockwise inference procedure}
\label{alg:mfm_hmm_sampler}
\begin{algorithmic}[1]
\Require Data $\mathcal D$; current
$(s_{1:T},r_{1:K_{\mathrm{act}}},\alpha_r,\beta_{1:T},\eta_{1:T})$; fixed threshold and
hyperparameters.
\Ensure One computational sweep.

\For{$t=1,\ldots,T$}
    \State Update $\beta_t$ by elliptical slice sampling, evaluating the
    projected likelihood in \eqref{eq:projected_loglikelihood}.
\EndFor

\For{$t=1,\ldots,T$}
    \State Update $\eta_t$ by Gaussian random-walk Metropolis--Hastings.
\EndFor

\For{each occupied state $k$}
    \State Update $r_k$ using \eqref{eq:radius_mh_ratio}.
\EndFor

\State Draw $\alpha_r$ from \eqref{eq:radius_rate_update}.
\State Construct the radius grid in \eqref{eq:radius_quantile_grid}.

\For{$t=1,\ldots,T$}
    \State Compute and normalize the state scores in
    \eqref{eq:existing_transition_factors}--\eqref{eq:adaptive_allocation_probability}.
    \State Draw the new allocation; if a new state is selected, initialize
    its radius using \eqref{eq:new_state_radius_initialization}.
    \State Remove empty states and compact the remaining labels.
\EndFor

\State Recompute
$\theta_t=\Pi_{\mathcal K_t(r_{s_t})}(\beta_t)$ for all seasons.
\State Store the draw after ordering state labels by increasing radius.
\end{algorithmic}
\end{algorithm}

\subsection{Computation of the projection step}
\label{subsec:computation_projection}

The projection is the convex quadratic program
\begin{equation}
\Pi_{\mathcal K_t(r)}(\beta_t)
=
\arg\min_{\vartheta}
\frac12\|\vartheta-\beta_t\|_2^2
\quad
\text{subject to}
\quad
\|A_t\vartheta\|_1\le r,\quad
\mathbf 1^\top\vartheta=0 .
\label{eq:computational_projection}
\end{equation}
In all analyses reported here the comparison graph is the unweighted
complete graph with each dyad counted once, and
\eqref{eq:computational_projection} has an exact solution. For every
$\vartheta$, with order statistics
$\vartheta_{(1)}\le\cdots\le\vartheta_{(N)}$,
$\|A_t\vartheta\|_1=\sum_{i=1}^N(2i-N-1)\vartheta_{(i)}$, and the projection
preserves the order of $\beta_t$. If $\|A_tH\beta_t\|_1\le r$ the solution is
$H\beta_t$. Otherwise, let $v$ be the values of $H\beta_t$ in increasing
order, $w_i=2i-N-1$, and $\vartheta(\lambda)$ the isotonic regression of
$v-\lambda w$. The sorted solution is $\vartheta(\lambda)$ with $\lambda>0$
solving $w^\top\vartheta(\lambda)=r$, found by a one-dimensional search, and
it is then returned to the original order. Because $\mathbf 1^\top v=0$,
$\mathbf 1^\top w=0$, and isotonic regression preserves the mean, the
zero-sum constraint holds automatically. Isotonic regression pools adjacent
values, which is how the exact ties of Proposition~\ref{prop:exact_ties}
arise in computation. For general comparison graphs, primal--dual or ADMM
algorithms can exploit the graph-fused structure.

\begin{proposition}[Almost-everywhere differentiability]
\label{prop:aew_differentiability}
Let $U(\beta_t) = -\ell_t(\beta_t;\eta_t,\kappa,r_{s_t}) +\|\beta_t\|_2^2/(2\sigma_\beta^2)$
be the negative log posterior kernel for $\beta_t$. For fixed
$(r_{s_t},\eta_t,\kappa)$, $U(\beta_t)$ is continuous and differentiable
almost everywhere with respect to Lebesgue measure on $\mathbb R^N$. See
Appendix~\ref{app:differentiability_proof} for a proof.
\end{proposition}

\subsection{Posterior summaries}
\label{subsec:post_mcmc}

The point estimates used in the simulations and in the applications are
computed from the retained draws of Algorithm~\ref{alg:mfm_hmm_sampler}. At each
retained iteration $m$, the occupied radii are sorted,
$r_{\sigma_m(1)}^{(m)} \le \cdots \le r_{\sigma_m(K^{(m)})}^{(m)}$, and the
state labels are permuted to match, so smaller labels always mean smaller
radii.

The number of states is estimated by the empirical mode $\widehat K$ of
$K^{(m)}$, with ties broken in favor of the smaller value. Over the draws
$\mathcal M_{\widehat K}=\{m:K^{(m)}=\widehat K\}$ we form the season
co-clustering matrix
\begin{equation}
\bar C_{tt'}
=
\frac{1}{|\mathcal M_{\widehat K}|}
\sum_{m\in\mathcal M_{\widehat K}}
\mathbb I\{s_t^{(m)}=s_{t'}^{(m)}\}
\label{eq:posterior_coclustering}
\end{equation}
and select the Dahl representative draw
\begin{equation}
m^\star
=
\arg\min_{m\in\mathcal M_{\widehat K}}
\|C^{(m)}-\bar C\|_F^2,
\qquad
C^{(m)}_{tt'}
=
\mathbb I\{s_t^{(m)}=s_{t'}^{(m)}\}.
\label{eq:dahl_representative}
\end{equation}
The state path and the radii both come from this draw,
$\hat s_t=s_t^{(m^\star)}$ and $\hat r_t=r_{\hat s_t}^{(m^\star)}$, so the
reported radius is the one paired with the representative path.

Team tiers and abilities are summarized season by season from the draws in
which season $t$ belongs to its Dahl state,
$\mathcal S_t=\{m\in\mathcal M_{\widehat K}:s_t^{(m)}=\hat s_t\}$. In each
such draw the tiers are the groups of teams with exactly equal abilities,
which the exact projection produces with positive probability
(Proposition~\ref{prop:exact_ties}); write $c_t^{(m)}$ for this partition
of the teams. The estimated tiers $\hat c_t$ are the SALSO estimate
\citep{dahl2022search} of $\{c_t^{(m)}:m\in\mathcal S_t\}$ under the variation of
information loss, with at most as many tiers as the most frequent number of
tiers among these partitions, and are labeled in decreasing order of
ability. Each team's reported ability is its posterior mean averaged over
its estimated tier,
\begin{equation}
\hat\theta_{ti}
=
\frac{1}{|\{j:\hat c_{tj}=\hat c_{ti}\}|}
\sum_{j:\hat c_{tj}=\hat c_{ti}}
\frac{1}{|\mathcal S_t|}\sum_{m\in\mathcal S_t}\theta_{tj}^{(m)}.
\label{eq:tier_averaged_ability}
\end{equation}
Home effects are summarized by their posterior means over all $M$
retained draws, $\hat\eta_t=M^{-1}\sum_{m=1}^M\eta_t^{(m)}$. The distribution of $K$, the
season co-clustering frequencies and the radius draws convey the uncertainty
around these point estimates.

The figures of the applications also show, for each season, the posterior
median and 90\% credible interval of the season radius and the posterior
mean and 90\% credible interval of each team's ability over $\mathcal S_t$.
These intervals average over regime allocations and are therefore
approximate in the sense of Section~\ref{sec:inference}.

\subsection{Diagnostics}
\label{subsec:mcmc_diagnostics}

For the two-chain real-data analyses, we inspect trace plots of the
season-assigned radii $r_{s_t}$, the home effects and selected components of
$\theta_t$, compare the distribution of the number of occupied states $K$
between chains, and compute split-$\widehat R$ and effective sample sizes for
the continuous parameters. The proposal scales of the two
Metropolis--Hastings updates are fixed within each run, and their acceptance
rates are recorded as a further check on mixing.

\section{Derivations for Appendix~\ref{app:computation}}
\label{app:derivations}

\subsection{Hastings correction for the log-scale radius proposal}
\label{app:radius_hastings}

The factor $r_k^\star/r_k$ in \eqref{eq:radius_mh_ratio} is the Hastings correction
for the asymmetric proposal on $r_k$ induced by a symmetric random walk on
$\log r_k$. Since $\log r_k^\star\mid\log r_k\sim N(\log r_k,\varsigma_r^2)$,
writing $r_k^\star=e^{\log r_k+\zeta_k}$ with $\zeta_k\sim
N(0,\varsigma_r^2)$ and applying the change-of-variables Jacobian
$|d\zeta_k/dr_k^\star|=1/r_k^\star$ gives proposal density
$q(r_k^\star\mid r_k)=\phi(\zeta_k/\varsigma_r)/(\varsigma_r r_k^\star)$. By
symmetry of $\phi$ in $\zeta_k\leftrightarrow-\zeta_k$,
\begin{equation*}
\frac{q(r_k\mid r_k^\star)}{q(r_k^\star\mid r_k)}
=
\frac{\phi(\zeta_k/\varsigma_r)/(\varsigma_r r_k)}
{\phi(\zeta_k/\varsigma_r)/(\varsigma_r r_k^\star)}
=
\frac{r_k^\star}{r_k},
\end{equation*}
which is exactly the factor appearing in \eqref{eq:radius_mh_ratio}.

\subsection{Proof of Proposition~\ref{prop:aew_differentiability}}
\label{app:differentiability_proof}

\begin{proof}[Proof of Proposition~\ref{prop:aew_differentiability}]
The Euclidean projection onto a closed convex polyhedron is Lipschitz
continuous and piecewise affine, the affine pieces corresponding to the
partition of $\mathbb R^N$ induced by the active face of the polyhedron at
each point; this is a standard consequence of parametric quadratic
programming applied to a fixed polyhedral feasible set. Since
$\mathcal K_t(r_{s_t})$ is polyhedral --- the constraint
$\|A_t\vartheta\|_1\le r_{s_t}$ admits a finite halfspace representation,
intersected with the hyperplane $\mathbf 1^\top\vartheta=0$ --- the map
$\beta_t\mapsto\Pi_{\mathcal K_t(r_{s_t})}(\beta_t)$ is differentiable
except on a finite union of lower-dimensional boundaries between affine
pieces, a set of Lebesgue measure zero. The likelihood is smooth in the
projected abilities away from these boundaries, and the Gaussian quadratic
term is smooth everywhere, so $U(\beta_t)$ is differentiable everywhere
except on this measure-zero set.
\end{proof}

\section{Simulations}\label{sec:simu}

This appendix gives the simulation designs summarized in
Section~\ref{sec:simulation} and the full results. The study asks whether
the model recovers the number of states, the state sequence, the radii, the
team abilities and the tiers, across two or three states, 20 or 40 seasons,
and one or two ability patterns per state. Each of the eight settings is
replicated 50 times.

\subsection{Simulation Settings}

Every season has $N=20$ teams playing a double round robin, so each pair
meets once at each venue and a season contains $2\binom{20}{2}=380$ matches.
When team $i$ hosts team $j$ in season $t$, the latent performance difference
is $Y_{ijt}^{*}=\theta_{ti}-\theta_{tj}+\eta_t+\xi_{ijt}$, where $\xi_{ijt}$ is
a standard logistic error with variance $\pi^2/3$ as in
Section~\ref{subsec:snr_gini}, and the observed outcome is
\[
Y_{ijt}
=
\begin{cases}
1, & Y_{ijt}^{*}>\kappa_2,\\
0, & -\kappa_2\leq Y_{ijt}^{*}\leq\kappa_2,\\
-1, & Y_{ijt}^{*}<-\kappa_2,
\end{cases}
\]
that is, a home win, draw or home loss. We set $\eta_t=0.30$ and
$\kappa_2=0.30$ in every season. This is the ordered likelihood
\eqref{eq:ordered_bt_likelihood} with $\kappa_1=-\kappa_2$, so the simulated
data satisfy the anchoring of Section~\ref{subsec:projection_model} by
construction. The radius of an ability vector is its total variation on the
complete graph, with each of the $\binom{20}{2}=190$ pairs counted once,
$r(\boldsymbol\theta)=\|A\boldsymbol\theta\|_1=\sum_{i<j}|\theta_i-\theta_j|$.

All designs are built from two kinds of ability vector. The two-tier vector
\[
\boldsymbol u_a
=
(\underbrace{-a,\ldots,-a}_{10},\underbrace{a,\ldots,a}_{10})
\]
has radius $r(\boldsymbol u_a)=200a$, so $\boldsymbol u_{1/2}$,
$\boldsymbol u_1$ and $\boldsymbol u_2$ have radii 100, 200 and 400. The
three-tier vector
\[
\boldsymbol v
=
(\underbrace{-4/3,\ldots,-4/3}_{5},\underbrace{0,\ldots,0}_{10},
\underbrace{4/3,\ldots,4/3}_{5})
\]
has radius 200, the same as $\boldsymbol u_1$, but a different tier
structure. In the baseline settings every state keeps a single ability
vector. In the heterogeneous settings the radius-200 state alternates between
$\boldsymbol u_1$ and $\boldsymbol v$ from one occurrence to the next,
starting with $\boldsymbol u_1$, so the model has to place two different
tier structures in the same state. Table~\ref{tab:sim_designs} lists the
eight settings.

\begin{table}[htbp]
\centering
\caption{Simulation designs. States with radius 100, 200 and 400 use
$\boldsymbol u_{1/2}$, $\boldsymbol u_1$ and $\boldsymbol u_2$, except that
in the heterogeneous settings the radius-200 state alternates between
$\boldsymbol u_1$ and $\boldsymbol v$, starting with $\boldsymbol u_1$.
$k^m$ denotes $m$ consecutive seasons in state $k$, and $(\cdot)^{\times 2}$
a sequence repeated twice.}
\label{tab:sim_designs}
\begin{tabular}{@{}cllll@{}}
\toprule
$K$ & Radius-200 vector & Radii & State sequence, $T=20$ & State sequence, $T=40$ \\
\midrule
2 & $\boldsymbol u_1$ & 200, 400 & $(1^5,2^5,1^5,2^5)$ & $(1^{10},2^{10},1^{10},2^{10})$ \\
3 & $\boldsymbol u_1$ & 100, 200, 400 & $(1^5,2^5,3^5,2^5)$ & $(1^5,2^5,3^5,2^5)^{\times 2}$ \\
2 & $\boldsymbol u_1$, $\boldsymbol v$ alternating & 200, 400 & $(1^5,2^5,1^5,2^5)$ & $(1^5,2^5,1^5,2^5)^{\times 2}$ \\
3 & $\boldsymbol u_1$, $\boldsymbol v$ alternating & 100, 200, 400 & $(1^5,2^5,3^5,2^5)$ & $(1^5,2^5,3^5,2^5)^{\times 2}$ \\
\bottomrule
\end{tabular}
\end{table}

In every design a state persists for several seasons and recurs after other
states have intervened. One comparison is not one-factor: at $T=40$ the
two-state heterogeneous design switches state every five seasons, whereas
the two-state baseline switches every ten, so it is a more dynamic design
rather than the baseline with only the ability patterns changed. The model
was fitted with the sampler and posterior summaries of
Section~\ref{sec:inference} and Appendix~\ref{subsec:post_mcmc}, with the
hyperparameters, allocation tuning parameters and summary rules of
Appendix~\ref{app:sim_details} held fixed across replicates.
\subsection{Simulation Results}
\label{subsec:simulation_results}

\begin{figure}[H]
    \centering
    \includegraphics[width=\textwidth]{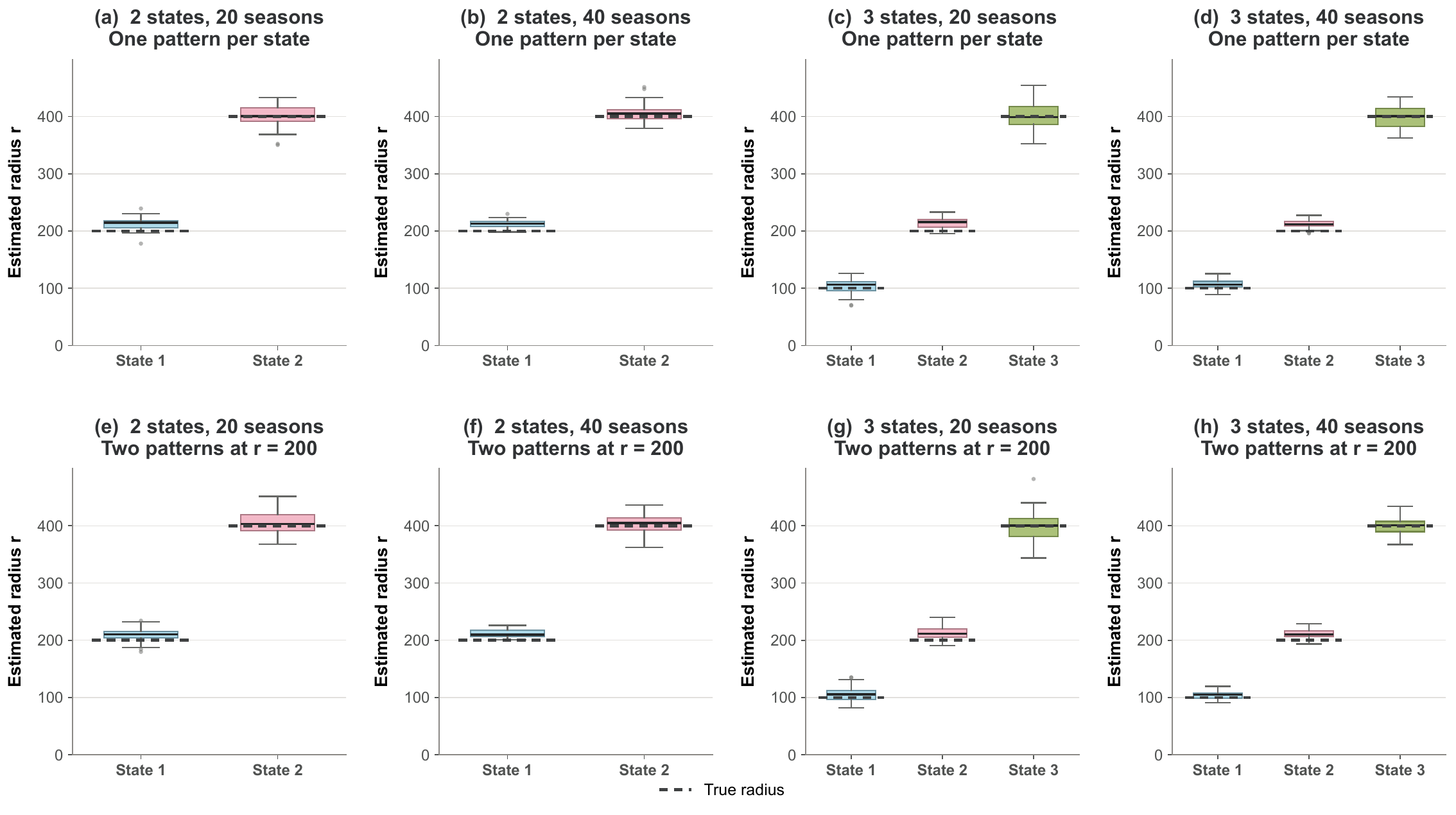}
    \caption{Radius estimates across the eight simulation settings. Dashed
    horizontal segments mark the true radii.}
    \label{fig:radius_recovery}
\end{figure}

Table~\ref{tab:simulation_results} reports means over the 50 replicates of
each setting, with Monte Carlo standard errors in parentheses, and
Figure~\ref{fig:radius_recovery} shows the radius estimates.

\begin{table}[H]
\centering
\caption{Simulation results over 50 independent replicates. $K$ is the true
number of states, $T$ the number of seasons, and $P$ the number of ability
patterns per state ($P=1$: baseline settings; $P=2$: heterogeneous
settings).}
\label{tab:simulation_results}
\setlength{\tabcolsep}{5pt}
\begin{tabular}{@{}ccc ccccc@{}}
\toprule
$K$ & $T$ & $P$ & $K$ recovery & State ARI & $\theta$-MSE & Class ARI & $\eta$-MSE \\
\midrule
2 & 20 & 1 & 0.98 & 0.9901 (0.0059) & 0.0260 (0.0014) & 0.9414 (0.0032) & 0.0154 (0.0006) \\
2 & 40 & 1 & 0.98 & 0.9990 (0.0010) & 0.0214 (0.0006) & 0.9441 (0.0019) & 0.0156 (0.0005) \\
2 & 20 & 2 & 0.92 & 0.9891 (0.0060) & 0.0434 (0.0014) & 0.9039 (0.0037) & 0.0155 (0.0006) \\
2 & 40 & 2 & 0.98 & 0.9970 (0.0022) & 0.0432 (0.0010) & 0.9027 (0.0025) & 0.0157 (0.0005) \\
3 & 20 & 1 & 0.90 & 0.9879 (0.0050) & 0.0340 (0.0012) & 0.8858 (0.0040) & 0.0152 (0.0008) \\
3 & 40 & 1 & 0.84 & 0.9894 (0.0036) & 0.0325 (0.0009) & 0.8847 (0.0031) & 0.0147 (0.0005) \\
3 & 20 & 2 & 0.90 & 0.9852 (0.0062) & 0.0549 (0.0018) & 0.8405 (0.0053) & 0.0146 (0.0008) \\
3 & 40 & 2 & 0.94 & 0.9905 (0.0041) & 0.0538 (0.0012) & 0.8404 (0.0039) & 0.0128 (0.0004) \\
\bottomrule
\end{tabular}
\end{table}

The state path is recovered well in every setting. Mean state ARI lies
between $0.985$ and $0.999$, and in each of the four pairs of designs it is
higher with 40 seasons than with 20. The number of states is recovered in
92\% to 98\% of two-state replicates and in 84\% to 94\% of three-state
replicates. Every miss overestimates the number of states by one, and the
extra state is small: it holds a single season in 22 of the 28 misses and at
most two seasons in 26, usually at the start or end of a block. In these
replicates the posterior is split between the two values, with typically
about a third of its mass on the true number. A longer series does not
reliably remove such misses, since recovery rises from 92\% to 98\% in the
two-state heterogeneous design but falls from 90\% to 84\% in the three-state
baseline.

Abilities and tiers are recovered closely. $\theta$-MSE is between $0.021$
and $0.055$ and is slightly lower at $T=40$ than at $T=20$ in every pair.
Class ARI ranges from $0.903$ to $0.944$ with two states and from $0.840$ to
$0.886$ with three. It is lowest in the three-state heterogeneous designs,
where the radius-200 state alternates between a two-tier and a three-tier
configuration. The alternation costs tier recovery but not regime recovery:
mean state ARI is $0.985$ to $0.997$ in the heterogeneous designs and
$0.988$ to $0.999$ in the corresponding baselines, so seasons that share a
radius but not a tier structure are almost always placed in the same state.

The radius estimates in Figure~\ref{fig:radius_recovery} lie close to the
truth, with the mean Dahl estimate within 8\% of the true radius in every
setting, and the radii of different states are clearly separated. Home
effects are estimated equally well throughout, with mean squared error
between $0.0128$ and $0.0157$, and the main error of the model across all
eight designs is an occasional extra state that absorbs one or two seasons.

\subsection{Robustness to Missing Games}
\label{subsec:missing_games}

To examine robustness to incomplete match records, we extended the baseline
two-state setting with \(T=20\) seasons by independently retaining each
scheduled match with probability $p\in\{1.00,0.90,0.80,0.70,0.60\}$.
The resulting missingness mechanism is fixture-level missing completely at
random (MCAR). The complete double round-robin schedule contains 380 matches
per season, so the five settings retain approximately 380, 342, 304, 266 and
228 matches per season. A new retention mask is generated independently for
every replicate and retention level. All other data-generating parameters,
prior specifications, MCMC settings and posterior-summary rules are those of
the complete-data two-state \(T=20\) setting, and each setting has 50
independent replicates.

\begin{figure}[htbp]
    \centering
    \includegraphics[width=0.8\textwidth]{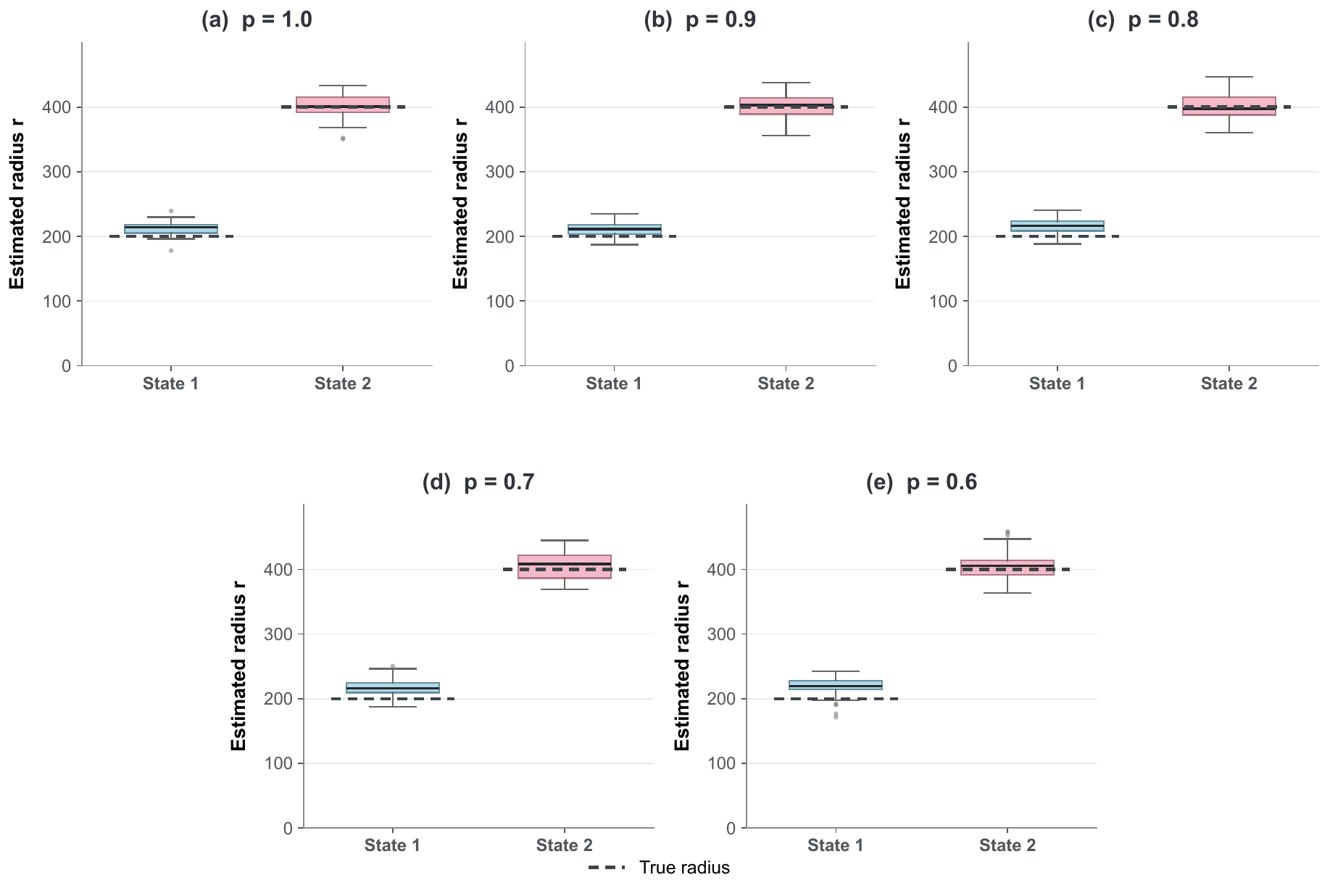}
    \caption{Recovery of the competitive-balance radius under fixture-level MCAR missingness. Each boxplot summarizes state-specific radius estimates across 50 independent replicates, using the median across seasons within each replicate and state. Dashed segments mark the true radii, \(r_1=200\) and \(r_2=400\).}
    \label{fig:missing_data_radius}
\end{figure}

Table~\ref{tab:missing_data_results} reports the Monte Carlo means, with one
row for each retention probability. The reduction column gives the expected
percentage of scheduled matches removed, \(100(1-p)\%\). For \(K\)
recovery, State ARI and Class ARI, the values in parentheses are absolute
differences from the complete-data baseline; for the MSE measures they are
percentage changes relative to the baseline. Radius recovery is shown in
Figure~\ref{fig:missing_data_radius}.

\begin{table}[htbp]
\centering
\caption{Robustness to fixture-level MCAR missingness over 50 independent
replicates. Values in parentheses give changes relative to the complete-data
baseline. Changes in recovery and ARI are absolute differences, whereas
changes in MSE are relative percentages.}
\label{tab:missing_data_results}
\small
\setlength{\tabcolsep}{3.5pt}
\renewcommand{\arraystretch}{1.45}
\begin{tabular}{@{}ccccccc@{}}
\toprule
Retention \(p\)
& Reduction
& \(K\) recovery
& State ARI
& \(\theta\)-MSE
& \(\eta\)-MSE
& Class ARI \\
\midrule
1.00
& 0\%
& 0.980
& 0.9901
& 0.0260
& 0.0154
& 0.9414 \\

0.90
& 10\%
& \shortstack{1.000\\(+0.020)}
& \shortstack{0.9880\\(-0.0021)}
& \shortstack{0.0310\\(+19.0\%)}
& \shortstack{0.0169\\(+10.3\%)}
& \shortstack{0.9355\\(-0.0059)} \\

0.80
& 20\%
& \shortstack{1.000\\(+0.020)}
& \shortstack{0.9920\\(+0.0019)}
& \shortstack{0.0401\\(+54.0\%)}
& \shortstack{0.0191\\(+24.3\%)}
& \shortstack{0.9245\\(-0.0170)} \\

0.70
& 30\%
& \shortstack{0.980\\(0.000)}
& \shortstack{0.9861\\(-0.0040)}
& \shortstack{0.0516\\(+98.1\%)}
& \shortstack{0.0220\\(+43.4\%)}
& \shortstack{0.9114\\(-0.0300)} \\

0.60
& 40\%
& \shortstack{0.980\\(0.000)}
& \shortstack{0.9828\\(-0.0072)}
& \shortstack{0.0736\\(+182.7\%)}
& \shortstack{0.0260\\(+69.3\%)}
& \shortstack{0.8859\\(-0.0555)} \\
\bottomrule
\end{tabular}
\end{table}

Removing matches has almost no effect on the regime structure. The number
of states is recovered in every replicate at $p=0.9$ and $p=0.8$ and in 49
of 50 at the other three retention levels, and mean state ARI stays above
$0.98$ even with 40\% of the matches removed. The misses at $p=0.7$ and
$p=0.6$ add one state, with one and four seasons respectively. Each season
still has about 228 matches at $p=0.6$, and the states persist for five
seasons at a time, so there is enough information to place each season. The
small changes in the recovery of the number of states across retention
levels are within Monte Carlo error.

The loss of information falls on the parameters that are estimated season
by season. Relative to the complete data, $\theta$-MSE rises by 19\% at
$p=0.9$, 54\% at $p=0.8$, 98\% at $p=0.7$ and 183\% at $p=0.6$. Over the
same range, mean class ARI drops from $0.941$ to $0.886$, and $\eta$-MSE
rises by 69\% at the lowest retention level.

Figure~\ref{fig:missing_data_radius} shows the same pattern for the radius.
The estimates stay near their true values, but they spread out as matches
are removed: the central 95\% of the radius-200 estimates spans 33 units
with complete data and 63 units at $p=0.6$.

Under independent fixture-level missingness of up to 40\% in this dense
double round robin, the model still identifies the regimes, while the
abilities, tiers and radii become less precise. These conclusions apply to
matches missing at random and not to missing teams, missing seasons or
missing connected parts of the schedule.
\subsection{Calibration and Sampling Diagnostics}
\label{subsec:simulation_calibration}

Table~\ref{tab:simulation_calibration} compares posterior uncertainty with
the truth in the eight main settings and in the four missing-data settings.
Equal-tailed 95\% credible intervals cover the true home effect in 93\% to
96\% of season replicates and the true team abilities in 97\% to 99\%. Because the allocation kernel is approximate, the regime frequencies below
are not exact posterior probabilities, and this table checks their
calibration directly. The smallest set of values of $K$ holding 95\% of the
draws contains the true number in at least 98\% of replicates, and on
average 0.77 to 0.95 of the draws have the true number. The frequency
with which two seasons share a state averages 0.96 to 0.99 for pairs
that truly share a state and at most 0.017 for pairs that do not. Grouping
the season pairs of the eight main settings by this probability, every pair
with probability above 0.95 truly shares a state and no pair below 0.05
does, and in each intermediate group the observed frequency is at least the
average probability.

\begin{table}[htbp]
\centering
\caption{Calibration over 50 replicates per setting; the bottom rows are
the missing-data settings with retention probability $p$. Columns 4--5:
coverage of equal-tailed 95\% credible intervals for the home effects and
team abilities. Column 6: proportion of replicates whose 95\% set of values of $K$
contains the true $K$. Column 7: mean frequency of the true $K$ in the
draws. Columns 8--9: mean frequency with which two seasons share a
state, for pairs in the same true state and in different true states.}
\label{tab:simulation_calibration}
\small
\setlength{\tabcolsep}{4pt}
\begin{tabular}{@{}ccc cccc cc@{}}
\toprule
$K$ & $T$ & $P$ & $\eta$ & $\theta$ & $K$ set & $\Pr(K_{\mathrm{true}})$
& Same state & Different states \\
\midrule
2 & 20 & 1 & 0.95 & 0.99 & 1.00 & 0.90 & 0.978 & 0.004 \\
2 & 40 & 1 & 0.95 & 0.99 & 1.00 & 0.95 & 0.994 & 0.000 \\
2 & 20 & 2 & 0.95 & 0.98 & 1.00 & 0.86 & 0.974 & 0.001 \\
2 & 40 & 2 & 0.95 & 0.98 & 1.00 & 0.89 & 0.981 & 0.001 \\
3 & 20 & 1 & 0.94 & 0.98 & 1.00 & 0.82 & 0.974 & 0.004 \\
3 & 40 & 1 & 0.93 & 0.99 & 0.98 & 0.77 & 0.980 & 0.003 \\
3 & 20 & 2 & 0.93 & 0.97 & 1.00 & 0.83 & 0.976 & 0.005 \\
3 & 40 & 2 & 0.96 & 0.97 & 0.98 & 0.80 & 0.975 & 0.003 \\
\midrule
\multicolumn{3}{c}{$p=0.9$} & 0.95 & 0.99 & 1.00 & 0.91 & 0.977 & 0.005 \\
\multicolumn{3}{c}{$p=0.8$} & 0.94 & 0.98 & 1.00 & 0.91 & 0.977 & 0.005 \\
\multicolumn{3}{c}{$p=0.7$} & 0.95 & 0.99 & 1.00 & 0.91 & 0.973 & 0.010 \\
\multicolumn{3}{c}{$p=0.6$} & 0.94 & 0.98 & 1.00 & 0.88 & 0.959 & 0.017 \\
\bottomrule
\end{tabular}
\end{table}

Table~\ref{tab:simulation_sampling} summarizes the behaviour of the
sampler. The radius and home-effect proposals are accepted at rates of 0.18
to 0.35 and 0.86 to 0.90. The number of occupied states changes 98 to 259
times on average among the 5{,}000 retained draws, so the chains move
between values of $K$ rather than settling on one, and the frequency of
the true $K$ in the first and second halves of the retained
draws differs by at most 0.03. The state path reaches an adjusted Rand
index of 0.99 with the truth after 305 to 639 iterations on average and
after at most 4{,}325 iterations in any replicate, within the
5{,}000-iteration burn-in. Two teams in the same true tier have exactly
equal abilities in 54\% to 63\% of the draws, and two teams in different
tiers in 0.1\% to 5.4\%.

\begin{table}[htbp]
\centering
\caption{Sampling diagnostics over 50 replicates per setting; the bottom
rows are the missing-data settings with retention probability $p$.
Acceptance rates of the radius and home-effect updates; mean number of
changes in the number of occupied states among the retained draws;
frequency of the true $K$ in the first and second halves of the
retained draws; first iteration at which the state path reaches an adjusted
Rand index of 0.99, mean (maximum) over replicates; and the proportion of
draws in which two teams have exactly equal abilities, for pairs in the
same true tier and in different true tiers.}
\label{tab:simulation_sampling}
\small
\setlength{\tabcolsep}{4pt}
\begin{tabular}{@{}ccc cc c c c cc@{}}
\toprule
$K$ & $T$ & $P$ & $r$ acc. & $\eta$ acc. & $K$ changes
& $\Pr(K_{\mathrm{true}})$: halves & ARI $\ge 0.99$
& Tie, same & Tie, different \\
\midrule
2 & 20 & 1 & 0.25 & 0.87 & 133 & 0.90 / 0.90 & 410 (904) & 0.63 & 0.001 \\
2 & 40 & 1 & 0.18 & 0.87 & 98 & 0.95 / 0.95 & 487 (1113) & 0.63 & 0.001 \\
2 & 20 & 2 & 0.26 & 0.87 & 149 & 0.86 / 0.87 & 457 (989) & 0.60 & 0.016 \\
2 & 40 & 2 & 0.19 & 0.87 & 115 & 0.90 / 0.89 & 520 (1099) & 0.60 & 0.017 \\
3 & 20 & 1 & 0.35 & 0.86 & 259 & 0.80 / 0.83 & 418 (2165) & 0.63 & 0.040 \\
3 & 40 & 1 & 0.28 & 0.86 & 237 & 0.77 / 0.76 & 464 (1058) & 0.63 & 0.039 \\
3 & 20 & 2 & 0.35 & 0.86 & 248 & 0.83 / 0.83 & 441 (1386) & 0.60 & 0.054 \\
3 & 40 & 2 & 0.28 & 0.86 & 232 & 0.79 / 0.81 & 639 (1507) & 0.60 & 0.053 \\
\midrule
\multicolumn{3}{c}{$p=0.9$} & 0.26 & 0.88 & 135 & 0.91 / 0.91 & 463 (4325) & 0.61 & 0.002 \\
\multicolumn{3}{c}{$p=0.8$} & 0.27 & 0.89 & 149 & 0.91 / 0.91 & 381 (838) & 0.59 & 0.005 \\
\multicolumn{3}{c}{$p=0.7$} & 0.29 & 0.89 & 154 & 0.91 / 0.90 & 327 (773) & 0.57 & 0.008 \\
\multicolumn{3}{c}{$p=0.6$} & 0.30 & 0.90 & 174 & 0.87 / 0.88 & 305 (724) & 0.54 & 0.013 \\
\bottomrule
\end{tabular}
\end{table}

\section{Implementation Details for the Simulation Study}
\label{app:sim_details}

\subsection{Prior Specification and Computation}
\label{subsec:simulation_computation}

All twelve simulation settings use the same prior and sampler settings. The
auxiliary abilities have prior $\beta_t\sim N(0,\sigma_\beta^2I_N)$ with
$\sigma_\beta=6$, each home effect has an independent $N(0,1)$ prior, and the
state-specific radii follow
\(r_k\mid\alpha_r\sim\operatorname{Exponential}(\alpha_r)\) with
\(\alpha_r\sim\operatorname{Gamma}(a_r,b_r)\) in the shape--rate
parameterization and $(a_r,b_r)=(10,200)$. The threshold is held at its
true value $\kappa_2=0.30$. The latent state sequence is updated with the
allocation rule of Section~\ref{subsec:adaptive_allocation}, with
$(a_{\mathrm{tr}},\gamma,\lambda)=(2.5,0.5,0.5)$, and new states are proposed
with the $J=15$ point radius grid \eqref{eq:radius_quantile_grid}; occupied
radii are then updated by the log-scale random walk
\eqref{eq:radius_mh_ratio} and are not restricted to grid values. The
proposal standard deviations are $\varsigma_r=0.16$ for $\log r_k$ and
$\varsigma_\eta=0.05$ for $\eta_t$.

Each chain starts from a single state with radius $1/\alpha_r$, where
$\alpha_r=a_r/b_r$, auxiliary abilities drawn from $N(0,0.1^2)$ and home
effects equal to zero. It runs for 10{,}000 iterations, and the first 5{,}000 are discarded as burn-in. With eight replicates run in parallel on an Apple M5 laptop, one replicate took about 15 minutes with $T=20$ and about 31 minutes with $T=40$, and the 600 replicates of the twelve settings took about 26 hours in total. The data of replicate $\ell$ are generated with
seed $s_0+10{,}000\,\ell$ and the chain is run with seed $s_0+10{,}000\,\ell+777$,
where $s_0=2026$ for the two-state settings, $s_0=20261001$ for the
three-state settings except the heterogeneous design with $T=40$, and
$s_0=20261201$ for that design. In the missing-data study the retention mask
uses seed $s_0+10{,}000\,\ell+500{,}000{,}000+1000p$ and the chain seed is
increased by $1{,}000{,}000\times100p$.

\subsection{Posterior Summaries}
\label{subsec:simulation_summaries}

The point estimates are those of Appendix~\ref{subsec:post_mcmc}: the
Dahl-draw state path and radii (which are the radii plotted in
Figure~\ref{fig:radius_recovery}), the SALSO tiers and tier-averaged
abilities \eqref{eq:tier_averaged_ability}, and the posterior-mean home
effects. SALSO is run with the R package \texttt{salso} using 64 random
initializations and a fixed seed. The $\theta$-MSE is computed from the
tier-averaged abilities, and class ARI is computed between the estimated and
true team partitions separately for each season and then averaged across
seasons within each replicate.

\section{National Basketball Association (NBA) (2004/05 to 2023/24)}
\label{app:nba_application}

The second application uses regular-season games from the 20 NBA seasons
from 2004/05 to 2023/24, with each season a comparison graph on the 30
franchises. Unlike the Premier League, the NBA is a closed league that
regulates parity directly through the draft, the salary cap and the luxury
tax. The study period begins with the 2004/05 expansion to 30 teams, so the
league size is the same in every season, and it includes the 2019 draft
lottery reform, which flattened the lottery odds of the worst teams to
discourage tanking, and the play-in tournament introduced in 2020.

NBA games cannot end in a tie, so we classify each game by the final margin,
the home score minus the away score: a home win if it exceeds five points, a
home loss if it is below $-5$, and a close game, treated as a draw, otherwise.
This allows the same three-category ordered-logit likelihood to be used in
both applications. The schedule is unbalanced, and a pair of franchises can
meet up to four times, or five for seven pairs in 2023/24, when in-season
tournament knockout games counted toward the regular season. Each meeting enters the likelihood separately, but the
graph that defines the radius constraint remains the unweighted complete
graph on the 30 franchises, with $\binom{30}{2}=435$ pairs, including
the five pairs that did not meet in the shortened 2019/20 season; the
results of Section~\ref{sec:geometric_framework} require only that this
graph be connected. The data contain
24,038 games. A full season has 1,230, but fewer were played in the
lockout-shortened 2011/12 season, in 2012/13, when one game was cancelled
and not replayed, and in the COVID-19-affected 2019/20 and 2020/21 seasons,
and we use only the games that were played. Games at the neutral 2020
Orlando venue keep their designated home team, so $h_{tg}=1$ for every game,
as in Section~\ref{subsec:projection_model}. Franchise labels
are harmonized across relocations and renamings, such as Seattle and Oklahoma
City or New Jersey and Brooklyn.

The model, MCMC settings, threshold $\kappa_2=0.65$, home-effect priors,
radius grid and posterior summaries are those of the Premier League analysis
in Section~\ref{sec:applications}. The radius hyperparameters were calibrated
separately with the same pilot chains and conditional WAIC, giving
$(\sigma_\beta,a_r,b_r)=(2,10,350)$. Convergence diagnostics are in
Appendix~\ref{app:realdata_diagnostics}.

\begin{figure}[H]
    \centering
    \includegraphics[width=\textwidth]{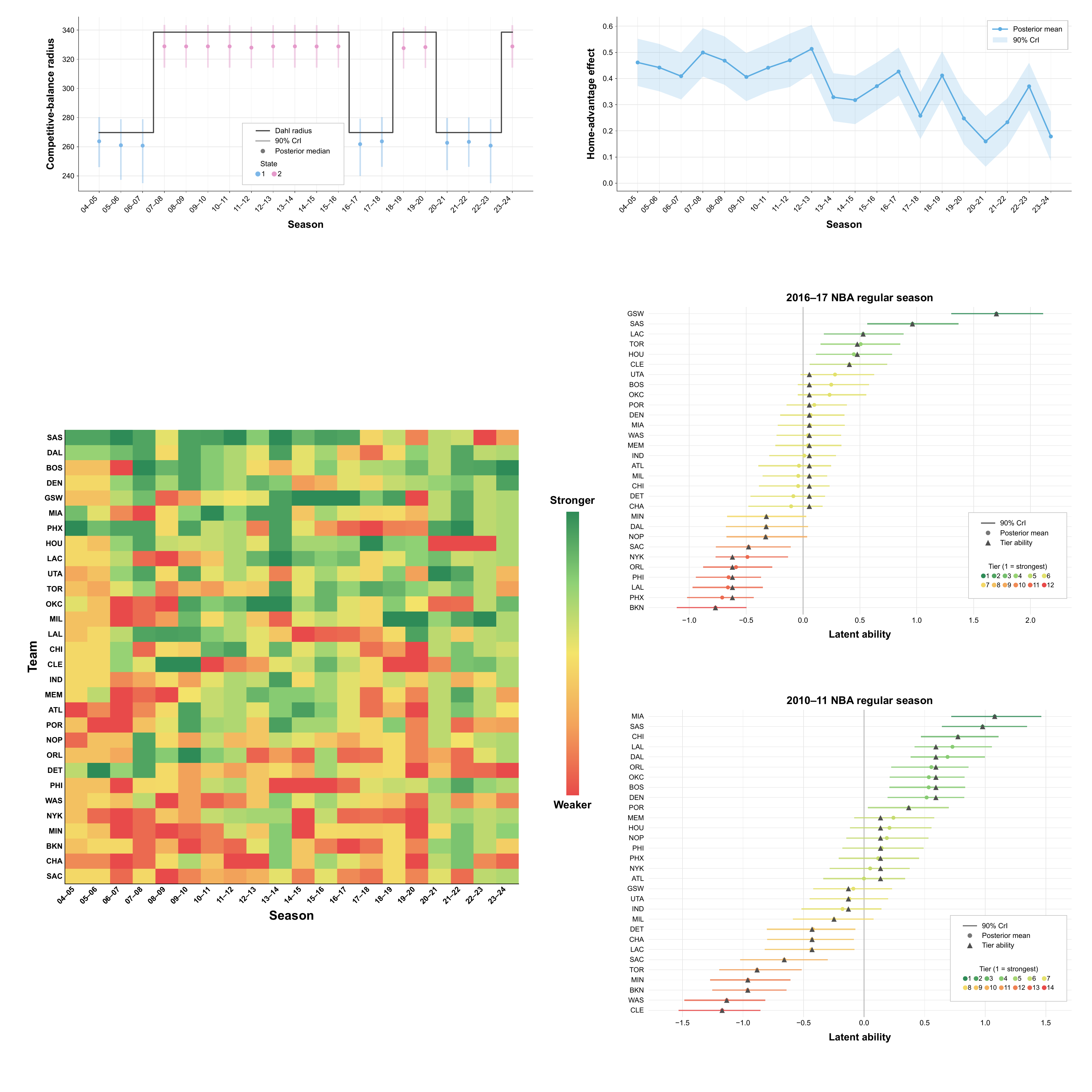}
    \caption{Posterior summaries for the NBA, 2004/05 to 2023/24. Upper
    panels: season-specific competitive-balance radius and home-advantage
    effect with 90\% credible intervals. Lower left: longitudinal team
    tiers. Right: team abilities and tiers for 2016/17 (State~1) and
    2010/11 (State~2); circles are posterior means, segments 90\% credible
    intervals, and triangles tier-level estimates.}
    \label{fig:nba_results}
\end{figure}

The sampler favors two states, which are occupied in $87.7\%$ of retained
draws. The Dahl draw gives radii of $269.8$ for the more
balanced State~1 and $338.5$ for the more stratified State~2, about 25\%
larger. Unlike the Premier League, the NBA switches back and forth between
the two. The seasons 2004/05 to 2006/07 are in State~1, followed by nine
seasons in State~2 from 2007/08 to 2015/16. The league then returns to
State~1 in 2016/17 and 2017/18, to State~2 in 2018/19 and 2019/20, to
State~1 from 2020/21 to 2022/23, and to State~2 in 2023/24, so 12 of the 20
seasons are in the stratified state. Among draws with two states, the average
self-transition frequencies are $0.52$ for State~1 and $0.78$ for State~2,
so stratified spells tend to last longer than balanced ones.

Within seasons, the tiers need not follow the order of total wins, because
close games count as draws and abilities are estimated from all
opponent-specific results. In 2010/11 Chicago had the best record at
62--20, ahead of San Antonio at 61--21 and Miami at 58--24, but the model
places Miami alone in the top tier, San Antonio second and Chicago third.
Miami had 52 decisive wins, 20 draws and 10 decisive losses, against 45, 25
and 12 for Chicago, and Chicago's three wins over Miami came by three, four
and one point, so all three count as draws. Miami later beat Chicago 4--1 in
the Eastern Conference Finals, which was not used in estimation. The season
followed LeBron James's move from Cleveland to Miami: Cleveland fell from
61--21 to 19--63 and is in the bottom tier. With fourteen tiers in total,
the large State~2 radius reflects separation across the whole league rather
than the dominance of one team.

The lower-radius 2016/17 season shows the opposite case. Golden State, with
67 wins after signing Kevin Durant, is alone in the top tier and San Antonio
is second, but below them many teams are close: the model finds twelve
tiers, and 14 of the 30 teams share one middle tier. A single dominant team
therefore does not imply a large radius, which measures how widely
abilities are spread across the league. The same season also shows the role
of head-to-head results. Boston finished first in the Eastern Conference at
53--29, ahead of Cleveland at 51--31, yet Cleveland is in the fifth tier and
Boston in the sixth. Cleveland had 40 decisive wins and 20 draws, against 36
and 25 for Boston, and won three of their four meetings by six, six and 23
points, while Boston's four-point win counts as a draw. Cleveland went on to
beat Boston 4--1 in the Eastern Conference Finals.

Across seasons, San Antonio stays in strong tiers for much of the earlier
period, and Golden State rises to a lasting place near the top, drops
sharply in 2019/20 and then recovers. As in the Premier League, seasons in
the same state need not have the same leading teams.

The posterior mean home effect is positive in every season, ranging from
about $0.16$ to $0.51$ with an average of $0.37$, and is smallest in
2020/21, when attendance was restricted. This is a season-specific effect
in the observation model rather than a causal effect of crowds. Finally, a
season-level measure of dispersion allows the timing of parity reforms to
be compared with changes in balance. The 2019 lottery reform and the
play-in tournament were followed by three seasons in State~1, from 2020/21
to 2022/23, before a return to State~2 in 2023/24. This timing is only
descriptive and does not show that the reforms caused the shift.
\section{Computational Diagnostics for the Real-Data Analyses}
\label{app:realdata_diagnostics}

For each dataset, we ran two independently initialized chains for 50,000 iterations, discarding the first 5,000 as burn-in. The allocation parameters are those of the simulations, $(a_{\mathrm{tr}},\gamma,\lambda)=(2.5,0.5,0.5)$, with the $J=15$ point radius grid, and the proposal standard deviations are $\varsigma_r=0.20$ for $\log r_k$ and $\varsigma_\eta=0.16$ for $\eta_t$. Run in parallel on the same laptop, each chain took about 56 minutes for the Premier League and about 61 minutes for the NBA.
All diagnostics were computed using the 45,000 retained draws per chain.
We monitored the number of occupied states $K$, the season-assigned
radius $r_{s_t}$, the home-advantage effect $\eta_t$, the shared
radius rate $\alpha_r$, and all team-season ability components
$\theta_{ti}$. The diagnostics used the full posterior samples,
allowing the number and assignments of states to vary across draws.

\begin{figure}[H]
\centering
\makebox[\textwidth][c]{%
    \includegraphics[width=\paperwidth]{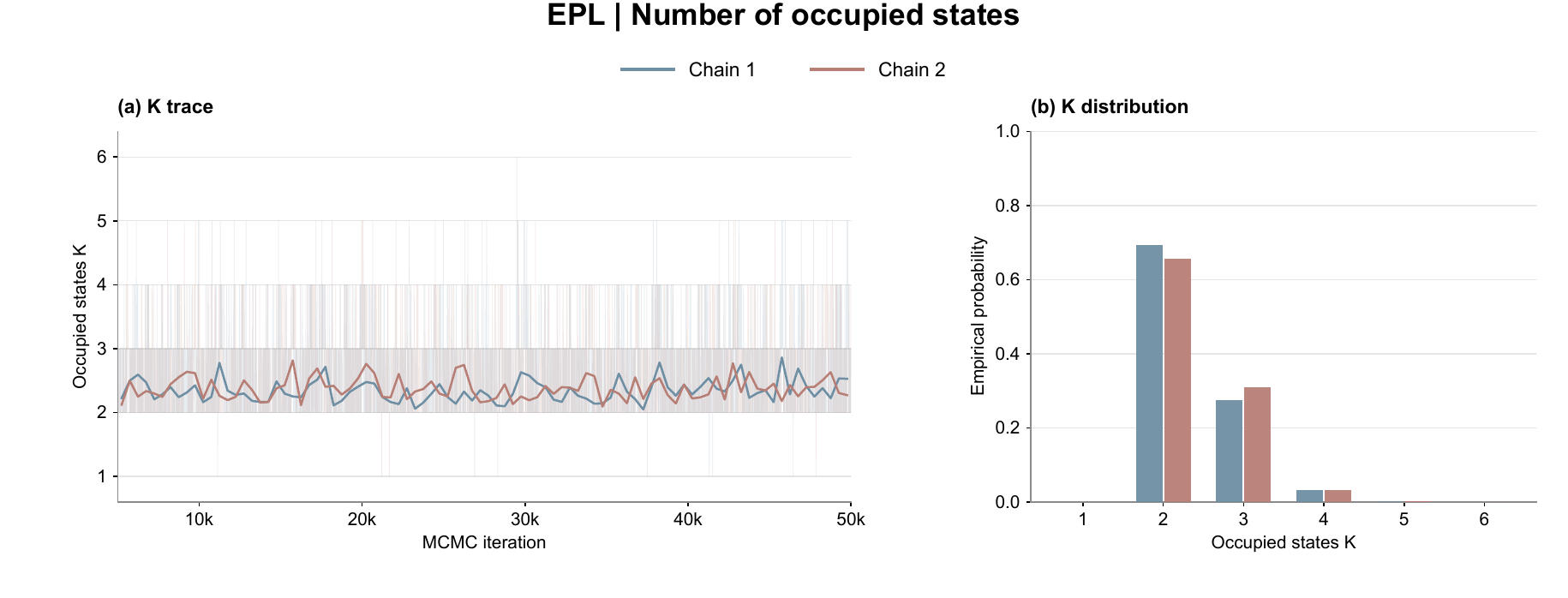}%
}
\caption{Diagnostics for the number of occupied states $K$ in the
Premier League. Left: retained traces and block means.
Right: empirical distributions normalized separately for each chain.}
\label{fig:epl_k_diagnostics}
\end{figure}

\begin{figure}[H]
\centering
\makebox[\textwidth][c]{%
    \includegraphics[width=\paperwidth]{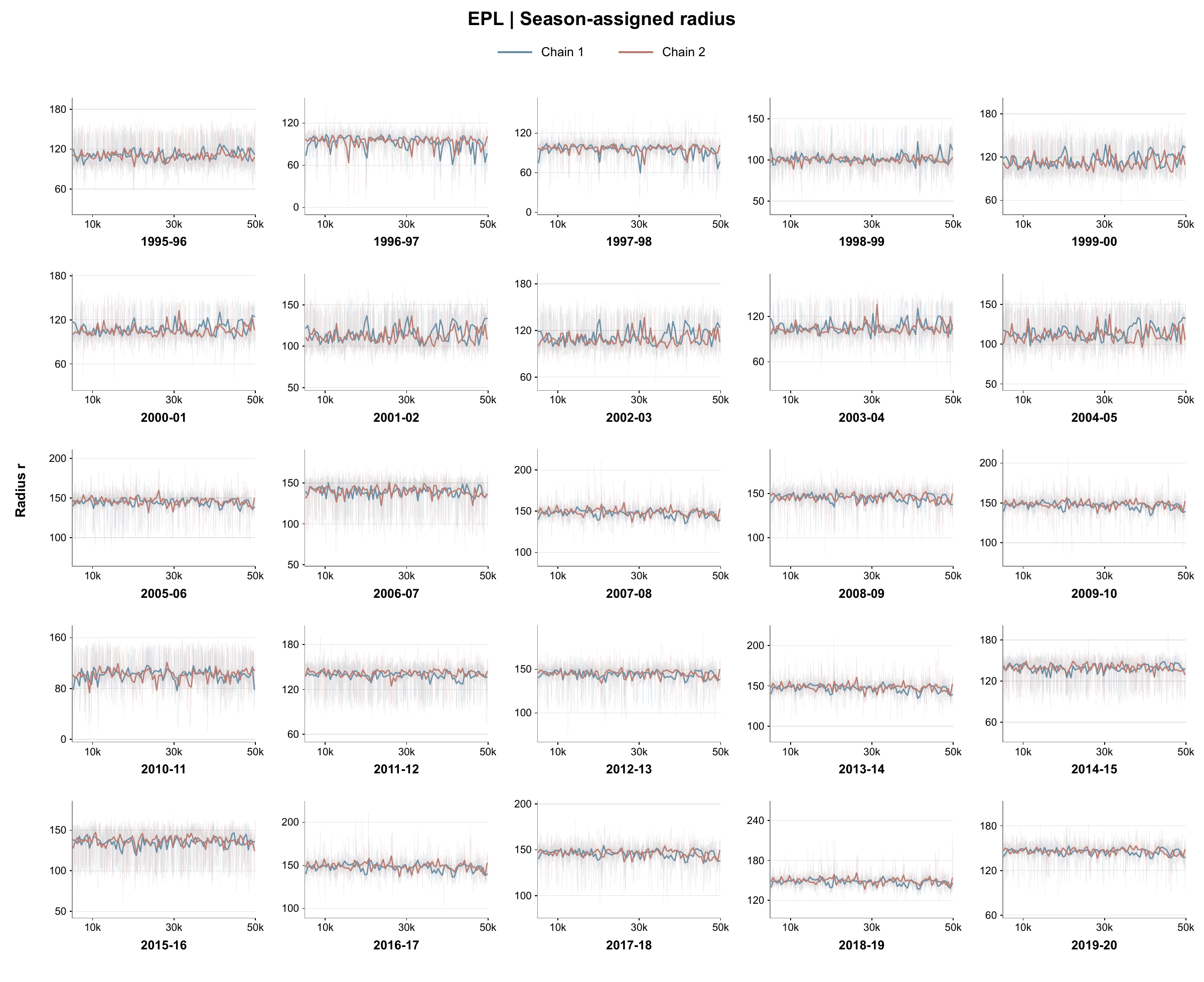}}
\caption{Season-assigned-radius traces for all 25 Premier League
seasons. The traces reflect variation in both state assignments
and radii across posterior draws.}
\label{fig:epl_radius_diagnostics}
\end{figure}

\begin{figure}[H]
\centering
\makebox[\textwidth][c]{%
    \includegraphics[width=\paperwidth]{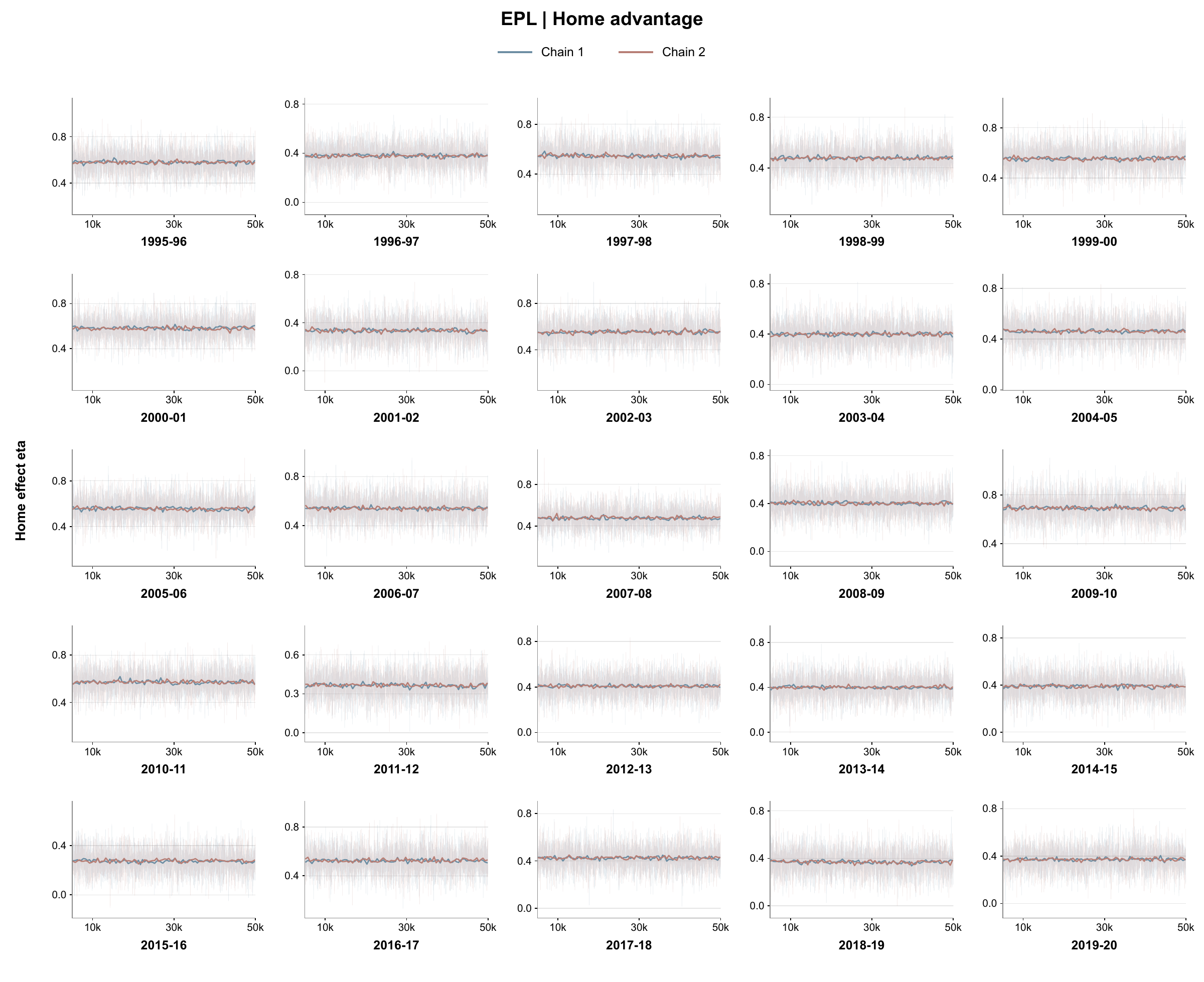}}
\caption{Home-advantage-effect traces for all 25 Premier League
seasons. Gray-blue and muted red indicate the two chains.}
\label{fig:epl_eta_diagnostics}
\end{figure}

\begin{figure}[H]
\centering
\makebox[\textwidth][c]{%
    \includegraphics[width=\paperwidth]{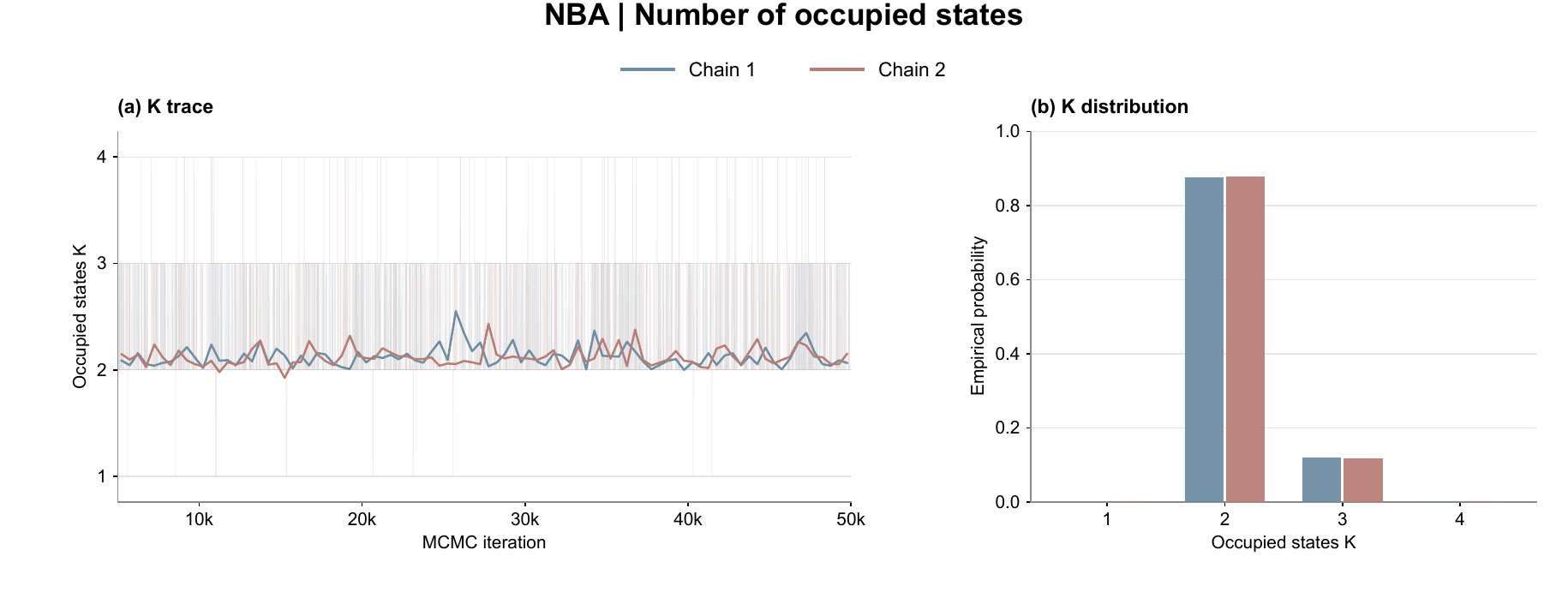}}
\caption{Diagnostics for the number of occupied states $K$ in the NBA.
Left: retained traces and block means.
Right: empirical distributions normalized separately for each chain.}
\label{fig:nba_k_diagnostics}
\end{figure}

\begin{figure}[H]
\centering
\makebox[\textwidth][c]{%
    \includegraphics[width=\paperwidth]{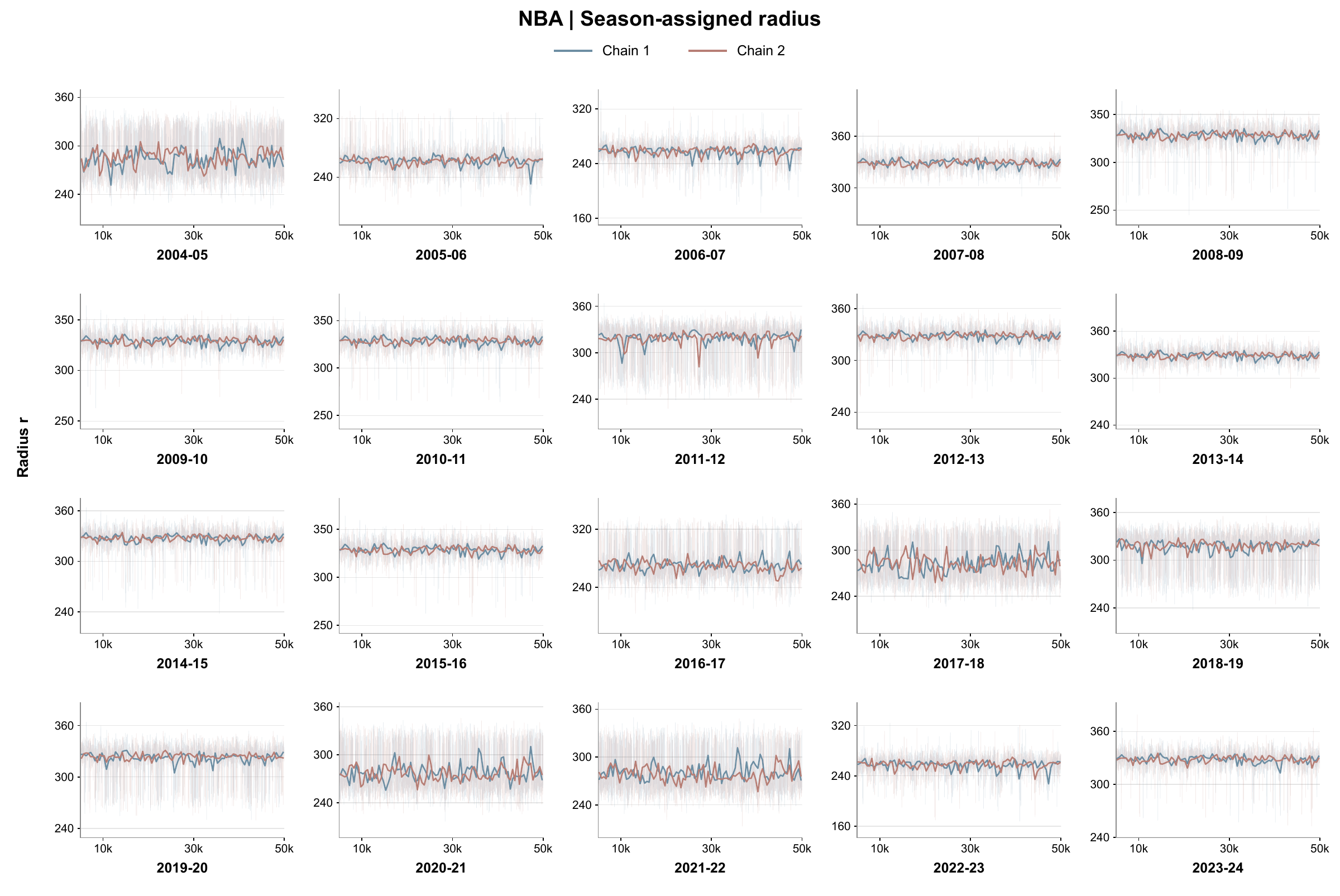}}
\caption{Season-assigned-radius traces for all 20 NBA seasons,
reflecting variation in both state assignments and radii.}
\label{fig:nba_radius_diagnostics}
\end{figure}

\begin{figure}[p]
\centering
\makebox[\textwidth][c]{%
    \includegraphics[width=\paperwidth]{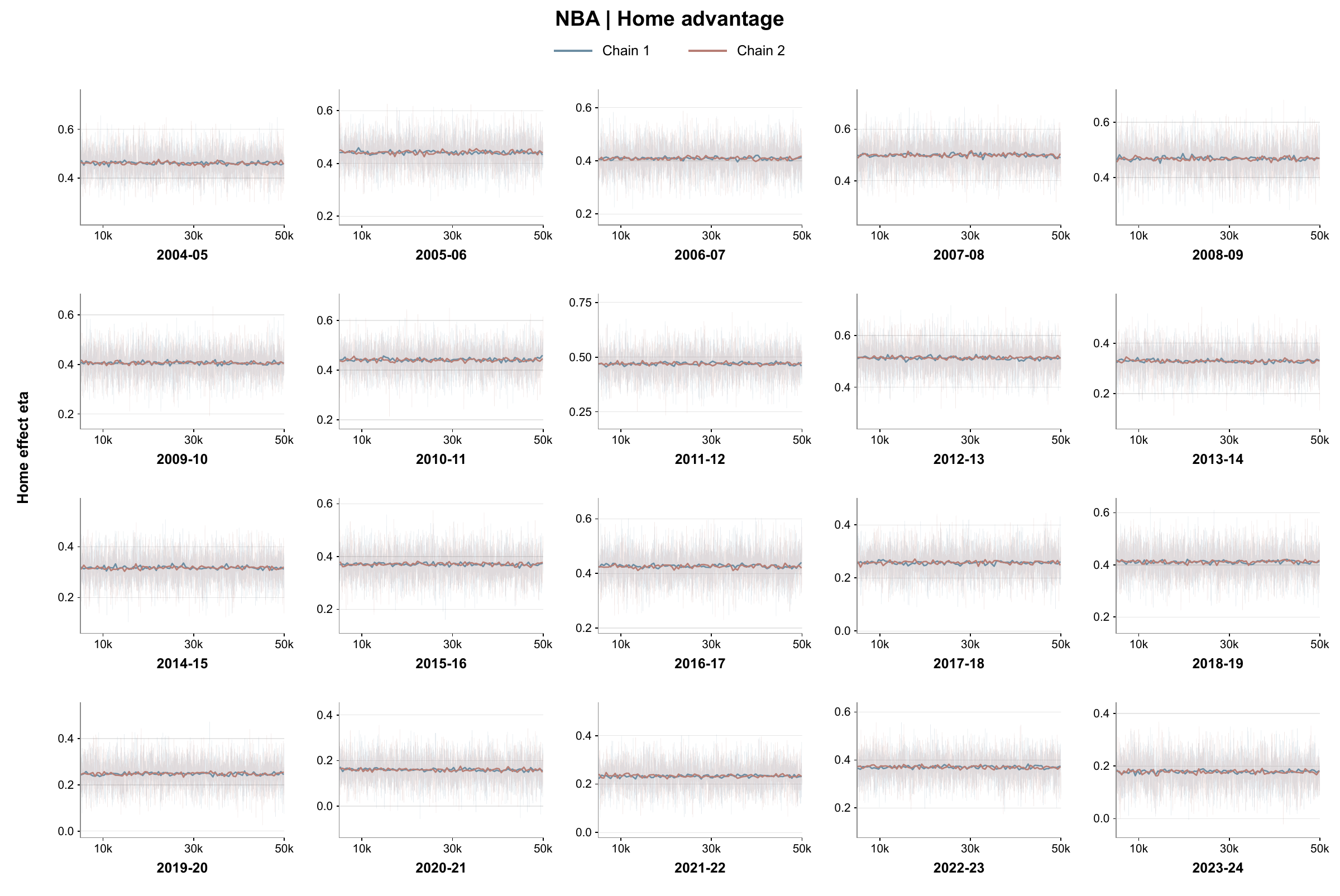}}
\caption{Home-advantage-effect traces for all 20 NBA seasons.
Gray-blue and muted red indicate the two chains.}
\label{fig:nba_eta_diagnostics}
\end{figure}

Figures~\ref{fig:epl_k_diagnostics}--\ref{fig:nba_eta_diagnostics}
show the traces and empirical distributions from both chains.
Faint lines represent individual draws, with every twentieth draw
shown for the radius and home-advantage traces to improve readability.
Solid lines represent means over non-overlapping blocks of 500 iterations.
All numerical diagnostics and empirical distributions of $K$ were
computed from the full, unthinned samples.

For the Premier League, $K=2$ was the most frequent outcome in both
chains, accounting for 69.2\% and 65.6\% of retained draws,
respectively. The proportions with $K=3$ were 27.4\% and 31.0\%.
For the NBA, the proportions with $K=2$ were 87.7\% and 87.8\%.
Thus, both datasets favored a two-state representation, with greater
uncertainty about a third state in the Premier League.
The rank-normalized split-$\widehat R$ values for $K$ were
1.0018 for the Premier League and 1.0009 for the NBA.

The home-advantage effects showed close agreement between chains,
with maximum $\widehat R$ values of 1.0007 for the Premier League
and 1.0004 for the NBA. For the season-assigned radii, the
corresponding maxima were 1.0134 and 1.0049. Four of the 25
Premier League radius components had $\widehat R>1.01$, compared
with none of the 20 NBA components. Among team abilities,
all 500 Premier League components had $\widehat R<1.01$
(maximum 1.0085), while 14 of the 600 NBA components exceeded
1.01 (maximum 1.0187). Across all monitored scalar quantities,
the minimum bulk and tail effective sample sizes were approximately
476 and 706 for the Premier League, and 231 and 214 for the NBA.

Overall, the diagnostics indicate good cross-chain agreement for
most monitored quantities. The relatively larger $\widehat R$
values were confined to a small number of Premier League radius
components and NBA team abilities.

\paragraph{Monte Carlo check of emission separation.}
Lemma~\ref{lem:small_radius_separation} establishes
Assumption~\ref{ass:emission_separation} only near $r=0$. As Monte Carlo
evidence at the fitted radii, and not as a proof of separation at all
radii, we evaluated
$\mathcal S_{\mathcal X}(r)$ under \eqref{eq:integrated_emission} at the
settings of both applications: the complete-graph constraint,
$\sigma_\beta=2$, $\eta\sim N(0,1)$ and $\kappa_2=0.65$. The home effect
was integrated by $60$-point Gauss--Hermite quadrature, and $\beta$ by
Monte Carlo with $20{,}000$ draws shared across radii. Under the
complete-graph constraint the pairwise contrasts are exchangeable, so
$\mathcal S_{\mathcal X}(r)=|\mathcal P|\,S_{\mathrm{pair}}(r)$, where
$|\mathcal P|=190$ in every Premier League season and $645$--$945$ in the
NBA. On grids of spacing $20$ over $[0,600]$ ($N=20$) and $46.7$ over
$[0,1400]$ ($N=30$), both including the fitted radii, every increment of
$S_{\mathrm{pair}}$ was positive, at least $4.9$ Monte Carlo standard errors
from zero. For the Premier League, $S_{\mathrm{pair}}=0.443$ at $r=97.1$
and $0.487$ at $r=155.4$, so the expected $\mathcal S_{\mathcal X}$ differs
by about $8$ between the two states, comparable to a binomial standard
deviation of about $7$ for a single season, so this statistic separates
the two states only weakly within any one season. For the NBA,
$S_{\mathrm{pair}}=0.456$ at $r=269.8$ and $0.478$ at $r=338.5$. The
increments shrink as the constraint stops binding: $\Pr(Z>r)$ falls below
$0.5$ near $r=430$ for the Premier League and near $r=980$ for the NBA,
whereas it exceeds $0.999$ at all four fitted radii.

\end{document}